\documentclass{aastex61}

\usepackage{graphicx}
\usepackage{amssymb}
\usepackage{amsmath}
\usepackage{epstopdf}
\usepackage{natbib}
\usepackage{float}
\usepackage{mathtools}

\begin{document}

\title{The Double Dust Envelopes of R Coronae Borealis Stars}
%The Cold Dust Envelopes of R Coronae Borealis Stars

\author{Edward J. Montiel}
\affiliation{Physics Department, University of California, Davis, CA 95616, USA; ejmontiel@ucdavis.edu}
\affiliation{Department of Physics \& Astronomy, Louisiana State University, Baton Rouge, LA 70803, USA}, 
\author{Geoffrey C. Clayton}
\affiliation{Department of Physics \& Astronomy, Louisiana State University, Baton Rouge, LA 70803, USA} 
\author{B.~E.~K. Sugerman}
\affiliation{Space Science Institute, 4750 Walnut St, Suite 205, Boulder, CO 80301}
\author{A. Evans}
\affiliation{Astrophysics Group, Lennard Jones Laboratory, Keele University, Keele, Staffordshire, ST5 5BG, UK}
\author{D.~A. Garcia-Hern{\' a}ndez}
\affiliation{Instituto de Astrof{\' i}sica de Canarias (IAC), E-38205 La Laguna, Tenerife, Spain}
\affiliation{Universidad de La Laguna (ULL), Departamento de Astrof{\' i}sica, E-38206, La Laguna, Tenerife, Spain}
\author{Kameswara Rao, N.}
\affiliation{Indian Institute of Astrophysics, Koramangala II Block, Bangalore, 560034, India}
\author{M. Matsuura}
\affiliation{School of Physics and Astronomy, Cardiff University, Queens Buildings, The Parade, Cardiff, CF24 3AA, UK}
\author{P. Tisserand}
\affiliation{Sorbonne Universit{\' e}s, UPMC Univ Paris 6 et CNRS, UMR 7095, Institut d'Astrophysique de Paris, 98 bis bd Arago, 75014 Paris, France}
%\altaffiltext{1}{Physics Department, University of California, Davis, CA 95616, USA; ejmontiel@ucdavis.edu}
%\altaffiltext{2}{Department of Physics \& Astronomy, Louisiana State University, Baton Rouge, LA 70803, USA}
%\altaffiltext{3}{Space Science Institute, 4750 Walnut St, Suite 205, Boulder, CO 80301}
%\altaffiltext{4}{Astrophysics Group, Lennard Jones Laboratory, Keele University, Keele, Staffordshire, ST5 5BG, UK}
%\altaffiltext{5}{Instituto de Astrof{\' i}sica de Canarias (IAC), E-38205 La Laguna, Tenerife, Spain}
%\altaffiltext{6}{Universidad de La Laguna (ULL), Departamento de Astrof{\' i}sica, E-38206, La Laguna, Tenerife, Spain}
%\altaffiltext{7}{Indian Institute of Astrophysics, Koramangala II Block, Bangalore, 560034, India}
%\altaffiltext{8}{School of Physics and Astronomy, Cardiff University, Queens Buildings, The Parade, Cardiff, CF24 3AA, UK}
%\altaffiltext{9}{Sorbonne Universit{\' e}s, UPMC Univ Paris 6 et CNRS, UMR 7095, Institut d'Astrophysique de Paris, 98 bis bd Arago, 75014 Paris, France}

\begin{abstract}
%Change for paper
The study of extended, cold dust envelopes surrounding R Coronae Borealis (RCB) stars began with their discovery by IRAS. RCB stars are carbon-rich supergiants characterized by their extreme hydrogen deficiency and for their irregular and spectacular declines in brightness (up to 9 mags). We have analyzed new and archival {\it Spitzer} Space Telescope and {\it Herschel} Space Observatory data of the envelopes of seven RCB stars to examine the morphology and investigate the origin of these dusty shells. {\it Herschel}, in particular, has revealed the first ever bow shock associated with an RCB star with its observations of SU~Tauri. These data have allowed the assembly of the most comprehensive spectral energy distributions (SEDs) of these stars with multi--wavelength data from the ultraviolet to the submillimeter. Radiative transfer modeling of the SEDs implies that the RCB stars in this sample are surrounded by an inner warm (up to 1,200 K) and an outer cold (up to 200 K) envelope. The outer shells are suggested to contain up to 10$^{-3}$ M$_\odot$ of dust and have existed for up to $10^5$ yr depending on the expansion rate of the dust. This age limit indicates that these structures have most likely been formed during the RCB phase.

\end{abstract}

%\keywords{editorials, notices --- 
%miscellaneous --- catalogs --- surveys}



\section{Introduction} \label{sec:intro}
R Coronae Borealis (RCB) stars provide an excellent opportunity to understand more about the advanced stages of stellar evolution \citep{1996PASP..108..225C,2012JAVSO..40..539C}. They form a rare class of hydrogen--poor, carbon--rich supergiants. Two formation scenarios have been proposed for their origin: the single degenerate final helium-shell flash (FF) model and the double degenerate (DD) white dwarf (WD) merger model \citep{1996ApJ...456..750I, 2002MNRAS.333..121S}. The latter involves the merger of a CO and a He WD \citep{1984ApJ...277..355W}, while the former takes the hot evolved central star of a planetary nebula (PN) and turns it into a cool supergiant \citep{Fujimoto:1977lr, 1979sss..meet..155R}.

The trademark behavior of RCB stars is their spectacular and irregular declines in brightness. These declines can take an RCB star up to 9 magnitudes fainter than its peak brightness, and are caused by the formation of discrete, thick clouds of carbon dust along the line of sight \citep{1935AN....254..151L, 1939ApJ....90..294O,1996PASP..108..225C}. All RCB stars show an infrared excess due to the presence of warm circumstellar material (CSM, \citealp[and references therein]{Feast:1997lr,2012JAVSO..40..539C}). Further, some RCB stars have been found to have cold, extended nebulosity \citep[e.g.,][]{Schaefer:1986lq,Walker:1985rr,1986ASSL..128..407W,2011MNRAS.414.1195B, 2011ApJ...743...44C}.  

The origins of this CSM material as well as the progenitors of the central RCB stars still remain shrouded in mystery. One important difference between the RCB stars formed in the two scenarios is that in the FF model, they would be surrounded by a fossil, neutral hydrogen-rich (HI-rich) PN shell \citep{Walker:1985rr, Gillett:1986cr, 1990MNRAS.247...91L, 1999ApJ...517L.143C, 2011ApJ...743...44C}. Three stars (Sakurai's Object, V605 Aquilae, and FG Sagittae) have been observed to undergo FF outbursts that transformed them from hot evolved stars into cool giants with spectroscopic properties similar to RCB stars \citep{1997AJ....114.2679C,1998ApJS..114..133G,1998A&A...332..651A,Asplund:1999bh,Asplund:2000qy,2006ApJ...646L..69C}. These FF stars are all surrounded by PNe which are still ionized. However, the cooler RCB central stars are no longer able to provide the needed ionizing radiation so the atoms in the shell have recombined. The velocity of the fossil PN shell would be similar to its ejection velocity, $\sim$ 20 -- 30 km s$^{-1}$. 

In the DD scenario, the stars may have had PN phases but they would have occurred so long ago, $\sim$10$^9$ years, that no structure resembling a fossil envelope would remain when the two WDs finally merge to form an RCB star. These shells could be material lost during the WD merger event, itself. This would have happened much more recently, $\lesssim$10$^4$ years ago, and would imply these structures are much less massive than previously estimated \citep{Gillett:1986cr, 2011ApJ...743...44C}. 

A third explanation for the observed shells is that they could have formed during the RCB phase. RCB stars are thought to produce dust at a rate of 10$^{-7}$ to 10$^{-6}$ M$_\odot$ yr$^{-1}$ \citep{2012JAVSO..40..539C}. \citet{2013AJ....146...23C} have found that newly forming clouds are propelled away from the central star at speeds up to 400 km s$^{-1}$. This also could result in the observed envelopes on a timescale of about 10$^4$ years.

We are now in an era where high spatial resolution and high sensitivity far-IR (FIR), submillimeter (sub-mm), and even radio observations exist of RCB stars in order to study their cold CSM material. We present unpublished {\it Spitzer} and {\it Herschel} observations of the RCB$/$HdC stars: MV Sagittarii (MV~Sgr), R Coronae Borealis (R~CrB), RY Sagittarii (RY~Sgr), SU Tauri (SU~Tau), UW Centauri (UW~Cen), V854 Centuari (V854~Cen), V Coronae Australis (V~CrA), and HD~173409. We have constructed multi--wavelength datasets ranging from the ultraviolet (UV) to sub-mm in order to better determine the mass, size, and morphology of the diffuse material surrounding these RCB stars. 

\section{Observations} \label{sec:obs}
We have combined multi--wavelength observations, which range from the ultraviolet (UV) to the submillimeter (sub-mm), in order to construct the most comprehensive spectral energy distributions (SEDs) of our sample RCB stars. SEDs for R~CrB and V605~Aql have been published previously in \citet{2011ApJ...743...44C} and \citet{2013ApJ...771..130C}, respectively. Stellar properties for our sample of RCB stars are presented in Table 1. Figures 1--4 show the light curves from the American Association of Variable Star Observers (AAVSO\footnote{https://www.aavso.org/data-download}) of our sample RCB stars with the epochs of the various observations that are included in our SED analysis marked.

%As seen in the AAVSO light curves (Figures 1--4) , we have made efforts to select to make sure that NIR observations and shorter were observed during maximum light. RCB stars are known to exhibit regular to semi--regular pulsations, $\delta$V $\lesssim$ 0.1 mag and periods of 40-100 days \citep{1990MNRAS.247...91L,Saio:2008qe}. This effect has minimal impact on our SED modeling. However, the IUE observations, despite being dereddened using CCM, are sensitive to small amounts of dust.

\subsection{Ultraviolet Spectra}
Many of the RCB stars were observed with the {\it International Ultraviolet Explorer} (IUE). Archival IUE data from the long wavelength spectrograph in the large aperture mode of MV~Sgr, UW~Cen, RY~Sgr, and V854~Cen were retrieved from the Barbara A. Mikulski Archive for Space Telescopes (MAST). The V854~Cen observation, LWP19951, was originally a part of the IUE program ``RCMBW'' (PI Barbara A. Whitney) and has been previously published in papers by \citet{1992ApJ...384L..19C} and \citet{1999AJ....117.3007L}. RY~Sgr, LWP30613, came from the IUE program ``HERAH'' (PI Albert V. Holm) and appeared in \citet{1999JAVSO..27..113H}. The IUE observation of MV~Sgr, LWR09008, came from Angelo Cassatella's program, AC414. The spectrum was included in a publication by \citet{1995A&A...299..135J}. Finally the observation for UW~Cen, LWR 13260, originated in a program by Aneurin Evans (EC228). This spectrum has not appeared in any refereed publications. All IUE spectra were corrected using the IDL routine CCM\_\_UNRED, which applies correction as described by \citet{1989ApJ...345..245C}. %{\bf the line of sight reddening}

\subsection{Optical Photometry}
RCB stars have been observed at all states between maximum and minimum light. Extensive ground-based monitoring in the optical was performed by multiple groups during the last century. Maximum light observations of SU~Tau and V~CrA were taken from \citet{1990MNRAS.247...91L}. SU~Tau was imaged in $BVR_{\rm C}I_{\rm C}$ photometric filters, while V~CrA only in $UBV$. RY~Sgr was observed near maximum light by \citet{1997MNRAS.285..358M}. They provide coverage with $UBVR_{\rm C}I_{\rm C}$ filters. Observations for MV~Sgr and UW~Cen at maximum light were retrieved from \citet{1990MNRAS.245..119G}. They observed both RCB stars with $UBVR_{\rm C}I_{\rm C}$ filters. Maximum light observations of V854~Cen were from \citet{Lawson:1989kx}. The observations were performed with $UBVR_{\rm C}I_{\rm C}$ filters. HD~173409 is the only hydrogen--deficient (HdC) star, which are spectroscopically similar to RCB stars but have neither been observed to have declines nor have an IR excess (see Section 5.8). Observations come from a monitoring campaign by \citet{1990SAAOC..14....1M}, who provide $UBVR_{\rm C}I_{\rm C}$ photometry. The photometry for our sample has been corrected for line of sight extinction by using the online\footnote{http://ned.ipac.caltech.edu/forms/calculator.html} extinction calculator provided by the NASA/IPAC Extragalatic Database (NED) and the method of \citet{2011ApJ...737..103S}.

\subsection{Near Infrared Photometry}
Ground--based monitoring campaigns in the near Infrared (NIR), in the $JHKLM$ bandpasses, have been conducted at a similar level to the optical. $JH$ observations are primarily of the stellar photosphere, so they follow the fluctuations between maximum and minimum light that distinguish the RCB class. $LM$ track the warm dust that has recently formed around an RCB star, even if the dust is not in the line of sight.

NIR photometry for UW~Cen also comes from \citet{1990MNRAS.245..119G}, who also provided $MN$ observations of MV~Sgr. These observations were taken simultaneously with their optical campaign described in the previous section.  $JHKLMN$ observations from \citet{1984MNRAS.208...25K} were used for MV~Sgr as well.  Long term NIR monitoring of HD~173409, RY~Sgr, SU~Tau, V854~Cen, and V~CrA were reported by \citet{Feast:1997lr} with photometry selected while the stars were at or near maximum light (see Figures 1--4). Additionally, photometry provided by the Two Micron All Sky Survey \citep[2MASS]{2006AJ....131.1163S} was used if the RCB star was at maximum light during observation by the survey. This applied to only two RCB stars in the sample, MV~Sgr and V854~Cen, as well as the HdC star, HD~173409.  The photometry was taken from the 2MASS Point Source Catalog \citep{2003tmc..book.....C}.

\subsection{Infrared Space Observatory} 
The Infrared Space Observatory (ISO) was a joint European Space Agency (ESA), Japanese Aerospace Exploration Agency (JAXA), and NASA mission launched November 17, 1995. One of its instruments, the Short Wave Spectrometer \citep[SWS][]{2003sws..bookR....L}, provided spectroscopy between 2.4 and 45~$\mu$m. Calibrated SWS spectra of RY~Sgr and R~CrB \citep{2003ApJS..147..379S} were retrieved from an ISO SWS science archive hosted by Gregory C. Sloan\footnote{https://isc.astro.cornell.edu/~sloan/library/swsatlas/aot1.html}.

\subsection{Spitzer Space Telescope}
%%For Nye -- PSF sizes
%%Spitzer/MIPS: 6, 18, & 38 arcsec (FWHM)
{\it Spitzer Space Telescope} ({\it Spitzer}, \citealp{2004ApJS..154....1W}) observations of RCB stars were acquired with all three instruments on board the satellite. These instruments are the Infrared Array Camera (IRAC, \citealp{2004ApJS..154...10F}), the Infrared Spectrograph (IRS, \citealp{Houck:2004lr}) and the Multiband Imaging Photometer for {\it Spitzer} (MIPS, \citealp{2004ApJS..154....1W}). IRAC was the NIR imager on {\it Spitzer} and provided simultaneous observations at 3.6, 4.5, 5.8, and 8.0 $\mu$m (central wavelengths). Only UW~Cen was observed with IRAC (PI A. Evans, ID 40061). 

IRS provided wavelength coverage in the range 5.3 to 38~$\mu$m with both low (R $\sim$ 90) and high (R $\sim$ 600) resolution (Houck et al. 2004). Archival IRS observations of RCB stars at both resolutions (PI D. Lambert, ID 50212) were previously published by \citet{2011ApJ...739...37G,2013ApJ...773..107G}. Low resolution IRS observations were retrieved from the Cornell Atlas of {\it Spitzer}/IRS Sources (CASSIS, \citealp{2011ApJS..196....8L}), which provides a standard reduction of all the sources observed with the IRS. This was performed using the Spectroscopy Modeling Analysis and Reduction Tool (SMART, \citealp{2004PASP..116..975H,2010PASP..122..231L}). 

MIPS was the FIR imager on {\it Spitzer} and observed at (central) wavelengths of 24, 70, and 160~$\mu$m with PSF full width at half maximum (FWHM) of 6\arcsec, 18\arcsec, and 40\arcsec, respectively. Archival MIPS observations of RCB stars come from two programs, PIs G. Clayton (ID 30029) and A. Evans (ID 3362). The raw data were processed using the MIPS DAT package \citep{2005PASP..117..503G}, which performs standard reductions for IR array detectors as well as MIPS specific routines. The output images were then calibrated according to the methods established by \citet{2007PASP..119..994E}, \citet{2007PASP..119.1019G}, and \citet{2007PASP..119.1038S} for the 24, 70, and 160~$\mu$m bands, respectively.

\subsection{Wide-field Infrared Survey Explorer}
The Wide-Field Infrared Survey Explorer (WISE, \citealp{2010AJ....140.1868W}) was a NASA medium-class explorer mission that was launched in December 2009. Its mission was to survey the entire sky over 10 months at 3.4, 4.5, 12, and 22~$\mu$m. Two catalogs of WISE sources were released in 2012 (WISE All-Sky, \citealp{2012wise.rept....1C}) and 2013 (ALLWISE, \citealp{2014yCat.2328....0C}), which encompass over 500 million and 700 million objects, respectively. The differences between the catalogs are detailed in \citet{2014yCat.2328....0C}. We have adopted the ALLWISE photometry for our SED analysis and this photometry can be found in the individual tables for our sample stars in Section 5. The WISE observations of R~CrB and V854~Cen are saturated, which makes the published photometry in both catalogs unreliable and not useable.

\subsection{AKARI}
AKARI was a JAXA satellite launched in February 2006 \citep{2007PASJ...59S.369M} and operated in two modes: an all-sky survey, similar to WISE, and a pointed mode for specific targets. It had two instruments: the Infrared Camera (IRC, \citealp{2007PASJ...59S.401O}) and the Far Infrared Surveyor (FIS, \citealp{2007PASJ...59S.389K}). The IRC contained three individual cameras observing at central wavelengths of 3.6, 9, and 18~$\mu$m. The FIS had two detectors arrays that enabled both wide and narrow band FIR imaging. The central wavelengths of the narrow band imaging were 65 and 160~$\mu$m, while for wide band imaging it was 90 and 140~$\mu$m. 

Two all-sky catalogs were released by the AKARI team. They are a MIR/IRC catalog \citep{2010A&A...514A...1I}, which published photometry at 9 and$/$or 18~$\mu$m for $\sim$870,000 individual sources, and a FIR/FIS catalog \citep{2009ASPC..418....3Y} containing the four FIS bands for $\sim$430,000 sources. AKARI photometry, in at least one of the six bands, was published for all of the RCB stars in our sample. 

\subsection{Infrared Astronomical Satellite}
The Infrared Astronomical Satellite (IRAS, \citealp{1984ApJ...278L...1N}), which was the first space-based observatory to survey the entire sky in the IR, operated in the MIR and FIR at central wavelengths of 12, 25, 60, and 100~$\mu$m. Two catalogs of IRAS photometry have been published and updated since the end of the mission. They are the IRAS Faint Source Catalog (FSC, \citealp{1990IRASF.C......0M}) and the Point Source Catalog (PSC, \citealp{1988iras....7.....H}). Both catalogs provide photometry in at least one of the four IRAS bands for a total of $\sim$300,000 individual sources. All of the RCB stars in our sample have IRAS observations in at least one of the four bands. 

\subsection{Herschel Space Observatory}
%%Herschel/PACS:  5.866, 6.969, 11.384 arcsec (FWHM)
%%Herschel/SPIRE: 18, 24, 37 arcsec (FWHM)
The {\it Herschel} Space Observatory ({\it Herschel}, \citealp{2010A&A...518L...1P}) has allowed for improved space-based resolution in both the FIR and sub-mm to detect and map cold dust surrounding stars. Our sample of RCB stars was observed with {\it Herschel} under an open time program led by PI G. Clayton (OT1\_\_gclayton\_\_1; 25.6 hrs). Observations were conducted with both the Photodetector Array Camera and Spectrometer (PACS) at 70, 100, and 160~$\mu$m \citep{2010A&A...518L...2P} and the Spectral and Photometric Imaging REceiver (SPIRE) at 250, 350, and 500~$\mu$m \citep{2010A&A...518L...3G}.

The IDL routine Scanamorphos (version 21.0, \citealp{2013PASP..125.1126R} was used for generating all of the final PACS and SPIRE maps for analysis. The map making process begins by downloading the raw satellite telemetry (Level 0 products) from the {\it Herschel} Science Archive (HSA). These products are then converted to physical units (Level 1 products), such as temperatures or voltages, with the {\it Herschel} Interactive Processing Environment (HIPE, version 12, \citealp{2010ASPC..434..139O}). HIPE is both a GUI and command line based software written in Jython (Java$+$Python). It is at these Level 1 products where the typical HIPE pipeline is interrupted (further processing with HIPE all the way to image products is possible) to generate FITS binary files that {\it Scanamorphos} can read and interact with. PACS maps are generated in the units of Jy pixel$^{-1}$ with 1\farcs0, 1\farcs4, and 2\farcs0 pixels at 70, 100, and 160~$\mu$m, respectively. The choice of these pixel sizes correspond to PSFs with FWHM of, in increasing wavelength, 6\arcsec, 7\arcsec, 11\arcsec. SPIRE maps are generated in the units of Jy beam$^{-1}$ and are then converted into Jy pixel$^{-1}$ through a multiplicative constant derived from the individual SPIRE beams for each wavelength band. The maps have pixel sizes of 6\farcs0, 10\farcs0, and 14\farcs0 at 250, 350, and 500~$\mu$m, respectively. The SPIRE PSFs have FWHM of, in increasing wavelength, 18\arcsec, 24\arcsec, 37\arcsec.

\section{Photometry}
Photometry was done on the {\it Spitzer} and {\it Herschel} images in order to generate SEDs for the stars in our sample. There are many different programs that have been written to perform automated aperture and PSF photometry. We used the automated aperture routine Source Extractor (SExtractor, \citealp{1996A&AS..117..393B}). The power of SExtractor is that there are many tunable parameters that allow the user to maximize the program to perform photometry on their desired objects, whether they be point source or extended. SExtractor also provides robust post--run ancillary products such as residual, background, object, and aperture images in addition to performing aperture photometry on any given input images. These diagnostics were used to judge the success of any run. Further, we chose to use the Interactive Data Language (IDL) routine StarFinder \citep{2000ASPC..216..623D,2000SPIE.4007..879D}, which performs PSF photometry. StarFinder, similar to SExtractor, provides a suite of post--run images for the purposes of diagnostics. In particular, the point source subtracted image is of great use for investigating the presence of any faint nebulosity. SExtractor was used for all the photometry except for the {\it Spitzer}$/$MIPS observations of V~CrA, which are from StarFinder. The photometry used in this study for the individual stars is listed in Table 2--10. %because of prior experience with the software \citep{2015AJ....149...57M} , also based on prior experience \citep{2015AJ....149...57M}

\section{SED Modeling}%Monte Carlo Radiative Transfer}
\subsection{Monte Carlo Radiative Transfer}
We performed Monte Carlo radiative transfer (MCRT) modeling of the SEDs for the stars in our sample to better constrain the morphology and physical parameters of the dust surrounding these objects. We used the fully 3D MOnte CArlo SimulationS of Ionized Nebulae (MOCASSIN; version 2.02.70) code \citep{2003MNRAS.340.1136E, 2005MNRAS.362.1038E, 2008ApJS..175..534E}. The code is written in Fortran 90 and is capable to be run with parallel processing, through message passage interface (MPI). MOCASSIN was compiled with Intel's ``ifort" compiler, because it decreases the run time per model over free compilers such as gfortran. We used Open MPI for the MPI implementation. MOCASSIN is run by first defining a series of user inputs, such as: number of dimensions, grid size, dust density, composition, and distribution. Interactions, whether absorption or scattering, between photons and dust grains are governed by Mie scattering theory \citep{2005MNRAS.362.1038E}. MOCASSIN returns temperature, mass, and opacity of the dust shells. 

% This reveals the robustness of the code as any arbitrary geometry and viewing angle can be modeled. Any  MOCASSIN is a fully self--consistent 3D Cartesian dust RT code \citep{2005MNRAS.362.1038E}, which built on the previous version \citep{2003MNRAS.340.1136E} and introduced the version 2.0 series. The code is run by first defining a series of user inputs, such as: number of dimensions, grid size, dust density, composition, and distribution. This reveals the robustness of the code as any arbitrary geometry and viewing angle can be modeled. Any interactions, whether absorption or scattering, between photons and dust grains are governed by Mie scattering theory \citep{2005MNRAS.362.1038E}. MOCASSIN returns temperature, mass, and opacity of the dust shells. 

For the sample, we chose to model these systems as a central point source surrounded by a gas--free dust shell. These shells are further assumed to be ``smooth'', which means that there are no inhomogeneities (``clumps''), with the dust density profile falling by r$^{-2}$ from the inner radius (R$_{\rm in}$) to the outer radius (R$_{\rm out}$). We further took advantage of axial symmetry to model only one--eighth of the envelope rather than a full envelope. The composition of the dust grains was determined by prior analysis of the spectra of RCB stars, which is consistent with amorphous Carbon (amC) grains \citep{1984ApJ...280..228H,Clayton:2011lr,2011ApJ...739...37G,2013ApJ...773..107G}. This is due to the extinction curve peaking between 2400 and 2500 \AA\ \citep{1984ApJ...280..228H} and the featureless nature of the spectra in the optical and IR \citep{2011ApJ...739...37G}. Thus, our MCRT models were performed with 100\% amC grains. The grain size distribution was motivated by the findings of \citet{1984ApJ...280..228H}, who used IUE observations of R~CrB and RY~Sgr to find that dust grain sizes appeared consistent with a distribution between 5--60 nm (0.005 to 0.06~$\mu$m). A power law distribution following \citet{1977ApJ...217..425M}, ${\rm a}^{-3.5}$, specifies the size distribution of the dust grains. Detailed discussion of the modeling of individual stars can be found in the next section.

%An MRN--like \citep{1977ApJ...217..425M} power law distribution, ${\rm a}^{-3.5}$, specifies the size distribution of the dust grains
%to
%A power law distribution following \citet{1977ApJ...217..425M}, ${\rm a}^{-3.5}$, specifies the size distribution of the dust grains

\subsection{Semi-Analytic Modeling}
%quickSAND models over spherical polar grid
We also modeled a subset of our SEDs (see Section 5.8.2) with a semi-analytic Fortran code called QuickSAND (Quick Semi-ANalytic Dust, \citealp{2012ApJ...749..170S}). The code computes an SED for a source surrounded by a spherical shell after being given: R$_{\rm in}$, R$_{\rm out}$, source luminosity, source temperature, density profile for the shell, number density at R$_{\rm in}$, dust composition, and distance to the object. The modeling is performed over a spherical polar grid. QuickSAND can be run to either generate a single SED, or to output a grid of SEDs over a predefined parameter space. We were provided a custom version which operates on an exponential grid to maximize resolution for shells that cover many orders of magnitude in size between R$_{\rm in}$ and R$_{\rm out}$. 

\section{Circumstellar Shells of R Coronae Borealis Stars}
A twofold approach was adopted for our investigation into the cold CSM of our sample RCB stars. First, the unpublished, archival {\it Spitzer} and {\it Herschel} images were examined by eye to identify morphological features. Next, the results from aperture and$/$or PSF photometry were used to fill in the FIR$/$sub-mm regime for the maximum light SEDs of these stars. The goal of these two methods is to achieve a better understanding of the CSM of RCB stars. This, by extension, allows for a more accurate picture of the mass loss history for these stars and clearer idea of their progenitors.

As seen in the AAVSO light curves (Figures 1--4) , we have made efforts to select to make sure that NIR observations and shorter were observed during maximum light. RCB stars are known to exhibit regular to semi--regular pulsations, $\delta$V $\lesssim$ 0.1 mag and periods of 40-100 days \citep{1990MNRAS.247...91L,Saio:2008qe}. This effect has minimal impact on our SED modeling. However, the IUE observations, despite being dereddened using CCM, are sensitive to small amounts of dust.

\subsection{MV~Sgr}
Variability in MV~Sgr was first discovered by \citet{1928BHarO.855...22W}. It would be another 30 years until it was identified as an RCB star \citep{1958AJ.....63Q..50H,1959AJ.....64..241H}. \citet{1959AJ.....64..241H} also discussed the results of early spectra reported by \citet{1964ApJ...140.1317H}, which confirmed the hydrogen--deficiency of MV~Sgr. However, what was unexpected was that the spectrum of MV Sgr revealed that it was similar in temperature to a B--type star. This extreme temperature makes this star a member of the unique subset of ``hot'' RCB stars (of which only 4 are known total) \citep{2002AJ....123.3387D}. \citet{1996MNRAS.282..889P} found the Li I 6708 \AA\ line in emission.% [OI] and [NII] emission lines were also found to be present in the MV~Sgr spectrum, which indicates the presence of a nebula around the star \citep{1996MNRAS.282..889P}.

\subsubsection{Image Inspection}
MV~Sgr was not observed with {\it Herschel}, so the {\it Spitzer}$/$MIPS observations are the only FIR images of this star. Postage stamp images from MIPS can be seen in Figure 5. MV~Sgr appears as a point source at 24~$\mu$m, as the warm dust remains unresolved. The emission at 70~$\mu$m measures colder dust, farther from the central star, but this dust is also unresolved. No emission is detected in the MIPS 160~$\mu$m observation.

\subsubsection{Radiative Transfer Modeling}
Archival photometry and spectroscopy were combined with new photometry from the {\it Spitzer}$/$MIPS observations to construct the SED for MV~Sgr. See Table 2 for the input values. The maximum--light SED is presented in Figure 6 along with the best--fit MOCASSIN models. The SED in the UV$/$optical is fit well by a T$_{eff} =$16,000 K blackbody as determined by \citet{1984ApJ...278..224D} and \citet{2002AJ....123.3387D}. The input luminosity for the MOCASSIN modeling was determined by assuming an absolute magnitude M$_{\rm V}$=$-$3.0 for the hot RCB stars \citep{Tisserand:2009fj}. This corresponds to a distance of 11.5 kpc and luminosity of $\sim$5,200 L$_\odot$.

The effect of a strong IR excess can be seen after 1.0~$\mu$m as the SED continues to rise as wavelength increases. The IR component was best--fit by two concentric, smooth shells with density falling as r$^{-2}$. This modeling strategy is reinforced with a ``by eye'' examination of the SED where the presence of two separate components in the IR can be easily seen. The first peak is at $\sim$4.6~$\mu$m, and corresponds to an envelope beginning at $3.45 \times 10^{14}$ cm and extending to $9.45 \times 10^{15}$ cm. The dust mass is $7.59 \times 10^{-8}$ M$_\odot$ while temperatures range from 1,000 K down to 200 K at the inner and outer radii, respectively. The second peak occurs at $\sim$ 25~$\mu$m with a best--fit envelope having an inner radius $3.25 \times 10^{16}$ cm and outer radius $9.45 \times 10^{17}$ cm. Dust temperatures in the shell range from 150 K to 50 K with a mass of $3.27 \times 10^{-4}$ M$_\odot$.

The shape of the SED in the IR regime was also examined by \citet{2011ApJ...739...37G, 2013ApJ...773..107G}. \citet{2011ApJ...739...37G} found that the two blackbody curves have temperatures 1500 and $\sim$200 K, which agree with temperatures from our MCRT modeling. MV~Sgr has been among the least active of RCB stars in terms of decline events. In all the years of monitoring this star there have only been 3 observed declines \citep{1959AJ.....64..241H, 2017JAVSO..45..159L}. In spite of this seemingly low level of activity, the dust mass in the outer envelope is about the same as other RCB stars in our sample. This is due to the puff like nature of dust formation events. Declines only happen when the cloud condenses along our line of sight with the RCB star. There can be any number of puffs, at any time, forming around the central RCB star that we are not able to detect in the visible \citep{2011ApJ...739...37G,2013ApJ...773..107G,2015MNRAS.447.3664R}. Thus, an appreciable envelope with a reservoir of cold dust can still be constructed even if a star is observed to remain at maximum light.

\subsection{R~CrB}
R~CrB is the eponymous member of the RCB class having been first discovered as variable in the late 18$^{\rm th}$ Century \citep{1797RSPT...87..133P}. \citet{1953ApJ...117...25B} was among the first to note the hydrogen deficient, but carbon--rich nature of R~CrB and RCB stars. \citet{Kennan:1963} first identified the star as having Li via the 6708 \AA\ feature. R~CrB has also been found to be enriched with $^{19}$F via lines at 6902.47 and 6834.26 \AA\ \citep{Pandey:2008eu}.
  
\subsubsection{Image Inspection}
Observations of R~CrB in the FIR and sub-mm were previously inspected and discussed by \citet{2011ApJ...743...44C}, which included both {\it Spitzer}$/$MIPS and {\it Herschel}$/$SPIRE. {\it Herschel}$/$PACS observations were taken after the paper was published and are presented here for the first time. Figure 7 contains the complete 9--panel postage stamp series of the MIPS, PACS, and SPIRE images of R~CrB.

Previous discussions of R~CrB's nebulosity point the spherical nature of its morphology \citep{Gillett:1986cr,2011ApJ...743...44C}. These works had at their disposal the highest sensitivity and angular resolution FIR$/$sub-mm observations for their time. The {\it Herschel}$/$PACS images reinforce the apparent spherical shape to the R~CrB nebulosity. %The diffuse structure seen prior at other wavelengths is also resolved at these wavelengths (as well as the background galaxy cluster).

\subsubsection{Radiative Transfer Modeling}
The maximum--light SED of R~CrB was originally modeled and presented by \citet{2011ApJ...743...44C}. New photometry from the {\it Herschel}$/$PACS observations of R~CrB were added to the \citet{2011ApJ...743...44C} SED and remodeled using MOCASSIN. The SED is displayed in Figure 8 with the input photometry held in Table 3. The best--fit MOCASSIN model is represented by the dashed line. Parameters from \citet{2011ApJ...743...44C} were adopted for our own MCRT modeling. These include an effective temperature of 6,750 K and distance of 1.40 kpc, which results in a luminosity of 9,150  L$_\odot$.

The R~CrB SED was best modeled using two concentric dust envelopes. The inner shell extends from $1.00 \times 10^{15}$ cm to $3.00 \times 10^{16}$ cm. The mass of this envelope was found to be $9.09 \times 10^{-7}$ M$_\odot$ with dust temperatures ranging from 700 K down to 180 K. A second envelope was modeled to account for the presence of additional colder material that one envelope cannot entirely account for. This outer shell has an inner radius of $3.40 \times 10^{17}$ cm and outer radius of $1.00 \times 10^{19}$ cm. The dust mass contained in this envelope is $2.42 \times 10^{-4}$ M$_\odot$ with temperatures ranging from 80 K to 20 K. 

R~CrB is, by far, the best studied of any RCB star, so it comes as no surprise that its SED has also been extensively studied \citep{Gillett:1986cr,1986MNRAS.222..357K,1990MNRAS.245..119G,1993ApJS...86..517Y,1993ApJ...409..725Y,1996MNRAS.281.1139N,1996A&A...315L.249W,2001ApJ...555..925L,2011ApJ...739...37G,2011ApJ...743...44C,2015MNRAS.447.3664R}. %The decade long gap following \citet{2001ApJ...555..925L} before any new works were published coincides with the launch of several space--based IR missions (i.e., {\it Spitzer}, {\it Herschel}, AKARI, WISE) that were able to provide additional wavelength coverage as well as overlap with IRAS. 

We have compared our MOCASSIN results to those of \citet{2011ApJ...739...37G} and \citet{2011ApJ...743...44C}. \citet{2015MNRAS.447.3664R} builds on the work presented by \citet{2011ApJ...739...37G} and focuses more on tracking changes in the brightnesses of RCB stars in the 30 years of space--based MIR observations. A two component (star $+$ single IR excess) blackbody fit was used by \citet{2011ApJ...739...37G} to describe the R~CrB SED. They described the stellar component with a blackbody of T$_{\rm star} =$ 6750 K (derived from \citealt{Asplund:2000qy}), and the IR excess with a blackbody that had a maximum dust temperature of 950 K, which was based on the {\it Spitzer}$/$IRS spectrum between 10 and 20~$\mu$m \citep{2011ApJ...739...37G}.

\citet{2011ApJ...743...44C} presented the results of their full 3D (spherical polar grid) MCRT code. The code included non--isotropic scattering, polarization, and thermal emission from dust \citep{2003ApJ...591.1049W,2003ApJ...598.1079W,2006ApJS..167..256R}. The best--fit model found that the observed SED could be explained by the presence of a dusty disk surrounded by a larger envelope. The disk extended from $6.28 \times 10^{14}$ cm to $2.24 \times 10^{15}$ cm and had a dust mass of $3.5 \times 10^{-6}$ M$_\odot$ \citep{2011ApJ...743...44C}. The shell had radii of $1.95 \times 10^{18}$ cm and $1.32 \times 10^{19}$ cm at the inner and outer boundaries, respectively \citep{2011ApJ...743...44C}. The dust mass of the Clayton et al. envelope was also found to be roughly two orders of magnitude higher ($\sim2.0 \times 10^{-2}$ M$_\odot$). This discrepancy is attributed to Clayton et al. using a luminosity of 5645 L$_\odot$, which is roughly a factor of two lower than our input luminosity, and a full MRN size distribution \citep{1977ApJ...217..425M}. %The dust was assumed to be 100\% amorphous carbon grains and followed an MRN size distribution \citep{1977ApJ...217..425M}. 

\subsection{RY~Sgr}
RY~Sgr was first suspected to be variable in 1893 while under observation by Colonel E. E. Markwick while he was stationed in Gibraltar \citep{1896ApJ.....4..138P,2011arXiv1109.4234S}. \citet{1896ApJ.....4..138P} also noted that the spectrum of the new variable was found to be peculiar after being discovered by Williamina Fleming. By the early 1950s, RY~Sgr was known to be hydrogen--deficient and classified as an RCB star \citep{1953ApJ...117...25B}. \citet{Lambert:1994uq} found no evidence for Li overabundance in the spectrum of RY~Sgr. The presence of $^{19}$F was found in RY~Sgr's atmosphere from absorption lines located at 6902 and 6834 \AA\ \citep{Pandey:2008eu}. RCB stars, as a class, are known to show brightness fluctuations via pulsations in addition to their spectacular declines. RY~Sgr was first discovered to be pulsating with 0.5 magnitude variations and a period of $\sim$39 days by \citet{CampbellJacchia:1946}.
 
\subsubsection{Image Inspection}
Diffuse nebulosity surrounding RY~Sgr was searched for in the unpublished, archival {\it Spitzer}$/$MIPS, {\it Herschel}$/$PACS, and {\it Herschel}$/$SPIRE observations. These observations provide the highest angular resolution and sensitivity for RY~Sgr from 24 to 500~$\mu$m. A 9--panel mosaic containing these images are found in Figure 9.

RY~Sgr appears as a point source in the {\it Spitzer}$/$MIPS 24~$\mu$m image. RY~Sgr begins to become more extended in the 70 and 160~$\mu$m observations, however the angular resolution at MIPS is not high enough to separate out the PSF from any diffuse nebulosity. {\it Herschel}$/$PACS was able to provide the necessary angular resolution to resolve the diffuse nebulosity surrounding RY~Sgr. The diffuse structure appears spherical. Yet, the density of the shell appears to be higher in the northern region when compared with the southern. This is reinforced with the {\it Herschel}$/$SPIRE observations at 250 and 350 ~$\mu$m where the angular resolution and sensitivity are still high enough to resolve the shell from the background. However, by 500~$\mu$m the emission from the envelope has become too weak to resolve anything more than a rough determination of where the nebulosity is.  % the resolution and sensitivity have degraded that only a rough determination of where the nebulosity lies can be made and nothing stronger.

\subsubsection{Radiative Transfer Modeling}
The available photometry and spectroscopy for RY~Sgr (while at maximum--light) were combined to construct its SED, which can be seen in Figure 10. The dashed line represents the best--fit MCRT model from MOCASSIN. The photometry is found in Table 4. A blackbody with T$_{eff} = 7250$ K was adopted from atmosphere modeling by \citet{Asplund:2000qy}. A distance of 1.5 kpc was determined by assuming an absolute V--band magnitude of $-5$ \citep{Tisserand:2009fj}. This results in an input luminosity of 8,900 L$_\odot$ for RY~Sgr. 

The SED begins to be dominated by the RY~Sgr CSM after 1.6~$\mu$m ($H$--band) due to contributions from warm dust close to the central star. A spherical envelope with inner radius at $8.62 \times 10^{14}$ cm and outer radius at $5.00 \times 10^{16}$ cm describes the SED between 1.6 and $\sim$25.0~$\mu$m. The dust mass of this envelope is $8.90 \times 10^{-7}$ M$_\odot$ with temperatures ranging from $\sim$500 K down to $\sim$200 K. This wavelength region of the RY~Sgr SED was also examined by \citet{2011ApJ...739...37G}. The central star was represented by a blackbody of 7200 K. The maximum temperature of the blackbody used to fit this dust component was found to be 675 K \citep{2011ApJ...739...37G}.

However, a second blackbody peak can clearly be seen in the photometry longer than 40~$\mu$m that does not lie on the Rayleigh--Jeans tail of the first IR excess. A second envelope was modeled with an inner radius at $5.15 \times 10^{17}$ cm and extending outward to $4.50 \times 10^{18}$ cm. The dust mass of this envelope is $7.25 \times 10^{-4}$ M$_\odot$ with temperatures ranging from $\sim$60 K down to $\sim$30 K.% The dust mass in the modeled outer envelope is the second highest in dust mass. This component was most likely not found by \citet{2011ApJ...739...37G}, since they restricted themselves to observations shorter than 40~$\mu$m.

\subsection{SU~Tau}
Variability in SU~Tau was first noted by \citet{1908HarCi.140....1C} with a note that it could be an RCB star. This classification was strengthened further in a later {\it Harvard College Observatory Bulletin} \citep{1916BHarO.617....1B}. SU~Tau, like R~CrB, has been found to be rich in Li and $^{19}$F \citep{Lambert:1994uq,Pandey:2008eu}.

\subsubsection{Image Inspection}
% PACS is displayed on the left and SPIRE on the right. The blue, green, and red correspond to the shortest, intermediate, and longest wavelengths within the specific instrument, respectively.  
Unpublished FIR and sub-mm observations of SU Tau exist from {\it Spitzer}$/$MIPS, {\it Herschel}$/$PACS, and {\it Herschel}$/$SPIRE. These observations are presented as a 9--panel postage image in Figure 11. The sensitivity with {\it Spitzer}$/$MIPS is enough to detect the presence of dust surrounding SU~Tau, however the angular resolution is not sufficient enough to separate nebulosity from the PSF. 

The need for improved angular resolution becomes quickly apparent when examining the {\it Herschel} observations. The galaxy, 2MFGC 4715 \citep{2004BSAO...57....5M}, and the SU~Tau CSM, which are blended in the {\it Spitzer}$/$MIPS 70~$\mu$m image, are well separated and can also be further distinguished in 3--color images, which can be found in Figure 12. The morphology of SU~Tau's CSM is unlike that found around any other RCB star. A bow shock type feature, which dominates the eastern half of the image, is clearly visible in observations with both {\it Herschel} instruments with definitive detections out to 350~$\mu$m. Diffuse nebulosity can be discerned in the western half of the PACS 3--color image. 

The outer edge of the bow shock extends $\sim$30$''$ to 50$''$ from the central position of SU~Tau (see Figure 13). This corresponds to a physical distance of 7.4 to 10.8 pc assuming the distance to SU~Tau is 3.3 kpc (see below). The flux of the large overdensity, located in the southeast, was sampled in the {\it Herschel}/$PACS$ observations with an elliptical region with semi--major axis of 2.9$''$ and semi--minor axis of 1.8$''$. The measured fluxes were 0.0284, 0.0328, and 0.0193 Jy  at 70, 100, and 160~$\mu$m, respectively. These values correspond to 5\%, 10\%, and 15\% of the calculated flux for the dust emission centered on SU~Tau (see below). Estimated dust temperatures at the location of the outer edge of the bow shock are about 30 K. This is consistent with a blackbody with a peak wavelength of 100~$\mu$m, which is exactly where the maximum flux value for the overdensity was found to be. 

It is difficult to determine for certain how long any interaction between the SU~Tau CSM and the ISM has been occurring. If we assume that any interaction is much less than the time for the material to expand outward to its current distance from the central star, then we can at least put some bound on its age. Should this material be part of a fossil PN structure, as predicted in the FF scenario, then it would take the material between $2.4 \times 10^5$ to $5.3 \times 10^5$ years to reach its current location. This assumes that the initial shell was expanding at typical PNe velocities (20--30 km s$^{-1}$). However, in the DD scenario, the dust would have outward velocities of at least 400 to 900 km s$^{-1}$. The lower limit comes from observations of the He I $\lambda$10830  line \citep{1992ApJ...397..652C,2003ApJ...595..412C,2013AJ....146...23C}, while the upper limit comes from simulations of WD mergers \citep{2015AJ....150...14M}. These velocities indicate the dust would reach the determined distances on the order of $10^4$ years.

A search for evidence of this feature in archival observations of SU~Tau found what could be diffuse emission associated with the bow shock in the 2MASS $J$--band (see Figure 14). The 2MASS $J$--band filter has a central wavelength of $1.235\pm0.006$~$\mu$m with a bandwidth of $0.162\pm0.001$~$\mu$m \citep{2003AJ....126.1090C}. The wavelength of the electron transition between the fifth to third energy levels of Hydrogen (Paschen--$\beta$) is 1.282~$\mu$m, which falls within the 2MASS J--band bandwidth.

Stellar bow shocks typically manifest due to interactions between the stellar wind of a rapidly moving star and the denser, slower ISM through which the star is currently moving. Bow shocks have been detected in the FIR around other stars (e.g, R~Hydrae, \citealp{2006ApJ...648L..39U}; Betelgeuse, \citealp{2012A&A...548A.113D,2012A&A...537A..35C}). The obvious difference between SU~Tau, as well as other RCB stars, and these other stars is the extreme hydrogen--deficiency that RCB stars are known to have. The bow shock seen around Betelgeuse is caused by material lost during the red supergiant phase \citep{2012A&A...548A.113D} running into the denser ISM in the direction of the star's motion. In the case of SU~Tau, this lends itself to the question: what is the composition of the material being shocked? If SU~Tau was formed via the FF scenario, then having H--rich material at the outskirts of its CSM is not surprising. However, we are only seeing emission from dust in the bow shock and are not able to comment on the composition of any gas associated with the SU~Tau bow shock. %Although if the dust envelopes of RCB stars are mostly helium filled, then the presence of shocked H--rich material could be from the ISM that SU~Tau is moving into. 

\subsubsection{Radiative Transfer Modeling}
SU~Tau's archival maximum--light photometry and spectroscopy were combined with photometry from the unpublished {\it Spitzer}$/$MIPS, {\it Herschel}$/$PACS, and {\it Herschel}$/$SPIRE ($3\sigma$ upper limits) observations to construct its SED, which can be seen in Figure 15. The bow shock feature was not included in the photometry aperture, and the background galaxy was masked out in all {\it Herschel} observations. Further, it is highly likely that the IRAS 60~$\mu$m, 100~$\mu$m, and AKARI 100~$\mu$m points are contaminated by flux from the background galaxy. These points are still included on the SED, but had no influence on determining the final fit model. The dashed line represents the best--fit MCRT model from MOCASSIN. The photometry is found in Table 5. A blackbody with T$_{eff} = 6500$ K was adopted from atmosphere modeling by \citet{Asplund:2000qy}. A distance of 3.3 kpc was determined by assuming an absolute $V$--band magnitude of $-5$ \citep{Tisserand:2009fj}. This results in an input luminosity of 10,450 L$_\odot$ for SU~Tau. 

The SU~Tau CSM begins to dominate the SED beginning around 2.2~$\mu$m ($K$--band), which indicates the presence of warm dust. A spherical envelope with inner radius at $2.10 \times 10^{15}$ cm and outer radius at $4.25 \times 10^{16}$ cm describes the SED from 1.6~$\mu$m out to barely before the 70.0~$\mu$m points. The dust mass of this envelope is $2.27 \times 10^{-6}$ M$_\odot$ with temperatures ranging from $\sim$600 K down to $\sim$150 K. This regime was also included in the blackbody fitting analysis by \citet{2011ApJ...739...37G}. They were able to fit a blackbody temperature of 6500 K, which agrees with the temperature determined by \citet{Asplund:2000qy}. The IR excess was fit with a 635 K blackbody \citep{2011ApJ...739...37G}, which is in agreement with the temperature range from our MOCASSIN RT modeling. 

However, a second excess can be seen to arise for photometry longer than 70~$\mu$m as the points lie above the single dust envelope model fits. A second envelope was added with an inner radius at $1.00 \times 10^{18}$ cm and extending outward to $9.00 \times 10^{18}$ cm. The dust mass of this envelope is $6.80 \times 10^{-4}$ M$_\odot$ with temperatures ranging from $\sim$50 K down to $\sim$25 K. 

\subsection{UW~Cen}
Brightness fluctuations in UW~Cen were first noticed by \citet{1906HarCi.122....1L} when it exhibited a 1.6 mag change. \citet{1952AnHar.115...61G} first suggested that UW~Cen is an RCB star. UW~Cen is one of two stars (the other V854 Cen--see below) that has been examined for $^{18}$O, $^{19}$F, and Li. Lithium has been known about in UW~Cen since \citet{Lambert:1994uq} found the Li resonance doublet at 6707~\AA\ in its spectrum. \citet{Pandey:2008eu} discovered $^{19}$F by absorption lines at 6834.26, 6902.47, 7398.68, and 7425.6~\AA. The search for $^{18}$O resulted in a null detection owing to UW~Cen being too warm to display molecular features \citep{2009ApJ...696.1733G}. %No comments as to what kind of variable were made by Leavitt \& Pickering. 

The CSM of UW~Cen is unique among all of the RCB stars. It is the only RCB star discovered to have a reflection nebula surrounding it \citep{1991MNRAS.248P...1P,1999ApJ...517L.143C}. The nebula is $\sim 15"$ in diameter. It is only visible either during deep declines when the dust along the line of sight serves as a ``natural'' coronagraph or an actual coronagraph is used to block the light from the central star. \citet{1999ApJ...517L.143C} found that the morphology of the nebula had changed significantly from year to year. These changes were too fast for any physical changes in the nebula to be occurring. \citet{1999ApJ...517L.143C} deduced that the changing pattern of new dust clouds condensing around the star resulted in variations in how the reflection nebula was illuminated.

\subsubsection{Image Inspection}
The longest wavelength photometric observations of UW~Cen that had been previously examined were the IRAS observations from the 1980s \citep{1986ASSL..128..407W,Schaefer:1986lq}. Archival, unpublished {\it Spitzer}$/$MIPS and {\it Herschel}$/$PACS images are presented in Figure 16. UW~Cen appears as a point source in the {\it Spitzer}$/$MIPS images. The higher angular resolution provided by {\it Herschel}$/$PACS allows the morphology of the UW~Cen CSM to be resolved. The nebula, seen in all three {\it Herschel}$/$PACS wavelengths, lies well beyond the reflection nebula (diameter $\sim 15''$) known to exist around UW~Cen. The nebula appears spherical at 100 and 160~$\mu$m.

\subsubsection{Radiative Transfer Modeling}
The maximum--light UW~Cen SED was made by combining archival photometry and spectroscopy with unpublished photometry from {\it Spitzer}$/$MIPS and {\it Herschel}$/$PACS observations. The SED can be found in Figure 17, and all input photometry in Table 6. The dashed line plotted over the SED represents the best--fit MCRT model from MOCASSIN. \citet{Asplund:2000qy} found from their modeling of spectra against line--blanketed models of stellar atmospheres that the effective temperature of UW~Cen is $\sim$7500 K. This temperature has been adopted for our MCRT modeling. The distance to UW~Cen, 3.5 kpc, was calculated from the relation between absolute $V$--band magnitude and $V-I$ color presented in \citet{Tisserand:2009fj}. This is a departure from the previous distance calculation of 5.5 kpc \citep{1990MNRAS.247...91L,1999ApJ...517L.143C}, due to underestimating the line of sight extinction. This new distance results in an input luminosity of 7,320 L$_\odot$.

The UW~Cen SED begins to show the influence from CSM after 2.2~$\mu$m ($K$--band). A spherical envelope with inner radius at $1.55 \times 10^{15}$ cm and outer radius at $4.50 \times 10^{16}$ cm describes the SED from 1.6 to $\sim$25~$\mu$m. The dust mass of this envelope is $2.40 \times 10^{-6}$ M$_\odot$ with temperatures ranging from $\sim$600 K down to $\sim$150 K. This part of the SED is dominated by the presence of warm dust surrounding the central star. The variability in measurements around 3~$\mu$m is due to changes in the amount of warm dust that has recently condensed around UW~Cen at the times the observations were taken.

UW~Cen's SED at wavelengths longer than $\sim$30~$\mu$m is unlike that of any other known RCB star. A clear second dust component can be seen as the long--low resolution IRS spectrum starts to rise again to a peak around 70~$\mu$m before falling again at wavelengths longer than 100~$\mu$m. This component was also modeled with a spherical envelope with an inner radius at $7.00 \times 10^{17}$ cm and outer radius of $2.50 \times 10^{18}$ cm. The dust mass of this envelope is $5.14 \times 10^{-3}$ M$_\odot$ with temperatures ranging from $\sim$70 K down to $\sim$40 K. 

Analysis of the UW~Cen SED was previously presented by both \citet{1999ApJ...517L.143C} and \citet{2011ApJ...739...37G}. \citet{1999ApJ...517L.143C} fit only the optical to MIR with two Planck functions of temperatures 6000$\pm$500 and 650$\pm$50 K (see their Figure 3). They did not fit wavelengths longer than 12~$\mu$m due to the possibility of contamination from IR bright cirrus clouds along the line of sight to UW~Cen. However, they do comment that this contribution can be fit with a Planck function of 100 K. A dust mass of $\sim 6 \times 10^{-4}$ M$_\odot$ is derived with a total mass of $\sim$ 0.2 M$_\odot$ assuming a normal gas--to--dust ratio \citep{1999ApJ...517L.143C}. 

A four component fit, stellar $+$ three to account for CSM contribution, was adopted by \citet{2011ApJ...739...37G}. Similar to \citet{1999ApJ...517L.143C}, their fits are only comprised of Planck functions. The temperatures of the blackbody fits were 7500, 630, 120, and 50 K \citep{2011ApJ...739...37G}. No estimate for the dust masses of any of the components were presented.  

\subsection{V854~Cen} 
In terms of RCB stars, V854~Cen was discovered relatively recently. In 1986 V854~Cen (NSV~6708) was found to be at 7.5 mag when the previous brightest known maximum for this star was at 9.7 mag \citep{1986IBVS.2928....1M}. Further analysis of archival plates and film by \citet{1986IBVS.2928....1M} found that the star appeared as faint as 15.5 mag. The peak $V$--band brightness 7 mag makes V854~Cen the third brightest RCB star in the entire sky after R~CrB and RY~Sgr. A star of that brightness would not have been overlooked by the community at-large. An examination of archival plates by \citet{1986IAUC.4245....2M} found that V854~Cen had been in decline since at least 1913, which implies that it had been continuously forming dust along the line of sight during the intervening years.

The abundances of V854~Cen are unusual even for an RCB star. \citet{Lawson:1989kx} found that V854~Cen was much more H--rich than R~CrB, and for that matter any other known RCB star. It is one of five RCB stars that compose the designation ``minority'' RCB stars \citep{Lambert:1994uq}. This classification is made primarily by the lower iron abundances of these RCB stars in relation to the rest of the class. \citet{2012ApJ...747..102H}  found evidence that V854~Cen might show enrichment of $^{13}$C from their analysis of high--resolution optical spectroscopy focused the $^{12}$C$^{13}$C Swan bandhead. They find a  $^{12}$C$^{13}$C ratio of 16 to 24, but comment that higher S$/$N observations are required to better determine the value. This isotope of carbon is found in FF objects like Sakurai's object, but not in the majority of RCB or HdC stars.  It has been examined for any signs of $^{18}$O, $^{19}$F, or Li and all have resulted in no detections \citep{Lambert:1994uq,Pandey:2008eu,2009ApJ...696.1733G}. Finally, it has been found to have C$_{60}$ emission from {\it Spitzer}$/$IRS observations \citep{2011ApJ...729..126G}. %  is also one four RCB stars that show appreciable levels of $^{13}$C from the analysis of the $^{12}$C$^{13}$C Swan bandhead \citep{2012ApJ...747..102H}

\subsubsection{Image Inspection}
In a similar fashion to MV~Sgr, the only FIR observations of V854~Cen are provided by {\it Spitzer}$/$MIPS. The 24, 70, and 160~$\mu$m observations, left to right, respectively, can be found in a 3--panel postage stamp displayed in Figure 18. Not much can be said about the morphology of the V854~Cen CSM from these images. %This is likely reflected due to a large reservoir of dust that should exist surrounding V854~Cen due to its nearly half--century decline.

\subsubsection{Radiative Transfer Modeling}
The inputs for the maximum--light SED of V854~Cen are similar to that of MV~Sgr. Specifically, this means that archival photometry and spectroscopy are combined with unpublished {\it Spitzer}$/$MIPS photometry. The SED can be seen in Figure 19 and all input photometry in Table 7. The dashed line represents the best--fit MCRT model from MOCASSIN. An input stellar blackbody with an effective temperature of 6,750 K was adopted from atmosphere modeling of \citet{Asplund:2000qy}. A distance of 2.28 kpc was determined by assuming an absolute $V$--band magnitude of $-5$ \citep{Tisserand:2009fj}. This results in an input luminosity of 11,760 L$_\odot$.

Warm dust in the V854~Cen CSM starts to influence the SED after 1.6~$\mu$m. This dust component was modeled with an envelope that has an inner radius at $4.88 \times 10^{14}$ cm and outer radius $1.00 \times 10^{16}$ cm. The dust mass of this envelope is $3.08 \times 10^{-7}$ M$_\odot$ with temperatures ranging from $\sim$1,200 K down to $\sim$300 K. A second envelope was also modeled with an inner radius at $3.45 \times 10^{16}$ cm and extending outward to $1.00 \times 10^{18}$ cm. The dust mass in this envelope is $2.60 \times 10^{-5}$ M$_\odot$ with temperatures ranging from $\sim$200 K down to $\sim$50 K. %In order to account for a slight excess in the photometry at wavelengths longer than 60~$\mu$m, which indicates the presence of a colder reservoir of dust. 

The maximum light SED was also in the blackbody fitting performed by \citet{2011ApJ...739...37G}. A three component (star + two IR excess) was found by Garc{\'{\i}}a-Hern{\' a}ndez et al. to best describe the SED. V854~Cen was fit with a 6750 K blackbody, while the two IR excess with blackbodies of 900 and 140 K \citep{2011ApJ...739...37G}. The stellar component agrees with the temperature derived by \citet{1998A&A...332..651A}. The temperatures for the IR excesses fall within the ranges for our two modeled envelopes. 

\subsection{V~CrA}
Changes of at least 1 mag in the brightness of V~CrA were first reported by \citet{1896ApJ.....4..234P}. V~CrA is also a minority RCB star in addition to having an enrichment of $^{13}$C by the detection of the $^{12}$C$^{13}$C Swan bandhead \citep{Rao:2008lr}. This makes it one of three RCB stars to be confirmed to show enrichment of this isotope of carbon. No appreciable level of Li was found to be in its photosphere \citep{Lambert:1994uq}. \citet{Pandey:2008eu} made a possible detection of $^{19}$F in the spectrum of V~CrA.

\subsubsection{Image Inspection}
{\it Spitzer}$/$MIPS, {\it Herschel}$/$PACS, and {\it Herschel}$/$SPIRE images of V~CrA were examined for the presence of diffuse CSM. These images displayed as postage stamps can be found in Figure 20. The improved angular resolution with {\it Herschel}$/$PACS allows accurate detections of V~CrA at all three wavelengths. There does appear to be a hint of nebulosity in the North--South direction in the {\it Herschel}$/$PACS images but V~CrA appears as a point source in the {\it Herschel}$/$SPIRE images.

\subsubsection{Radiative Transfer Modeling}
The V~CrA SED was made from available maximum--light photometry and spectroscopy, which were combined with photometry from unpublished {\it Spitzer}$/$MIPS, {\it Herschel}$/$PACS, and {\it Herschel}$/$SPIRE. The photometry can be found in Table 8, while the SED is displayed in Figure 21. The {\it Spitzer}$/$MIPS 160~$\mu$m and {\it Herschel}$/$SPIRE are all $3\sigma$ upper limits. The dashed line represents the best--fit MCRT model from MOCASSIN. A blackbody with T$_{eff} = 6250$ K was adopted from atmosphere modeling by \citet{Asplund:2000qy}. A distance of 5.5 kpc was determined by assuming an absolute V--band magnitude of $-5$ \citep{Tisserand:2009fj}. This results in an input luminosity of 6,550 L$_\odot$. 

The influence from warm dust in the V~CrA CSM starts to become prominent after 1.6~$\mu$m ($H$--band). This material was modeled by a spherical envelope with inner radius at $1.70 \times 10^{15}$ cm and outer radius at $4.90 \times 10^{16}$ cm. The dust mass of this envelope is $4.00 \times 10^{-6}$ M$_\odot$ with temperatures ranging from $\sim$500 K down to $\sim$150 K. 

As with the other RCB stars, the V~CrA SED cannont be fit with a single dust envelope. Thus a second envelope was modeled to describe the presence of colder CSM surrounding V~CrA. This envelope was modeled with an inner radius at $1.00 \times 10^{17}$ cm and extending outward to $1.00 \times 10^{18}$ cm. The dust mass of this envelope is $5.90 \times 10^{-5}$ M$_\odot$ with temperatures ranging from $\sim$130 K down to $\sim$50 K. 

The blackbody fitting to the V~CrA maximum light SED is among the most complex in the sample of \citet{2011ApJ...739...37G}. A fit that includes only optical and {\it Spitzer} observations was best described by three components (star + two IR excesses); while the fitting of optical, $KLMN$, and IRAS 25$\mu$m observations yielded four components (star + three IR excesses) \citep{2011ApJ...739...37G}. V~CrA was fit with a 6500 K blackbody in both models, which agrees with the spectroscopic effective temperature \citep{2008MNRAS.384..477R}, and is not very different than the 6250 K \citep{Asplund:2000qy} used in our modeling. The temperatures in the two IR excess scenarios agree with the ranges determined by the two modeled envelopes. Our MOCASSIN modeling also appears to point to the need of a small reservoir of hot dust close to V~CrA to account for the flux values in $MN$ and first two channels of WISE. 

\subsection{Sample Properties}
The sample presented here is small, but it is of interest to look for trends among the RCB dust shells that have been studied. The results of the MOCASSIN modeling can be found in Table 11. The average properties are also be compared against the HdC star, HD~173409, and the final flash object, V605~Aql.  

\subsubsection{CSM Morphology}
%The SEDs of R~CrB, SU~Tau, V854~Cen, and V~CrA appear like a single, continuous envelope should have been able to describe their SEDs. The SEDs of MV~Sgr, RY~Sgr, and UW~Cen are all clear cases that a second dust component is necessary.  

A major surprise of the MCRT SED modeling of the individual RCB stars is that they {\it all} required the modeling of two discrete, thick dust shells. We employed QuickSAND (see Section 4.2) in order to better examine the possibility of fitting the SEDs with a single, continuous envelope. QuickSAND models were performed on the SEDs of R~CrB, SU~Tau, V854~Cen, and V~CrA using the best fit parameters from our MOCASSIN modeling, but with one continuous shell. These stars were selected because the shape of their SEDs appear as if they could be fit by one envelope.

This modeling was accomplished by using R$_{\rm in}$ of the inner envelope, R$_{\rm out}$ of the outer envelope. The output SEDs from the QuickSAND modeling (solid red line) are overplotted on the MOCASSIN best fit (dashed black line) and maximum light SEDs (squares and solid black lines) in Figures 22 to 25. The resulting QuickSAND SEDs are close to the MOCASSIN best fit models, but are overall a poorer fit to the maximum light SEDs. Additional modeling found that better fits could be achieved by decreasing R$_{\rm out}$. However, this particular constraint is harder to control (see discussion below).

The results of our MCRT agree with those of \citet{1996MNRAS.281.1139N}, who performed analytic modeling and radiative transfer modeling of the R~CrB SED. Nagendra \& Leung used the available IRAS data \citep{Gillett:1986cr} in their modeling and they determined that a double shell was the optimal way to fit the SED. Their models were unable to describe emission longer than 60~$\mu$m with only one shell. However, Nagendra \& Leung did investigate how to model the R~CrB SED with a single dusty envelope. They had to greatly increase the contribution from the interstellar radiation field, by a factor of 3 to 30 (depending on the density profile of the shell) times the normal value, in order to accomplish this \citep{1996MNRAS.281.1139N}. %(ISRF)

We also investigated whether or not the maximum--light SEDs could be described by ``thin'' dust envelopes. The qualification for an envelope as being ``thin'' was that R$_{\rm out} = 2.0 \times $R$_{\rm in}$. The maximum light SED of UW~Cen was modeled again (see Figure 26). The black dashed line is the same best--fit model as presented earlier, while the red dashed line represents the thin shell model. The modeling with two thin shells has good agreement to the best--fit model up through 10~$\mu$m. However, beyond 10~$\mu$m the thin model does not describe the SED well. %Overall the fit is worse especially the region between the two maxima in the SED. It fits the blue maximum of the SED well but lies a little below the red maximum. %The overall behavior of the SED is recovered with thin shells, especially the location of the local maxima of both IR excesses. Nevertheless, the limitation of this modeling strategy is that it adequately describes neither the region between the maxima nor the SED.

The location for R$_{\rm out}$ can be calculated from the {\it Spitzer} or {\it Herschel} observations. This is derived from the angular diameter distance relation (for small angles): $\theta \approx x/D$, where $\theta$ is the angular size of the extended object, x is the physical size of the object, and D is the distance to the object. The average ratio of R$_{\rm out-Outer}$/R$_{\rm out-Measured}$, excluding UW~Cen (see below), is 1.30 with minimum and maximum values of 0.68 (V854~Cen) and 3.31(RY~Sgr), respectively (see Table 11 for the entire sample). R$_{\rm out-Outer}$ is the outer radius of the outer envelope derived from our MCRT modeling, while R$_{\rm out-Measured}$ is the value of the same parameter as calculated from the FIR imaging.

The differences between the modeled and measured outer radii are likely due to the uncertainties in the distance to the Galactic RCB stars. The {\it Gaia} second data release (DR2) was published April 2018 containing accurate parallaxes for nearly 1.3 billion stars. Six stars, five RCB stars and one HdC star, from our sample were included in the {\it Gaia} DR2. The distances and $1\sigma$\ uncertainties, accounting for systematics, to these stars are found in Table 1. While {\it Gaia} represents the largest collection of precise distance measurements, V854~Cen and V~CrA are not included in the DR2 release. Further, the uncertainties are still large for MV~Sgr and UW~Cen. This leaves R~CrB, RY~Sgr, and SU~Tau, for which the distances used for modeling in this paper and the {\it Gaia} DR2 distances are essentially the same. The largest difference is for SU~Tau which is still within a factor of no more than 2.4 times the upper {\it Gaia} DR2 uncertainty. Therefore, we have chosen to keep our calculated distances for our modeling. The uncertainties in our calculated distances are tied to the effective temperatures chosen for our MCRT modeling \citep{1997A&A...318..521A,Asplund:2000qy,2002AJ....123.3387D} and that the absolute brightness of the sample RCB stars range between M$_V = -3$ and M$_V = -5$ \citep{Alcock:2001lr,Tisserand:2009fj}. These assumptions are not independent of each other. A different choice of temperature would lead to a different absolute brightness, estimated distance, and measured outer radius. %The large differences between the modeled and measured outer radii at the minimum, 30\% below the expected size, and maximum, 330\% higher than the expected size, are likely due to the uncertainties in the distance to the Galactic RCB stars for V854~Cen and RY~Sgr || estimated to be up to a factor The differences between our calculated distances and the {\it Gaia} DR2 are no more than a factor of two for any individual star. When this uncertainty, \bf which  is on the order of 2, is taken into account then the differences between the modeled and calculated values become less severe. The distances used for modeling in this paper and the {\it Gaia} DR2 distances are the same within {\bf a factor of no more than 2.4 times the {\it Gaia} DR2 uncertainties for their respective RCB star. We have chosen to keep our calculated distances for our modeling rather than adopt the {\it Gaia} distances because the uncertainties are still quite large for not of

UW~Cen is the only case where the outer radius of the inner shell can be calculated because of its reflection nebula, which can be seen at optical wavelengths. The diameter of the reflection nebula has been measured at 15$''$ \citep{1991MNRAS.248P...1P,1999ApJ...517L.143C}. \citet{1999ApJ...517L.143C} calculated R$_{\rm out-Measured}$ as being $6.00 \times 10^{17}$ cm assuming a distance of 5.5 kpc \citep{1990MNRAS.247...91L}. We derived a distance of 3.5 kpc from a higher E(B--V) than Clayton et al. assumed was present for the line of sight to UW~Cen. This corresponds to the slightly smaller value of $4.01 \times 10^{17}$ cm for R$_{\rm out-Measured}$. The value of R$_{\rm out-Measured}$ derived from the FIR observations is $2.62 \times 10^{18}$ cm. 

First, it is important to note that the best fit inner envelope lies entirely within R$_{\rm out-Measured}$ for the reflection nebula. Second, that the two values for R$_{\rm out-Measured}$ roughly correspond to the modeled values for R$_{\rm in-Outer}$ and R$_{\rm out-Outer}$. R$_{\rm in-Outer}$ is $7.00 \times 10^{17}$ cm, which is larger than the derived value but in good agreement with the \citet{1999ApJ...517L.143C} calculation. This indicates that the outer edge of the reflection nebula possibly represents the beginning of the second envelope containing the large reservoir of cold dust predicted by the MCRT. Further, there is an excellent agreement ($<$5\%) between the modeled value of R$_{\rm out-Outer}$ and the calculation of R$_{\rm out-Measured}$ from the FIR observations. 

\subsubsection{Envelope Masses \& Decline Activity}
The possibility that these large, diffuse shells could have formed during the RCB phase was first examined for R~CrB by \citet{2015AJ....150...14M}. Hence, we searched for a relationship between the physical size and dust mass of the inner and outer envelopes and frequency of declines. It is commonly accepted that the declines are caused when a cloud of carbon dust condenses, along our line of sight, near the central RCB star. Over time radiation pressure from the central star acts on the cloud driving it outward into the larger circumstellar environment. Thus, the frequency and$/$or length of time an RCB star spends near minimum light is evidence for the formation of fresh dust, at least along the line of sight. Long term monitoring at 3.4~$\mu$m (L--band) is able to follow the creation of new clouds out of the line of sight \citep[i.e.][]{Feast:1997lr,1997MNRAS.285..339F,2010ARep...54..620B}. The formation of individual clouds cannot be seen in flux increases, but changes in L--band brightness by a factor of 2 over roughly 2 or 3 years indicates higher and lower dust formation activity. The rapid outward expansion of these new dust clouds may produce the large observed envelopes during the RCB phase.
  
\citet{1996AcA....46..325J} examined the frequency of declines and the average time between declines in a sample of RCB stars (see her Table 1). All seven of the RCB stars in this paper are in the Jurcsik sample. The minimum inner envelope mass, as determined from our MOCASSIN modeling, is $7.59 \times 10^{-8}$ M$_\odot$ (MV~Sgr) while the maximum is $4.00 \times 10^{-6}$ M$_\odot$ (V~CrA). MV~Sgr is among the least active RCB stars with $\Delta$T$_{\rm fades}$ quoted at 6900 days from 2 declines in, at the time, 38 years of observations \citep{1996AcA....46..325J}. On the more active side, UW~Cen has had 13 declines in 40 years of observations, which corresponds to $\Delta$T$_{\rm fades} =$ 1100 days, \citet{1996AcA....46..325J}. The mass of UW~Cen's inner shell is $2.40 \times 10^{-6}$ M$_\odot$, which is the second largest in our sample. It has also experienced at least two deep decline events since \citet{1996AcA....46..325J} was published (see the bottom panel of Figure 2). 

This implies that either MV~Sgr has been producing dust at rate that is $\sim 1/10$ of the other RCB stars or that we could be viewing it more pole-on. Spectropolarimetric observations of R~CrB taken near minimum light suggest that the clouds of dust are more likely to form around the equatorial region of an RCB star than the polar regions \citep{1988ApJ...325L...9S}. Thus, if MV~Sgr appears to be more pole-on it can have a large IR excess from dust production events while only being observed to have a few declines. 

We next compared the derived dust masses to the modeled outer radius for the warm and cold shells -- see Figure 27. A power law trend between the dust mass and outer radius appears to stand out when examining the inner envelopes. However, the origin of this trend arises from the outer radius and volume of the modeled inner envelopes. The envelope with the highest average density is V854~Cen and the lowest average density is UW~Cen. When the same properties are plotted for the outer envelopes, no obvious correlation stands out when this sample is treated as broadly all being RCB stars. However, a slight trend seems to be revealed when individual stars are separated by being either ``majority'' or ``minority'' RCB stars. \citet{Lambert:1994uq} define minority RCB stars as being more iron deficient relative to both other RCB stars and the Sun. In Figure 27, the majority RCB stars are represented by black squares and text while the minority RCB stars are represented by red squares and text. The warm shells do not reveal any insight even when divided into majority and minority. The cold shells of the minority RCB stars seem to be both smaller and less massive than the stars of the majority group. Several other chemical factors complicate whether this difference is entirely owed to being minority RCB stars. V854~Cen and V~CrA are also both known to be enriched with $^{13}$C as well as being the two most hydrogen-rich RCB stars.

V854~Cen, in particular, highlights a peculiar case for establishing whether the envelopes are produced during the current RCB phase. The dust masses of the inner and outer envelopes are $3.08 \times 10^{-7}$ M$_\odot$ and $2.60 \times 10^{-5}$ M$_\odot$, respectively. These values are the second lowest and lowest for our sample. This seems paradoxical since V854~Cen was in decline for nearly half of a century \citep{1986IAUC.4245....2M}. It has also been extremely active since its return to maximum light in the 1980s with 9 declines in 9 years of monitoring ($\Delta$T$_{\rm fades} = $ 370 days) \citep{1996AcA....46..325J} and more since then (see the upper panel in Figure 3). 

One possible resolution for this discrepancy is that the total time each star has been in the RCB phase is unknown (i.e., what are the relative ages of the different RCB stars to each other?). This issue cannot be resolved by our work, but should V854~Cen be younger than the rest of the RCB stars then the derived smaller masses of its envelopes would make sense, even with its near half century decline. Analysis of wind features via the He I $\lambda$10830 line, which has been used an indicator of dust expansion velocities (see further below), has found that the velocities seen in V854~Cen can be as strong as 700 km s$^{-1}$, which is a factor of two higher than has been measured in other RCB stars \citep{2013AJ....146...23C}. 

An important value for this analysis is knowing the true expansion velocity of the dust. Estimates for this motion range from tens to hundreds of km s$^{-1}$. The case for slower moving dust has been attributed to either the natural expansion of a PN shell \citep{2011ApJ...743...44C} or from high resolution (R $\sim$ 30,000), high S$/$N spectrum of scattered star light during deep declines \citep[and references therein]{2011ApJ...739...37G}. Observations of the He I $\lambda$10830 line suggest that the dust is rapidly accelerated up to 400 km s$^{-1}$ \citep{1992ApJ...397..652C,2003ApJ...595..412C,2013AJ....146...23C}. 

Therefore dust forming at 2 R$_{*}$ (170 R$_\odot$ or $1.2 \times 10^{13}$ cm) would take 2 to 20 years to reach 10$^{14}$ to 10$^{15}$ cm, typical values of R$_{in}$ for the inner shell from our RT modeling, at 20 km s$^{-1}$, respectively. Dust moving at higher implied velocities would cover the same distances in 3 to 9 months. These timescales are much shorter than the lower limit on the lifetime of an RCB star: $\sim$200 years from R~CrB \citep{1797RSPT...87..133P}. This seems to indicate that at least the inner envelopes could arise from dust ejected during the RCB phase. 

The critical issue to resolve is whether or not the continued outward expansion of this dust is also responsible for the observed cold envelopes. Dust moving at 400 km s$^{-1}$ would take about $10^3$ to $10^4$ years to reach anywhere from 10$^{18}$ cm to 10$^{19}$ cm, respectively. It would take an order of magnitude longer for dust moving at 20 km s$^{-1}$ to reach those same distances. The slower dust expansion has also led \citet{1986MNRAS.222..357K} to suggest that these envelopes could be remnant material from the initial red giant phase of these stars. However, in a DD scenario this phase of stellar evolution would have taken place billions of years before the WD binary would merge. 

\subsubsection{Comparison to an HdC Star \& Final Flash Stars}
The FF object, V605~Aql, experienced an event in the early 20th century that took it from below the limits of photographic plates (m = 15) all the way up to a peak magnitude of 10.2 in 1919 \citep{1920AN....211..119W}. \citet{1921BHarO.753R...2W} found that the rise to its maximum brightness was a slow climb over the preceding two years. The languid nature of this outburst originally earned V605~Aql a classification as a slow nova \citep{1921PASP...33..314L}. V605~Aql did not spend a significant time at its peak brightness. It began to fade quickly and within a year had fallen below 15th magnitude only to return in 1921, before ultimately fading for good in 1923 \citep{1997AJ....114.2679C,2006ApJ...646L..69C,2013ApJ...771..130C}.

During V605~Aql's 1921 re-brightening spectra were acquired at the 0.91--m Crossley telescope \citep{1921PASP...33..314L}. Lundmark discovered that the spectra pointed to V605~Aql as a cool carbon (R0) star and not a classical nova in the late stages of an outburst. This was the last major study of V605~Aql for nearly 50 years. Deep observations obtained independently and published simultaneously by \citet{1971ApJ...170..547F} and \citet{1971PASP...83..819V} revealed that V605~Aql lies the center of the old PN Abell 58 \citep{1966ApJ...144..259A}. Further, re-analysis of the \citet{1921PASP...33..314L} spectrum showed that the V605~Aql looked like a cool RCB star \citep{1973BAAS....5..442B,1997AJ....114.2679C}.

\citet{2013ApJ...771..130C} presented the modern SED of V605~Aql, which is reproduced in Figure 28. The SED was made from ground--based NIR photometry \citep{2001A&A...367..250H}, ground--based MIR spectroscopy, and IR photometry from several satellites. Optical photometry is not available due to obscuration by the material ejected during the 1919 outburst. In addition to this obscuring dust, a large reservoir of cold dust associated with V605~Aql is immediately apparent since the SED continues to rise to a maximum around 40~$\mu$m. 

The SED was fit with emission curves of amorphous carbon dust with temperatures and masses of: 810 K, $1.0 \times 10^{-11} {\rm M}_\odot$; 235K, $9.0 \times 10^{-6} {\rm M}_\odot$; 75 K, $2.0 \times 10^{-3} {\rm M}_\odot$ \citep{2013ApJ...771..130C}. These temperatures are in agreement with dust temperatures found in either the first or second envelope of the MCRT for our sample RCB stars. The dust masses calculated by Clayton et al. from the green (235 K) and blue (75 K) components correspond with those derived for the inner and outer shells, respectively, of our RCB sample. However, the evolution of V605~Aql, itself, has been too rapid when compared to RCB stars. \citet{1997AJ....114.2679C} commented that in 1921 the spectrum of V605~Aql resembled a cool RCB star with T$_{eff} \simeq 5000$ K. However, spectra obtained in 2001 revealed the presence of C IV in emission, which indicates that V605~Aql has evolved horizontally back across the HR diagram and is now consistent with T$_{eff} \sim 95,000$ K \citep{2006ApJ...646L..69C}, an increase of 90,000 K in only 80 years. This change in temperature is more rapid than the minimum lifetime of an RCB star (200 years from R~CrB) and from the estimated lifetimes from population synthesis of the RCB stars ($10^4 - 10^5$ yr; \citealp{2015ApJ...809..184K}).% which is a temperature change not seen in any RCB stars including the hot RCB stars.

HD~173409 is not an RCB star as normally defined, but designated as a hydrogen--deficient carbon (HdC) star. These stars are spectroscopically similar to RCB stars, but have neither the characteristic declines in brightness nor display any evidence for IR excess \citep{Warner:1967lr,2010ApJ...723L.238G,2012A&A...539A..51T}. The spectrum of HD~173409 was first noted as being different from the majority of other stars by \citet{1896ApJ.....4..142P} and identified as being hydrogen--deficient by \citet{1953ApJ...117...25B}. The HdC stars are also known to have an overabundance of $^{18}$O \citep{2005ApJ...623L.141C,Clayton:2007ve,2009ApJ...696.1733G,Garcia-Hernandez:2010fk}, however the effective temperature of HD~173409 is too high for molecular bands to be detected in its spectrum. 

HD~173409 was included in the {\it Herschel} observing campaign to learn if any cold dust could be surrounding the central star that might have gone previously undetected. This is an excellent test to determine if HdC stars are, in fact, RCB stars that are in an extended period of low decline activity. The {\it Herschel} PACS and SPIRE observations of HD~173409 can be seen in Figure 29. No nebulosity is visible around HD~173409 in these images.

HD~173409 was in the \citet{2012A&A...539A..51T} sample, which examined the early data release by the WISE science team.  No excess was found in the NIR$/$MIR from the WISE observations. We have constructed the HD~173409 SED from archival photometry and photometry from the ALLWISE catalog. Additionally, 3$\sigma$ upper limits were determined for the {\it Herschel} observations using a 30$''$--diameter aperture centered on the position of HD~173409. These upper limits were then included in the HD~173409 SED, which can be found in Figure 30 with a 7000K blackbody overplotted. The photometry that has gone into the HD~173409 SED can be found in Table 10. The absence of an infrared excess in the SED suggests that the CSM of HD~173409 is relatively dust free. One HdC star, HD~175893, was discovered to have an IR excess from WISE photometry \citep{2012A&A...539A..51T}. This HdC star could be an example of an RCB star in a phase of low activity in terms of dust production. %The SED reinforces that there is no detectable dust surrounding this HdC star. 

\section{Discussion and Conclusions}
\citet{2015AJ....150...14M} presented and explored three possible interpretations for the origins of the diffuse, dusty nebulosity that surrounds some RCB stars: (1) they are fossil planetary nebulae (PNe); (2) they are remnant material from the merger of a CO and a He white dwarf binary, (3) they have been constructed from dust ejection events during the current RCB phase. We will now examine the results presented here in the context of these three scenarios. The results of the MOCASSIN models are presented in Table 11. 

The MCRT modeling of these SEDs suggests the existence of two discrete, concentric spherical shells around each of our sample RCB stars. The construction of these shells during the current RCB phase is critically tied to the number of dust puffs produced, the expansion velocity of the dust puffs, and the lifetime of RCB stars. It has been suggested that during a decline a single puff contains $\sim$10$^{-8}$ M$_{\sun}$ of dust \citep{1992ApJ...397..652C,2011ApJ...743...44C}. Then, $\sim$10$^{-7}$ M$_{\sun}$ of dust would form per year if a dust puff forms somewhere around the star every 50 days. Thus, an RCB inner envelope would be produced in about 10 years and it is unlikely that any of the inner shells are the remnant material of a WD merger or fossil PN.

The origin of the outer shells is of greater uncertainty. The data seem to suggest that at some point dust formation ceased and then restarted, or that the inner and outer shells have different origins. For example, the inner shell could be from the RCB phase and the outer shell could be a fossil PN shell or remnant material from WD merger. A knowledge of the hydrogen abundance in these shells would help determine whether they are fossil PN shells or not. If the envelopes are fossil PNe then they should be H--rich. H I measurements at 21--cm of R~CrB put lower limits on any hydrogen in its dust shell \citep{2015AJ....150...14M}. Assuming R~CrB is a typical RCB star, then it is unlikely its dust shell is a fossil PN. Further, recent modeling of the merger rates of WD binaries by \citet{2015ApJ...809..184K} found that typically the merger will not take place for at least 500 Myr after both stars become WD. This is reinforced by the discovery that the nearby system WD 1242-105 is a binary white dwarf expected to merge in 740 Myr \citep{2015AJ....149..176D}. After these lengths of time, it is very unlikely that any PN material would still be around an RCB star.

Hydrodynamic modeling of the material that remains following a WD merger suggests that these envelopes would contain M$_{\rm Dust} \leq 10^{-6}$ M$_\odot$ \citep{2015AJ....150...14M}. The mean mass of the outer envelopes in this sample is $10^{-3}$ M$_\odot$. The least massive envelope (V854~Cen) is implied to contain $2.60 \times 10^{-5}$ M$_\odot$ of dust. This is still an order of magnitude higher than predicted for remnant material from a WD merger. 

The velocity of the expanding dust has been estimated as being tens to hundreds of km s$^{-1}$. Slower expansion velocities have been suggested by \citet[and references therein]{2011ApJ...739...37G}. Faster outward movement is suggested by the He I $\lambda$10830 line, which suggests that the dust is rapidly accelerated up to 400 km s$^{-1}$ \citep{1992ApJ...397..652C,2003ApJ...595..412C,2013AJ....146...23C}. The outer envelopes in our sample have implied outer radii that range from 10$^{18}$ cm to 10$^{19}$ cm (see Table 11). Material at these distances represents the oldest material to be shed by RCB stars. Dust moving with slower velocities, 20 km s$^{-1}$, would take about $10^4$ to $10^5$ years to reach anywhere from 10$^{18}$ cm to 10$^{19}$ cm, respectively. These times drop by an order of magnitude if the dust velocities agree more with the results of the He I $\lambda$10830 analysis. These timescales are both much longer than we have known about the RCB phenomenon \citep{1797RSPT...87..133P}. 

We have compared the observations of the RCB stars to the hydrogen--deficient carbon (HdC) stars and stars that have been observed to undergo a final flash (FF). HdC stars are essentially spectroscopic twins of RCB stars. HdC stars, however, do not experience decline events and lack any IR excess. The HdC star HD~173409 was observed with both PACS and SPIRE on {\it Herschel}. No emission associated with HD~173409 was detected in any of the {\it Herschel} observations. The SED for this star also shows no evidence for any IR excess when fit by a single 7000 K blackbody. Recently, one HdC star, HD~175893, was found to have an IR excess from analysis of WISE colors and could either represent a missing link between the two classes of objects or an RCB star going through an extended period of low dust formation \citep{2012A&A...539A..51T}.

The results of our sample were compared to the FF star, V605~Aql, and the findings of \citet{2013ApJ...771..130C}. \citet{2013ApJ...771..130C} presented the SED for V605~Aql, which indicates the presence of $\sim$10$^{-3}$ M$_\odot$ of dust associated with its 1919 ejecta. This is on a similar level to the dust masses derived from our MOCASSIN modeling for the outer shells. In this scenario, these envelopes would have been created in the recent past. However, the rapid evolution in the effective temperature of V605~Aql from 5,000 K to 95,00 K in around 80 years \citep{2006ApJ...646L..69C} has not been found in any RCB star.

The {\it Herschel} observations of SU~Tau with the PACS and SPIRE instruments have led to the discovery of a bow shock like structure. This is the first known RCB star to exhibit this type of feature, which represents interactions between the SU~Tau CSM and the local interstellar medium (ISM). The bow shock extends between 30$''$ to 50$''$ from the central position of SU~Tau with a brighter feature in the southeast possibly indicating a location where more material is beginning to pile up. 

RCB stars are among the most uncommon and bizarre objects discovered in the Universe. However, they provide the opportunity to greatly advance our knowledge in areas such as stellar evolution and stellar chemistry. Additional examination of these objects, especially at 21--cm, is needed to determine the origin of the cold, diffuse CSM seen around the RCB stars.

\acknowledgments
We thank the anonymous referee for feedback that has improved this paper. This work is based in part on observations made with the {\it Spitzer Space Telescope}, which is operated by the Jet Propulsion Laboratory, California Institute of Technology under a contract with NASA, and the {\it Herschel} Space Observatory operated by ESA space with science instruments provided by European--led Principal Investigator consortia and with important participation from NASA. Support for this work was provided by NASA through {\it Spitzer} contract No. 1287678 and {\it Herschel} contract No. 1485231 issued by JPL$/$Caltech. DAGH acknowledges support provided by the Spanish Ministry of Economy and Competitiveness (MINECO) under grant AYA--2017--88254--P.

\software{Scanamorphos (version 21.0; \citealp{2013PASP..125.1126R}), HIPE (v12; \citealp{2010ASPC..434..139O}), SExtractor \citep{1996A&AS..117..393B}, StarFinder \citep{2000ASPC..216..623D,2000SPIE.4007..879D}, MOCASSIN (v2.02.70; \citealp{2003MNRAS.340.1136E, 2005MNRAS.362.1038E, 2008ApJS..175..534E}), QuickSAND \citep{2012ApJ...749..170S}, MIPS DAT \citep{2005PASP..117..503G}}

\bibliography{thesis}

\begin{thebibliography}{152}
\expandafter\ifx\csname natexlab\endcsname\relax\def\natexlab#1{#1}\fi

\bibitem[{{Abell}(1966)}]{1966ApJ...144..259A}
{Abell}, G.~O. 1966, \apj, 144, 259

\bibitem[{{Alcock} {et~al.}(2001){Alcock}, {Allsman}, {Alves}, {Axelrod},
  {Becker}, {Bennett}, {Clayton}, {Cook}, {Dalal}, {Drake}, {Freeman}, {Geha},
  {Gordon}, {Griest}, {Kilkenny}, {Lehner}, {Marshall}, {Minniti}, {Misselt},
  {Nelson}, {Peterson}, {Popowski}, {Pratt}, {Quinn}, {Stubbs}, {Sutherland},
  {Tomaney}, {Vandehei}, \& {Welch}}]{Alcock:2001lr}
{Alcock}, C., {et~al.} 2001, ApJ, 554, 298

\bibitem[{{Asplund} {et~al.}(1998){Asplund}, {Gustafsson}, {Kameswara Rao}, \&
  {Lambert}}]{1998A&A...332..651A}
{Asplund}, M., {Gustafsson}, B., {Kameswara Rao}, N., \& {Lambert}, D.~L. 1998,
  A\&A, 332, 651

\bibitem[{{Asplund} {et~al.}(1997){Asplund}, {Gustafsson}, {Kiselman}, \&
  {Eriksson}}]{1997A&A...318..521A}
{Asplund}, M., {Gustafsson}, B., {Kiselman}, D., \& {Eriksson}, K. 1997, \aap,
  318, 521

\bibitem[{{Asplund} {et~al.}(2000){Asplund}, {Gustafsson}, {Lambert}, \&
  {Rao}}]{Asplund:2000qy}
{Asplund}, M., {Gustafsson}, B., {Lambert}, D.~L., \& {Rao}, N.~K. 2000, A\&A,
  353, 287

\bibitem[{{Asplund} {et~al.}(1999){Asplund}, {Lambert}, {Kipper}, {Pollacco},
  \& {Shetrone}}]{Asplund:1999bh}
{Asplund}, M., {Lambert}, D.~L., {Kipper}, T., {Pollacco}, D., \& {Shetrone},
  M.~D. 1999, A\&A, 343, 507

\bibitem[{{Barnard}(1916)}]{1916BHarO.617....1B}
{Barnard}, F.~A.~P. 1916, Harvard College Observatory Bulletin, 617, 1

\bibitem[{{Bertin} \& {Arnouts}(1996)}]{1996A&AS..117..393B}
{Bertin}, E., \& {Arnouts}, S. 1996, \aaps, 117, 393

\bibitem[{{Bidelman}(1953)}]{1953ApJ...117...25B}
{Bidelman}, W.~P. 1953, \apj, 117, 25

\bibitem[{{Bidelman}(1973)}]{1973BAAS....5..442B}
{Bidelman}, W.~P. 1973, in \baas, Vol.~5, Bulletin of the American Astronomical
  Society, 442

\bibitem[{{Bogdanov} {et~al.}(2010){Bogdanov}, {Taranova}, \&
  {Shenavrin}}]{2010ARep...54..620B}
{Bogdanov}, M.~B., {Taranova}, O.~G., \& {Shenavrin}, V.~I. 2010, Astronomy
  Reports, 54, 620

\bibitem[{{Bright} {et~al.}(2011){Bright}, {Chesneau}, {Clayton}, {de Marco},
  {Le{\~a}o}, {Nordhaus}, \& {Gallagher}}]{2011MNRAS.414.1195B}
{Bright}, S.~N., {Chesneau}, O., {Clayton}, G.~C., {de Marco}, O., {Le{\~a}o},
  I.~C., {Nordhaus}, J., \& {Gallagher}, J.~S. 2011, MNRAS, 414, 1195

\bibitem[{{Campbell} \& {Jacchia}(1946)}]{CampbellJacchia:1946}
{Campbell}, L., \& {Jacchia}, L. 1946, The Story of Variable Stars (Blakiston,
  MA: Harvard Monographs)

\bibitem[{{Cannon} \& {Pickering}(1908)}]{1908HarCi.140....1C}
{Cannon}, A.~J., \& {Pickering}, E.~C. 1908, Harvard College Observatory
  Circular, 140, 1

\bibitem[{{Cardelli} {et~al.}(1989){Cardelli}, {Clayton}, \&
  {Mathis}}]{1989ApJ...345..245C}
{Cardelli}, J.~A., {Clayton}, G.~C., \& {Mathis}, J.~S. 1989, \apj, 345, 245

\bibitem[{{Clayton}(1996)}]{1996PASP..108..225C}
{Clayton}, G.~C. 1996, PASP, 108, 225

\bibitem[{{Clayton}(2012)}]{2012JAVSO..40..539C}
---. 2012, Journal of the American Association of Variable Star Observers
  (JAAVSO), 40, 539

\bibitem[{{Clayton} \& {De Marco}(1997)}]{1997AJ....114.2679C}
{Clayton}, G.~C., \& {De Marco}, O. 1997, AJ, 114, 2679

\bibitem[{{Clayton} {et~al.}(2003){Clayton}, {Geballe}, \&
  {Bianchi}}]{2003ApJ...595..412C}
{Clayton}, G.~C., {Geballe}, T.~R., \& {Bianchi}, L. 2003, ApJ, 595, 412

\bibitem[{{Clayton} {et~al.}(2007){Clayton}, {Geballe}, {Herwig}, {Fryer}, \&
  {Asplund}}]{Clayton:2007ve}
{Clayton}, G.~C., {Geballe}, T.~R., {Herwig}, F., {Fryer}, C., \& {Asplund}, M.
  2007, ApJ, 662, 1220

\bibitem[{{Clayton} {et~al.}(2013{\natexlab{a}}){Clayton}, {Geballe}, \&
  {Zhang}}]{2013AJ....146...23C}
{Clayton}, G.~C., {Geballe}, T.~R., \& {Zhang}, W. 2013{\natexlab{a}}, \aj,
  146, 23

\bibitem[{{Clayton} {et~al.}(2005){Clayton}, {Herwig}, {Geballe}, {Asplund},
  {Tenenbaum}, {Engelbracht}, \& {Gordon}}]{2005ApJ...623L.141C}
{Clayton}, G.~C., {Herwig}, F., {Geballe}, T.~R., {Asplund}, M., {Tenenbaum},
  E.~D., {Engelbracht}, C.~W., \& {Gordon}, K.~D. 2005, ApJL, 623, L141

\bibitem[{{Clayton} {et~al.}(1999){Clayton}, {Kerber}, {Gordon}, {Lawson},
  {Wolff}, {Pollacco}, \& {Furlan}}]{1999ApJ...517L.143C}
{Clayton}, G.~C., {Kerber}, F., {Gordon}, K.~D., {Lawson}, W.~A., {Wolff},
  M.~J., {Pollacco}, D.~L., \& {Furlan}, E. 1999, ApJL, 517, L143

\bibitem[{{Clayton} {et~al.}(2006){Clayton}, {Kerber}, {Pirzkal}, {De Marco},
  {Crowther}, \& {Fedrow}}]{2006ApJ...646L..69C}
{Clayton}, G.~C., {Kerber}, F., {Pirzkal}, N., {De Marco}, O., {Crowther},
  P.~A., \& {Fedrow}, J.~M. 2006, ApJL, 646, L69

\bibitem[{{Clayton} {et~al.}(1992{\natexlab{a}}){Clayton}, {Whitney},
  {Stanford}, \& {Drilling}}]{1992ApJ...397..652C}
{Clayton}, G.~C., {Whitney}, B.~A., {Stanford}, S.~A., \& {Drilling}, J.~S.
  1992{\natexlab{a}}, \apj, 397, 652

\bibitem[{{Clayton} {et~al.}(1992{\natexlab{b}}){Clayton}, {Whitney},
  {Stanford}, {Drilling}, \& {Judge}}]{1992ApJ...384L..19C}
{Clayton}, G.~C., {Whitney}, B.~A., {Stanford}, S.~A., {Drilling}, J.~S., \&
  {Judge}, P.~G. 1992{\natexlab{b}}, ApJL, 384, L19

\bibitem[{{Clayton} {et~al.}(2011{\natexlab{a}}){Clayton}, {Sugerman},
  {Stanford}, {Whitney}, {Honor}, {Babler}, {Barlow}, {Gordon}, {Andrews},
  {Geballe}, {Bond}, {De Marco}, {Lawson}, {Sibthorpe}, {Olofsson},
  {Polehampton}, {Gomez}, {Matsuura}, {Hargrave}, {Ivison}, {Wesson}, {Leeks},
  {Swinyard}, \& {Lim}}]{2011ApJ...743...44C}
{Clayton}, G.~C., {et~al.} 2011{\natexlab{a}}, \apj, 743, 44

\bibitem[{{Clayton} {et~al.}(2011{\natexlab{b}}){Clayton}, {De Marco},
  {Whitney}, {Babler}, {Gallagher}, {Nordhaus}, {Speck}, {Wolff}, {Freeman},
  {Camp}, {Lawson}, {Roman-Duval}, {Misselt}, {Meade}, {Sonneborn}, {Matsuura},
  \& {Meixner}}]{Clayton:2011lr}
---. 2011{\natexlab{b}}, AJ, 142, 54

\bibitem[{{Clayton} {et~al.}(2013{\natexlab{b}}){Clayton}, {Bond}, {Long},
  {Meyer}, {Sugerman}, {Montiel}, {Sparks}, {Meakes}, {Chesneau}, \& {De
  Marco}}]{2013ApJ...771..130C}
---. 2013{\natexlab{b}}, \apj, 771, 130

\bibitem[{{Cohen} {et~al.}(2003){Cohen}, {Wheaton}, \&
  {Megeath}}]{2003AJ....126.1090C}
{Cohen}, M., {Wheaton}, W.~A., \& {Megeath}, S.~T. 2003, \aj, 126, 1090

\bibitem[{{Cox} {et~al.}(2012){Cox}, {Kerschbaum}, {van Marle}, {Decin},
  {Ladjal}, {Mayer}, {Groenewegen}, {van Eck}, {Royer}, {Ottensamer}, {Ueta},
  {Jorissen}, {Mecina}, {Meliani}, {Luntzer}, {Blommaert}, {Posch},
  {Vandenbussche}, \& {Waelkens}}]{2012A&A...537A..35C}
{Cox}, N.~L.~J., {et~al.} 2012, \aap, 537, A35

\bibitem[{{Cutri}(2014)}]{2014yCat.2328....0C}
{Cutri}, R.~M. 2014, VizieR Online Data Catalog, 2328

\bibitem[{{Cutri} {et~al.}(2003){Cutri}, {Skrutskie}, {van Dyk}, {Beichman},
  {Carpenter}, {Chester}, {Cambresy}, {Evans}, {Fowler}, {Gizis}, {Howard},
  {Huchra}, {Jarrett}, {Kopan}, {Kirkpatrick}, {Light}, {Marsh}, {McCallon},
  {Schneider}, {Stiening}, {Sykes}, {Weinberg}, {Wheaton}, {Wheelock}, \&
  {Zacarias}}]{2003tmc..book.....C}
{Cutri}, R.~M., {et~al.} 2003, {2MASS All Sky Catalog of point sources.}

\bibitem[{{Cutri} {et~al.}(2012){Cutri}, {Wright}, {Conrow}, {Bauer},
  {Benford}, {Brandenburg}, {Dailey}, {Eisenhardt}, {Evans}, {Fajardo-Acosta},
  {Fowler}, {Gelino}, {Grillmair}, {Harbut}, {Hoffman}, {Jarrett},
  {Kirkpatrick}, {Leisawitz}, {Liu}, {Mainzer}, {Marsh}, {Masci}, {McCallon},
  {Padgett}, {Ressler}, {Royer}, {Skrutskie}, {Stanford}, {Wyatt}, {Tholen},
  {Tsai}, {Wachter}, {Wheelock}, {Yan}, {Alles}, {Beck}, {Grav}, {Masiero},
  {McCollum}, {McGehee}, {Papin}, \& {Wittman}}]{2012wise.rept....1C}
---. 2012, {Explanatory Supplement to the WISE All-Sky Data Release Products},
  Tech. rep.

\bibitem[{{De Marco} {et~al.}(2002){De Marco}, {Clayton}, {Herwig}, {Pollacco},
  {Clark}, \& {Kilkenny}}]{2002AJ....123.3387D}
{De Marco}, O., {Clayton}, G.~C., {Herwig}, F., {Pollacco}, D.~L., {Clark},
  J.~S., \& {Kilkenny}, D. 2002, AJ, 123, 3387

\bibitem[{{Debes} {et~al.}(2015){Debes}, {Kilic}, {Tremblay},
  {L{\'o}pez-Morales}, {Anglada-Escude}, {Napiwotzki}, {Osip}, \&
  {Weinberger}}]{2015AJ....149..176D}
{Debes}, J.~H., {Kilic}, M., {Tremblay}, P.-E., {L{\'o}pez-Morales}, M.,
  {Anglada-Escude}, G., {Napiwotzki}, R., {Osip}, D., \& {Weinberger}, A. 2015,
  \aj, 149, 176

\bibitem[{{Decin} {et~al.}(2012){Decin}, {Cox}, {Royer}, {Van Marle},
  {Vandenbussche}, {Ladjal}, {Kerschbaum}, {Ottensamer}, {Barlow}, {Blommaert},
  {Gomez}, {Groenewegen}, {Lim}, {Swinyard}, {Waelkens}, \&
  {Tielens}}]{2012A&A...548A.113D}
{Decin}, L., {et~al.} 2012, \aap, 548, A113

\bibitem[{{Diolaiti} {et~al.}(2000{\natexlab{a}}){Diolaiti}, {Bendinelli},
  {Bonaccini}, {Close}, {Currie}, \& {Parmeggiani}}]{2000ASPC..216..623D}
{Diolaiti}, E., {Bendinelli}, O., {Bonaccini}, D., {Close}, L., {Currie}, D.,
  \& {Parmeggiani}, G. 2000{\natexlab{a}}, in Astronomical Society of the
  Pacific Conference Series, Vol. 216, Astronomical Data Analysis Software and
  Systems IX, ed. N.~{Manset}, C.~{Veillet}, \& D.~{Crabtree}, 623

\bibitem[{{Diolaiti} {et~al.}(2000{\natexlab{b}}){Diolaiti}, {Bendinelli},
  {Bonaccini}, {Close}, {Currie}, \& {Parmeggiani}}]{2000SPIE.4007..879D}
{Diolaiti}, E., {Bendinelli}, O., {Bonaccini}, D., {Close}, L.~M., {Currie},
  D.~G., \& {Parmeggiani}, G. 2000{\natexlab{b}}, in \procspie, Vol. 4007,
  Adaptive Optical Systems Technology, ed. P.~L. {Wizinowich}, 879--888

\bibitem[{{Drilling} {et~al.}(1984){Drilling}, {Schonberner}, {Heber}, \&
  {Lynas-Gray}}]{1984ApJ...278..224D}
{Drilling}, J.~S., {Schonberner}, D., {Heber}, U., \& {Lynas-Gray}, A.~E. 1984,
  \apj, 278, 224

\bibitem[{{Engelbracht} {et~al.}(2007){Engelbracht}, {Blaylock}, {Su}, {Rho},
  {Rieke}, {Muzerolle}, {Padgett}, {Hines}, {Gordon}, {Fadda},
  {Noriega-Crespo}, {Kelly}, {Latter}, {Hinz}, {Misselt}, {Morrison},
  {Stansberry}, {Shupe}, {Stolovy}, {Wheaton}, {Young}, {Neugebauer},
  {Wachter}, {P{\'e}rez-Gonz{\'a}lez}, {Frayer}, \&
  {Marleau}}]{2007PASP..119..994E}
{Engelbracht}, C.~W., {et~al.} 2007, \pasp, 119, 994

\bibitem[{{Ercolano} {et~al.}(2005){Ercolano}, {Barlow}, \&
  {Storey}}]{2005MNRAS.362.1038E}
{Ercolano}, B., {Barlow}, M.~J., \& {Storey}, P.~J. 2005, MNRAS, 362, 1038

\bibitem[{{Ercolano} {et~al.}(2003){Ercolano}, {Barlow}, {Storey}, \&
  {Liu}}]{2003MNRAS.340.1136E}
{Ercolano}, B., {Barlow}, M.~J., {Storey}, P.~J., \& {Liu}, X.-W. 2003, \mnras,
  340, 1136

\bibitem[{{Ercolano} {et~al.}(2008){Ercolano}, {Young}, {Drake}, \&
  {Raymond}}]{2008ApJS..175..534E}
{Ercolano}, B., {Young}, P.~R., {Drake}, J.~J., \& {Raymond}, J.~C. 2008, ApJS,
  175, 534

\bibitem[{{Fazio} {et~al.}(2004){Fazio}, {Hora}, {Allen}, {Ashby}, {Barmby},
  {Deutsch}, {Huang}, {Kleiner}, {Marengo}, {Megeath}, {Melnick}, {Pahre},
  {Patten}, {Polizotti}, {Smith}, {Taylor}, {Wang}, {Willner}, {Hoffmann},
  {Pipher}, {Forrest}, {McMurty}, {McCreight}, {McKelvey}, {McMurray}, {Koch},
  {Moseley}, {Arendt}, {Mentzell}, {Marx}, {Losch}, {Mayman}, {Eichhorn},
  {Krebs}, {Jhabvala}, {Gezari}, {Fixsen}, {Flores}, {Shakoorzadeh}, {Jungo},
  {Hakun}, {Workman}, {Karpati}, {Kichak}, {Whitley}, {Mann}, {Tollestrup},
  {Eisenhardt}, {Stern}, {Gorjian}, {Bhattacharya}, {Carey}, {Nelson},
  {Glaccum}, {Lacy}, {Lowrance}, {Laine}, {Reach}, {Stauffer}, {Surace},
  {Wilson}, {Wright}, {Hoffman}, {Domingo}, \& {Cohen}}]{2004ApJS..154...10F}
{Fazio}, G.~G., {et~al.} 2004, \apjs, 154, 10

\bibitem[{{Feast}(1997)}]{1997MNRAS.285..339F}
{Feast}, M.~W. 1997, \mnras, 285, 339

\bibitem[{{Feast} {et~al.}(1997){Feast}, {Carter}, {Roberts}, {Marang}, \&
  {Catchpole}}]{Feast:1997lr}
{Feast}, M.~W., {Carter}, B.~S., {Roberts}, G., {Marang}, F., \& {Catchpole},
  R.~M. 1997, \mnras, 285, 317

\bibitem[{{Ford}(1971)}]{1971ApJ...170..547F}
{Ford}, H.~C. 1971, \apj, 170, 547

\bibitem[{{Fujimoto}(1977)}]{Fujimoto:1977lr}
{Fujimoto}, M.~Y. 1977, PASJ, 29, 331

\bibitem[{{Gaposchkin}(1952)}]{1952AnHar.115...61G}
{Gaposchkin}, S. 1952, Annals of Harvard College Observatory, 115, 61

\bibitem[{{Garc{\'{\i}}a-Hern{\'a}ndez}
  {et~al.}(2009){Garc{\'{\i}}a-Hern{\'a}ndez}, {Hinkle}, {Lambert}, \&
  {Eriksson}}]{2009ApJ...696.1733G}
{Garc{\'{\i}}a-Hern{\'a}ndez}, D.~A., {Hinkle}, K.~H., {Lambert}, D.~L., \&
  {Eriksson}, K. 2009, \apj, 696, 1733

\bibitem[{{Garc{\'{\i}}a-Hern{\'a}ndez}
  {et~al.}(2011{\natexlab{a}}){Garc{\'{\i}}a-Hern{\'a}ndez}, {Kameswara Rao},
  \& {Lambert}}]{2011ApJ...729..126G}
{Garc{\'{\i}}a-Hern{\'a}ndez}, D.~A., {Kameswara Rao}, N., \& {Lambert}, D.~L.
  2011{\natexlab{a}}, \apj, 729, 126

\bibitem[{{Garc{\'{\i}}a-Hern{\'a}ndez}
  {et~al.}(2011{\natexlab{b}}){Garc{\'{\i}}a-Hern{\'a}ndez}, {Kameswara Rao},
  \& {Lambert}}]{2011ApJ...739...37G}
---. 2011{\natexlab{b}}, \apj, 739, 37

\bibitem[{{Garc{\'{\i}}a-Hern{\'a}ndez}
  {et~al.}(2010){Garc{\'{\i}}a-Hern{\'a}ndez}, {Lambert}, {Kameswara Rao},
  {Hinkle}, \& {Eriksson}}]{Garcia-Hernandez:2010fk}
{Garc{\'{\i}}a-Hern{\'a}ndez}, D.~A., {Lambert}, D.~L., {Kameswara Rao}, N.,
  {Hinkle}, K.~H., \& {Eriksson}, K. 2010, \apj, 714, 144

\bibitem[{{Garc{\'{\i}}a-Hern{\'a}ndez}
  {et~al.}(2013){Garc{\'{\i}}a-Hern{\'a}ndez}, {Rao}, \&
  {Lambert}}]{2013ApJ...773..107G}
{Garc{\'{\i}}a-Hern{\'a}ndez}, D.~A., {Rao}, N.~K., \& {Lambert}, D.~L. 2013,
  \apj, 773, 107

\bibitem[{{Gillett} {et~al.}(1986){Gillett}, {Backman}, {Beichman}, \&
  {Neugebauer}}]{Gillett:1986cr}
{Gillett}, F.~C., {Backman}, D.~E., {Beichman}, C., \& {Neugebauer}, G. 1986,
  \apj, 310, 842

\bibitem[{{Goldsmith} {et~al.}(1990){Goldsmith}, {Evans}, {Albinson}, \&
  {Bode}}]{1990MNRAS.245..119G}
{Goldsmith}, M.~J., {Evans}, A., {Albinson}, J.~S., \& {Bode}, M.~F. 1990,
  \mnras, 245, 119

\bibitem[{{Gonzalez} {et~al.}(1998){Gonzalez}, {Lambert}, {Wallerstein}, {Rao},
  {Smith}, \& {McCarthy}}]{1998ApJS..114..133G}
{Gonzalez}, G., {Lambert}, D.~L., {Wallerstein}, G., {Rao}, N.~K., {Smith},
  V.~V., \& {McCarthy}, J.~K. 1998, ApJS, 114, 133

\bibitem[{{Gordon} {et~al.}(2005){Gordon}, {Rieke}, {Engelbracht}, {Muzerolle},
  {Stansberry}, {Misselt}, {Morrison}, {Cadien}, {Young}, {Dole}, {Kelly},
  {Alonso-Herrero}, {Egami}, {Su}, {Papovich}, {Smith}, {Hines}, {Rieke},
  {Blaylock}, {P{\'e}rez-Gonz{\'a}lez}, {Le Floc'h}, {Hinz}, {Latter},
  {Hesselroth}, {Frayer}, {Noriega-Crespo}, {Masci}, {Padgett}, {Smylie}, \&
  {Haegel}}]{2005PASP..117..503G}
{Gordon}, K.~D., {et~al.} 2005, \pasp, 117, 503

\bibitem[{{Gordon} {et~al.}(2007){Gordon}, {Engelbracht}, {Fadda},
  {Stansberry}, {Wachter}, {Frayer}, {Rieke}, {Noriega-Crespo}, {Latter},
  {Young}, {Neugebauer}, {Balog}, {Beeman}, {Dole}, {Egami}, {Haller}, {Hines},
  {Kelly}, {Marleau}, {Misselt}, {Morrison}, {P{\'e}rez-Gonz{\'a}lez}, {Rho},
  \& {Wheaton}}]{2007PASP..119.1019G}
---. 2007, \pasp, 119, 1019

\bibitem[{{Goswami} {et~al.}(2010){Goswami}, {Karinkuzhi}, \&
  {Shantikumar}}]{2010ApJ...723L.238G}
{Goswami}, A., {Karinkuzhi}, D., \& {Shantikumar}, N.~S. 2010, \apjl, 723, L238

\bibitem[{{Griffin} {et~al.}(2010){Griffin}, {Abergel}, {Abreu}, {Ade},
  {Andr{\'e}}, {Augueres}, {Babbedge}, {Bae}, {Baillie}, {Baluteau}, {Barlow},
  {Bendo}, {Benielli}, {Bock}, {Bonhomme}, {Brisbin}, {Brockley-Blatt},
  {Caldwell}, {Cara}, {Castro-Rodriguez}, {Cerulli}, {Chanial}, {Chen},
  {Clark}, {Clements}, {Clerc}, {Coker}, {Communal}, {Conversi}, {Cox},
  {Crumb}, {Cunningham}, {Daly}, {Davis}, {de Antoni}, {Delderfield}, {Devin},
  {di Giorgio}, {Didschuns}, {Dohlen}, {Donati}, {Dowell}, {Dowell}, {Duband},
  {Dumaye}, {Emery}, {Ferlet}, {Ferrand}, {Fontignie}, {Fox}, {Franceschini},
  {Frerking}, {Fulton}, {Garcia}, {Gastaud}, {Gear}, {Glenn}, {Goizel},
  {Griffin}, {Grundy}, {Guest}, {Guillemet}, {Hargrave}, {Harwit}, {Hastings},
  {Hatziminaoglou}, {Herman}, {Hinde}, {Hristov}, {Huang}, {Imhof}, {Isaak},
  {Israelsson}, {Ivison}, {Jennings}, {Kiernan}, {King}, {Lange}, {Latter},
  {Laurent}, {Laurent}, {Leeks}, {Lellouch}, {Levenson}, {Li}, {Li},
  {Lilienthal}, {Lim}, {Liu}, {Lu}, {Madden}, {Mainetti}, {Marliani}, {McKay},
  {Mercier}, {Molinari}, {Morris}, {Moseley}, {Mulder}, {Mur}, {Naylor},
  {Nguyen}, {O'Halloran}, {Oliver}, {Olofsson}, {Olofsson}, {Orfei}, {Page},
  {Pain}, {Panuzzo}, {Papageorgiou}, {Parks}, {Parr-Burman}, {Pearce},
  {Pearson}, {P{\'e}rez-Fournon}, {Pinsard}, {Pisano}, {Podosek}, {Pohlen},
  {Polehampton}, {Pouliquen}, {Rigopoulou}, {Rizzo}, {Roseboom}, {Roussel},
  {Rowan-Robinson}, {Rownd}, {Saraceno}, {Sauvage}, {Savage}, {Savini},
  {Sawyer}, {Scharmberg}, {Schmitt}, {Schneider}, {Schulz}, {Schwartz},
  {Shafer}, {Shupe}, {Sibthorpe}, {Sidher}, {Smith}, {Smith}, {Smith},
  {Spencer}, {Stobie}, {Sudiwala}, {Sukhatme}, {Surace}, {Stevens}, {Swinyard},
  {Trichas}, {Tourette}, {Triou}, {Tseng}, {Tucker}, {Turner}, {Vaccari},
  {Valtchanov}, {Vigroux}, {Virique}, {Voellmer}, {Walker}, {Ward}, {Waskett},
  {Weilert}, {Wesson}, {White}, {Whitehouse}, {Wilson}, {Winter}, {Woodcraft},
  {Wright}, {Xu}, {Zavagno}, {Zemcov}, {Zhang}, \&
  {Zonca}}]{2010A&A...518L...3G}
{Griffin}, M.~J., {et~al.} 2010, \aap, 518, L3

\bibitem[{{Hecht} {et~al.}(1984){Hecht}, {Holm}, {Donn}, \&
  {Wu}}]{1984ApJ...280..228H}
{Hecht}, J.~H., {Holm}, A.~V., {Donn}, B., \& {Wu}, C.-C. 1984, \apj, 280, 228

\bibitem[{{Helou} \& {Walker}(1988)}]{1988iras....7.....H}
{Helou}, G., \& {Walker}, D.~W., eds. 1988, {Infrared astronomical satellite
  (IRAS) catalogs and atlases. Volume 7: The small scale structure catalog},
  Vol.~7

\bibitem[{{Hema} {et~al.}(2012){Hema}, {Pandey}, \&
  {Lambert}}]{2012ApJ...747..102H}
{Hema}, B.~P., {Pandey}, G., \& {Lambert}, D.~L. 2012, \apj, 747, 102

\bibitem[{{Herbig}(1964)}]{1964ApJ...140.1317H}
{Herbig}, G.~H. 1964, \apj, 140, 1317

\bibitem[{{Higdon} {et~al.}(2004){Higdon}, {Devost}, {Higdon}, {Brandl},
  {Houck}, {Hall}, {Barry}, {Charmandaris}, {Smith}, {Sloan}, \&
  {Green}}]{2004PASP..116..975H}
{Higdon}, S.~J.~U., {et~al.} 2004, \pasp, 116, 975

\bibitem[{{Hinkle} {et~al.}(2001){Hinkle}, {Joyce}, \&
  {Hedden}}]{2001A&A...367..250H}
{Hinkle}, K.~H., {Joyce}, R.~R., \& {Hedden}, A. 2001, \aap, 367, 250

\bibitem[{{Hinkle} {et~al.}(2008){Hinkle}, {Lebzelter}, {Joyce}, {Ridgway},
  {Close}, {Hron}, \& {Andre}}]{2008A&A...479..817H}
{Hinkle}, K.~H., {Lebzelter}, T., {Joyce}, R.~R., {Ridgway}, S., {Close}, L.,
  {Hron}, J., \& {Andre}, K. 2008, \aap, 479, 817

\bibitem[{{Hoffleit}(1958)}]{1958AJ.....63Q..50H}
{Hoffleit}, D. 1958, \aj, 63, 50

\bibitem[{{Hoffleit}(1959)}]{1959AJ.....64..241H}
---. 1959, \aj, 64, 241

\bibitem[{{Holm}(1999)}]{1999JAVSO..27..113H}
{Holm}, A.~V. 1999, Journal of the American Association of Variable Star
  Observers (JAAVSO), 27, 113

\bibitem[{{Houck} {et~al.}(2004){Houck}, {Roellig}, {van Cleve}, {Forrest},
  {Herter}, {Lawrence}, {Matthews}, {Reitsema}, {Soifer}, {Watson}, {Weedman},
  {Huisjen}, {Troeltzsch}, {Barry}, {Bernard-Salas}, {Blacken}, {Brandl},
  {Charmandaris}, {Devost}, {Gull}, {Hall}, {Henderson}, {Higdon}, {Pirger},
  {Schoenwald}, {Sloan}, {Uchida}, {Appleton}, {Armus}, {Burgdorf},
  {Fajardo-Acosta}, {Grillmair}, {Ingalls}, {Morris}, \&
  {Teplitz}}]{Houck:2004lr}
{Houck}, J.~R., {et~al.} 2004, \apjs, 154, 18

\bibitem[{{Iben} {et~al.}(1996){Iben}, {Tutukov}, \&
  {Yungelson}}]{1996ApJ...456..750I}
{Iben}, Jr., I., {Tutukov}, A.~V., \& {Yungelson}, L.~R. 1996, ApJ, 456, 750

\bibitem[{{Ishihara} {et~al.}(2010){Ishihara}, {Onaka}, {Kataza}, {Salama},
  {Alfageme}, {Cassatella}, {Cox}, {Garc{\'{\i}}a-Lario}, {Stephenson},
  {Cohen}, {Fujishiro}, {Fujiwara}, {Hasegawa}, {Ita}, {Kim}, {Matsuhara},
  {Murakami}, {M{\"u}ller}, {Nakagawa}, {Ohyama}, {Oyabu}, {Pyo}, {Sakon},
  {Shibai}, {Takita}, {Tanab{\'e}}, {Uemizu}, {Ueno}, {Usui}, {Wada},
  {Watarai}, {Yamamura}, \& {Yamauchi}}]{2010A&A...514A...1I}
{Ishihara}, D., {et~al.} 2010, \aap, 514, A1

\bibitem[{{Jeffery}(1995)}]{1995A&A...299..135J}
{Jeffery}, C.~S. 1995, \aap, 299, 135

\bibitem[{{Jurcsik}(1996)}]{1996AcA....46..325J}
{Jurcsik}, J. 1996, Acta Astronomica, 46, 325

\bibitem[{{Karakas} {et~al.}(2015){Karakas}, {Ruiter}, \&
  {Hampel}}]{2015ApJ...809..184K}
{Karakas}, A.~I., {Ruiter}, A.~J., \& {Hampel}, M. 2015, \apj, 809, 184

\bibitem[{{Kawada} {et~al.}(2007){Kawada}, {Baba}, {Barthel}, {Clements},
  {Cohen}, {Doi}, {Figueredo}, {Fujiwara}, {Goto}, {Hasegawa}, {Hibi}, {Hirao},
  {Hiromoto}, {Jeong}, {Kaneda}, {Kawai}, {Kawamura}, {Kester}, {Kii},
  {Kobayashi}, {Kwon}, {Lee}, {Makiuti}, {Matsuo}, {Matsuura}, {M{\"u}ller},
  {Murakami}, {Nagata}, {Nakagawa}, {Narita}, {Noda}, {Oh}, {Okada}, {Okuda},
  {Oliver}, {Ootsubo}, {Pak}, {Park}, {Pearson}, {Rowan-Robinson}, {Saito},
  {Salama}, {Sato}, {Savage}, {Serjeant}, {Shibai}, {Shirahata}, {Sohn},
  {Suzuki}, {Takagi}, {Takahashi}, {Thomson}, {Usui}, {Verdugo}, {Watabe},
  {White}, {Wang}, {Yamamura}, {Yamauchi}, \& {Yasuda}}]{2007PASJ...59S.389K}
{Kawada}, M., {et~al.} 2007, \pasj, 59, S389

\bibitem[{{Keenan} \& {Greenstein}(1963)}]{Kennan:1963}
{Keenan}, P.~C., \& {Greenstein}, J.~L. 1963, Contr. Perk. Obs., 2

\bibitem[{{Kilkenny} \& {Whittet}(1984)}]{1984MNRAS.208...25K}
{Kilkenny}, D., \& {Whittet}, D.~C.~B. 1984, \mnras, 208, 25

\bibitem[{{Lambert} \& {Rao}(1994)}]{Lambert:1994uq}
{Lambert}, D.~L., \& {Rao}, N.~K. 1994, Journal of Astrophysics and Astronomy,
  15, 47

\bibitem[{{Lambert} {et~al.}(2001){Lambert}, {Rao}, {Pandey}, \&
  {Ivans}}]{2001ApJ...555..925L}
{Lambert}, D.~L., {Rao}, N.~K., {Pandey}, G., \& {Ivans}, I.~I. 2001, \apj,
  555, 925

\bibitem[{{Landolt} \& {Clem}(2017)}]{2017JAVSO..45..159L}
{Landolt}, A.~U., \& {Clem}, J.~L. 2017, Journal of the American Association of
  Variable Star Observers (JAAVSO), 45, 159

\bibitem[{{Lawson} \& {Cottrell}(1989)}]{Lawson:1989kx}
{Lawson}, W.~A., \& {Cottrell}, P.~L. 1989, \mnras, 240, 689

\bibitem[{{Lawson} {et~al.}(1990){Lawson}, {Cottrell}, {Kilmartin}, \&
  {Gilmore}}]{1990MNRAS.247...91L}
{Lawson}, W.~A., {Cottrell}, P.~L., {Kilmartin}, P.~M., \& {Gilmore}, A.~C.
  1990, MNRAS, 247, 91

\bibitem[{{Lawson} {et~al.}(1999){Lawson}, {Maldoni}, {Clayton}, {Valencic},
  {Jones}, {Kilkenny}, {van Wyk}, {Roberts}, \& {Marang}}]{1999AJ....117.3007L}
{Lawson}, W.~A., {et~al.} 1999, \aj, 117, 3007

\bibitem[{{Leavitt} \& {Pickering}(1906)}]{1906HarCi.122....1L}
{Leavitt}, H.~S., \& {Pickering}, E.~C. 1906, Harvard College Observatory
  Circular, 122, 1

\bibitem[{{Lebouteiller} {et~al.}(2011){Lebouteiller}, {Barry}, {Spoon},
  {Bernard-Salas}, {Sloan}, {Houck}, \& {Weedman}}]{2011ApJS..196....8L}
{Lebouteiller}, V., {Barry}, D.~J., {Spoon}, H.~W.~W., {Bernard-Salas}, J.,
  {Sloan}, G.~C., {Houck}, J.~R., \& {Weedman}, D.~W. 2011, \apjs, 196, 8

\bibitem[{{Lebouteiller} {et~al.}(2010){Lebouteiller}, {Bernard-Salas},
  {Sloan}, \& {Barry}}]{2010PASP..122..231L}
{Lebouteiller}, V., {Bernard-Salas}, J., {Sloan}, G.~C., \& {Barry}, D.~J.
  2010, \pasp, 122, 231

\bibitem[{{Leech} {et~al.}(2003){Leech}, {Kester}, {Shipman}, {Beintema},
  {Feuchtgruber}, {Heras}, {Huygen}, {Lahuis}, {Lutz}, {Morris}, {Roelfsema},
  {Salama}, {Schaeidt}, {Valentijn}, {Vandenbussche}, {Wieprecht}, \& {de
  Graauw}}]{2003sws..bookR....L}
{Leech}, K., {et~al.} 2003, {The ISO Handbook, Volume V - SWS - The Short
  Wavelength Spectrometer}

\bibitem[{{Loreta}(1935)}]{1935AN....254..151L}
{Loreta}, E. 1935, Astronomische Nachrichten, 254, 151

\bibitem[{{Lundmark}(1921)}]{1921PASP...33..314L}
{Lundmark}, K. 1921, \pasp, 33, 314

\bibitem[{{Marang} {et~al.}(1990){Marang}, {Kilkenny}, {Menzies}, \&
  {Jones}}]{1990SAAOC..14....1M}
{Marang}, F., {Kilkenny}, D., {Menzies}, J.~W., \& {Jones}, J.~H.~S. 1990,
  South African Astronomical Observatory Circular, 14, 1

\bibitem[{{Mathis} {et~al.}(1977){Mathis}, {Rumpl}, \&
  {Nordsieck}}]{1977ApJ...217..425M}
{Mathis}, J.~S., {Rumpl}, W., \& {Nordsieck}, K.~H. 1977, \apj, 217, 425

\bibitem[{{McNaught}(1986)}]{1986IAUC.4245....2M}
{McNaught}, R. 1986, \iaucirc, 4245

\bibitem[{{McNaught} \& {Dawes}(1986)}]{1986IBVS.2928....1M}
{McNaught}, R.~H., \& {Dawes}, G. 1986, Information Bulletin on Variable Stars,
  2928

\bibitem[{{Menzies} \& {Feast}(1997)}]{1997MNRAS.285..358M}
{Menzies}, J.~W., \& {Feast}, M.~W. 1997, \mnras, 285, 358

\bibitem[{{Mitronova} {et~al.}(2004){Mitronova}, {Karachentsev},
  {Karachentseva}, {Jarrett}, \& {Kudrya}}]{2004BSAO...57....5M}
{Mitronova}, S.~N., {Karachentsev}, I.~D., {Karachentseva}, V.~E., {Jarrett},
  T.~H., \& {Kudrya}, Y.~N. 2004, Bulletin of the Special Astrophysics
  Observatory, 57, 5

\bibitem[{{Montiel} {et~al.}(2015){Montiel}, {Clayton}, {Marcello}, \&
  {Lockman}}]{2015AJ....150...14M}
{Montiel}, E.~J., {Clayton}, G.~C., {Marcello}, D.~C., \& {Lockman}, F.~J.
  2015, \aj, 150, 14

\bibitem[{{Moshir} \& {et al.}(1990)}]{1990IRASF.C......0M}
{Moshir}, M., \& {et al.} 1990, in IRAS Faint Source Catalogue, version 2.0
  (1990)

\bibitem[{{Murakami} {et~al.}(2007){Murakami}, {Baba}, {Barthel}, {Clements},
  {Cohen}, {Doi}, {Enya}, {Figueredo}, {Fujishiro}, {Fujiwara}, {Fujiwara},
  {Garcia-Lario}, {Goto}, {Hasegawa}, {Hibi}, {Hirao}, {Hiromoto}, {Hong},
  {Imai}, {Ishigaki}, {Ishiguro}, {Ishihara}, {Ita}, {Jeong}, {Jeong},
  {Kaneda}, {Kataza}, {Kawada}, {Kawai}, {Kawamura}, {Kessler}, {Kester},
  {Kii}, {Kim}, {Kim}, {Kobayashi}, {Koo}, {Kwon}, {Lee}, {Lorente}, {Makiuti},
  {Matsuhara}, {Matsumoto}, {Matsuo}, {Matsuura}, {M{\"u}ller}, {Murakami},
  {Nagata}, {Nakagawa}, {Naoi}, {Narita}, {Noda}, {Oh}, {Ohnishi}, {Ohyama},
  {Okada}, {Okuda}, {Oliver}, {Onaka}, {Ootsubo}, {Oyabu}, {Pak}, {Park},
  {Pearson}, {Rowan-Robinson}, {Saito}, {Sakon}, {Salama}, {Sato}, {Savage},
  {Serjeant}, {Shibai}, {Shirahata}, {Sohn}, {Suzuki}, {Takagi}, {Takahashi},
  {Tanab{\'e}}, {Takeuchi}, {Takita}, {Thomson}, {Uemizu}, {Ueno}, {Usui},
  {Verdugo}, {Wada}, {Wang}, {Watabe}, {Watarai}, {White}, {Yamamura},
  {Yamauchi}, \& {Yasuda}}]{2007PASJ...59S.369M}
{Murakami}, H., {et~al.} 2007, \pasj, 59, 369

\bibitem[{{Nagendra} \& {Leung}(1996)}]{1996MNRAS.281.1139N}
{Nagendra}, K.~N., \& {Leung}, C.~M. 1996, \mnras, 281, 1139

\bibitem[{{Neugebauer} {et~al.}(1984){Neugebauer}, {Habing}, {van Duinen},
  {Aumann}, {Baud}, {Beichman}, {Beintema}, {Boggess}, {Clegg}, {de Jong},
  {Emerson}, {Gautier}, {Gillett}, {Harris}, {Hauser}, {Houck}, {Jennings},
  {Low}, {Marsden}, {Miley}, {Olnon}, {Pottasch}, {Raimond}, {Rowan-Robinson},
  {Soifer}, {Walker}, {Wesselius}, \& {Young}}]{1984ApJ...278L...1N}
{Neugebauer}, G., {et~al.} 1984, \apjl, 278, L1

\bibitem[{{O'Keefe}(1939)}]{1939ApJ....90..294O}
{O'Keefe}, J.~A. 1939, \apj, 90, 294

\bibitem[{{Onaka} {et~al.}(2007){Onaka}, {Matsuhara}, {Wada}, {Fujishiro},
  {Fujiwara}, {Ishigaki}, {Ishihara}, {Ita}, {Kataza}, {Kim}, {Matsumoto},
  {Murakami}, {Ohyama}, {Oyabu}, {Sakon}, {Tanab{\'e}}, {Takagi}, {Uemizu},
  {Ueno}, {Usui}, {Watarai}, {Cohen}, {Enya}, {Ootsubo}, {Pearson}, {Takeyama},
  {Yamamuro}, \& {Ikeda}}]{2007PASJ...59S.401O}
{Onaka}, T., {et~al.} 2007, \pasj, 59, S401

\bibitem[{{Ott}(2010)}]{2010ASPC..434..139O}
{Ott}, S. 2010, in Astronomical Society of the Pacific Conference Series, Vol.
  434, Astronomical Data Analysis Software and Systems XIX, ed. Y.~{Mizumoto},
  K.-I. {Morita}, \& M.~{Ohishi}, 139

\bibitem[{{Pandey} {et~al.}(1996){Pandey}, {Kameswara Rao}, \&
  {Lambert}}]{1996MNRAS.282..889P}
{Pandey}, G., {Kameswara Rao}, N., \& {Lambert}, D.~L. 1996, \mnras, 282, 889

\bibitem[{{Pandey} {et~al.}(2008){Pandey}, {Lambert}, \& {Rao}}]{Pandey:2008eu}
{Pandey}, G., {Lambert}, D.~L., \& {Rao}, N.~K. 2008, ApJ, 674, 1068

\bibitem[{{Pickering}(1896)}]{1896ApJ.....4..138P}
{Pickering}, E.~C. 1896, \apj, 4

\bibitem[{{Pickering} \& {Fleming}(1896)}]{1896ApJ.....4..142P}
{Pickering}, E.~C., \& {Fleming}, W.~P. 1896, \apj, 4

\bibitem[{{Pickering} \& {Leland}(1896)}]{1896ApJ.....4..234P}
{Pickering}, E.~C., \& {Leland}, E.~F. 1896, \apj, 4

\bibitem[{{Pigott} \& {Englefield}(1797)}]{1797RSPT...87..133P}
{Pigott}, E., \& {Englefield}, H.~C. 1797, Royal Society of London
  Philosophical Transactions Series I, 87, 133

\bibitem[{{Pilbratt} {et~al.}(2010){Pilbratt}, {Riedinger}, {Passvogel},
  {Crone}, {Doyle}, {Gageur}, {Heras}, {Jewell}, {Metcalfe}, {Ott}, \&
  {Schmidt}}]{2010A&A...518L...1P}
{Pilbratt}, G.~L., {et~al.} 2010, \aap, 518, L1

\bibitem[{{Poglitsch} {et~al.}(2010){Poglitsch}, {Waelkens}, {Geis},
  {Feuchtgruber}, {Vandenbussche}, {Rodriguez}, {Krause}, {Renotte}, {van
  Hoof}, {Saraceno}, {Cepa}, {Kerschbaum}, {Agn{\`e}se}, {Ali}, {Altieri},
  {Andreani}, {Augueres}, {Balog}, {Barl}, {Bauer}, {Belbachir}, {Benedettini},
  {Billot}, {Boulade}, {Bischof}, {Blommaert}, {Callut}, {Cara}, {Cerulli},
  {Cesarsky}, {Contursi}, {Creten}, {De Meester}, {Doublier}, {Doumayrou},
  {Duband}, {Exter}, {Genzel}, {Gillis}, {Gr{\"o}zinger}, {Henning},
  {Herreros}, {Huygen}, {Inguscio}, {Jakob}, {Jamar}, {Jean}, {de Jong},
  {Katterloher}, {Kiss}, {Klaas}, {Lemke}, {Lutz}, {Madden}, {Marquet},
  {Martignac}, {Mazy}, {Merken}, {Montfort}, {Morbidelli}, {M{\"u}ller},
  {Nielbock}, {Okumura}, {Orfei}, {Ottensamer}, {Pezzuto}, {Popesso},
  {Putzeys}, {Regibo}, {Reveret}, {Royer}, {Sauvage}, {Schreiber}, {Stegmaier},
  {Schmitt}, {Schubert}, {Sturm}, {Thiel}, {Tofani}, {Vavrek}, {Wetzstein},
  {Wieprecht}, \& {Wiezorrek}}]{2010A&A...518L...2P}
{Poglitsch}, A., {et~al.} 2010, \aap, 518, L2

\bibitem[{{Pollacco} {et~al.}(1991){Pollacco}, {Hill}, {Houziaux}, \&
  {Manfroid}}]{1991MNRAS.248P...1P}
{Pollacco}, D.~L., {Hill}, P.~W., {Houziaux}, L., \& {Manfroid}, J. 1991,
  MNRAS, 248, 1P

\bibitem[{{Rao} \& {Lambert}(2008{\natexlab{a}})}]{Rao:2008lr}
{Rao}, N.~K., \& {Lambert}, D.~L. 2008{\natexlab{a}}, \mnras, 384, 477

\bibitem[{{Rao} \& {Lambert}(2008{\natexlab{b}})}]{2008MNRAS.384..477R}
---. 2008{\natexlab{b}}, \mnras, 384, 477

\bibitem[{{Rao} \& {Lambert}(2015)}]{2015MNRAS.447.3664R}
---. 2015, \mnras, 447, 3664

\bibitem[{{Rao} \& {Nandy}(1986)}]{1986MNRAS.222..357K}
{Rao}, N.~K., \& {Nandy}, K. 1986, \mnras, 222, 357

\bibitem[{{Renzini}(1979)}]{1979sss..meet..155R}
{Renzini}, A. 1979, in ASSL Vol. 75: Stars and star systems, ed. B.~E.
  {Westerlund}, 155

\bibitem[{{Robitaille} {et~al.}(2006){Robitaille}, {Whitney}, {Indebetouw},
  {Wood}, \& {Denzmore}}]{2006ApJS..167..256R}
{Robitaille}, T.~P., {Whitney}, B.~A., {Indebetouw}, R., {Wood}, K., \&
  {Denzmore}, P. 2006, \apjs, 167, 256

\bibitem[{{Roussel}(2013)}]{2013PASP..125.1126R}
{Roussel}, H. 2013, \pasp, 125, 1126

\bibitem[{{Saio}(2008)}]{Saio:2008qe}
{Saio}, H. 2008, in A. S. P. Conf. Ser., Vol. 391, Hydrogen-Deficient Stars,
  ed. {A.~Werner \& T.~Rauch}, 69

\bibitem[{{Saio} \& {Jeffery}(2002)}]{2002MNRAS.333..121S}
{Saio}, H., \& {Jeffery}, C.~S. 2002, MNRAS, 333, 121

\bibitem[{{Schaefer}(1986)}]{Schaefer:1986lq}
{Schaefer}, B.~E. 1986, \apj, 307, 644

\bibitem[{{Schlafly} \& {Finkbeiner}(2011)}]{2011ApJ...737..103S}
{Schlafly}, E.~F., \& {Finkbeiner}, D.~P. 2011, \apj, 737, 103

\bibitem[{{Shears}(2011)}]{2011arXiv1109.4234S}
{Shears}, J. 2011, ArXiv e-prints

\bibitem[{{Skrutskie} {et~al.}(2006){Skrutskie}, {Cutri}, {Stiening},
  {Weinberg}, {Schneider}, {Carpenter}, {Beichman}, {Capps}, {Chester},
  {Elias}, {Huchra}, {Liebert}, {Lonsdale}, {Monet}, {Price}, {Seitzer},
  {Jarrett}, {Kirkpatrick}, {Gizis}, {Howard}, {Evans}, {Fowler}, {Fullmer},
  {Hurt}, {Light}, {Kopan}, {Marsh}, {McCallon}, {Tam}, {Van Dyk}, \&
  {Wheelock}}]{2006AJ....131.1163S}
{Skrutskie}, M.~F., {et~al.} 2006, \aj, 131, 1163

\bibitem[{{Sloan} {et~al.}(2003){Sloan}, {Kraemer}, {Price}, \&
  {Shipman}}]{2003ApJS..147..379S}
{Sloan}, G.~C., {Kraemer}, K.~E., {Price}, S.~D., \& {Shipman}, R.~F. 2003,
  \apjs, 147, 379

\bibitem[{{Stanford} {et~al.}(1988){Stanford}, {Clayton}, {Meade}, {Nordsieck},
  {Whitney}, {Murison}, {Nook}, \& {Anderson}}]{1988ApJ...325L...9S}
{Stanford}, S.~A., {Clayton}, G.~C., {Meade}, M.~R., {Nordsieck}, K.~H.,
  {Whitney}, B.~A., {Murison}, M.~A., {Nook}, M.~A., \& {Anderson}, C.~M. 1988,
  \apjl, 325, L9

\bibitem[{{Stansberry} {et~al.}(2007){Stansberry}, {Gordon}, {Bhattacharya},
  {Engelbracht}, {Rieke}, {Marleau}, {Fadda}, {Frayer}, {Noriega-Crespo},
  {Wachter}, {Young}, {M{\"u}ller}, {Kelly}, {Blaylock}, {Henderson},
  {Neugebauer}, {Beeman}, \& {Haller}}]{2007PASP..119.1038S}
{Stansberry}, J.~A., {et~al.} 2007, \pasp, 119, 1038

\bibitem[{{Sugerman} {et~al.}(2012){Sugerman}, {Andrews}, {Barlow}, {Clayton},
  {Ercolano}, {Ghavamian}, {Kennicutt}, {Krause}, {Meixner}, \&
  {Otsuka}}]{2012ApJ...749..170S}
{Sugerman}, B.~E.~K., {et~al.} 2012, \apj, 749, 170

\bibitem[{{Tisserand}(2012)}]{2012A&A...539A..51T}
{Tisserand}, P. 2012, \aap, 539, A51

\bibitem[{{Tisserand} {et~al.}(2009){Tisserand}, {Wood}, {Marquette}, {Afonso},
  {Albert}, {Andersen}, {Ansari}, {Aubourg}, {Bareyre}, {Beaulieu}, {Charlot},
  {Coutures}, {Ferlet}, {Fouqu{\'e}}, {Glicenstein}, {Goldman}, {Gould},
  {Gros}, {de Kat}, {Lesquoy}, {Loup}, {Magneville}, {Maurice}, {Maury},
  {Milsztajn}, {Moniez}, {Palanque-Delabrouille}, {Perdereau}, {Rich},
  {Schwemling}, {Spiro}, \& {Vidal-Madjar}}]{Tisserand:2009fj}
{Tisserand}, P., {et~al.} 2009, A\&A, 501, 985

\bibitem[{{Ueta} {et~al.}(2006){Ueta}, {Speck}, {Stencel}, {Herwig}, {Gehrz},
  {Szczerba}, {Izumiura}, {Zijlstra}, {Latter}, {Matsuura}, {Meixner},
  {Steffen}, \& {Elitzur}}]{2006ApJ...648L..39U}
{Ueta}, T., {et~al.} 2006, \apjl, 648, L39

\bibitem[{{van den Bergh}(1971)}]{1971PASP...83..819V}
{van den Bergh}, S. 1971, \pasp, 83, 819

\bibitem[{{Walker}(1985)}]{Walker:1985rr}
{Walker}, H.~J. 1985, \aap, 152, 58

\bibitem[{{Walker}(1986)}]{1986ASSL..128..407W}
{Walker}, H.~J. 1986, in Astrophysics and Space Science Library, Vol. 128, IAU
  Colloq. 87: Hydrogen Deficient Stars and Related Objects, ed. K.~{Hunger},
  D.~{Schoenberner}, \& N.~{Kameswara Rao}, 407

\bibitem[{{Walker} {et~al.}(1996){Walker}, {Heinrichsen}, {Richards}, {Klaas},
  \& {Rasmussen}}]{1996A&A...315L.249W}
{Walker}, H.~J., {Heinrichsen}, I., {Richards}, P.~J., {Klaas}, U., \&
  {Rasmussen}, I.~L. 1996, \aap, 315, L249

\bibitem[{{Warner}(1967)}]{Warner:1967lr}
{Warner}, B. 1967, \mnras, 137, 119

\bibitem[{{Webbink}(1984)}]{1984ApJ...277..355W}
{Webbink}, R.~F. 1984, ApJ, 277, 355

\bibitem[{{Werner} {et~al.}(2004){Werner}, {Roellig}, {Low}, {Rieke}, {Rieke},
  {Hoffmann}, {Young}, {Houck}, {Brandl}, {Fazio}, {Hora}, {Gehrz}, {Helou},
  {Soifer}, {Stauffer}, {Keene}, {Eisenhardt}, {Gallagher}, {Gautier}, {Irace},
  {Lawrence}, {Simmons}, {Van Cleve}, {Jura}, {Wright}, \&
  {Cruikshank}}]{2004ApJS..154....1W}
{Werner}, M.~W., {et~al.} 2004, \apjs, 154, 1

\bibitem[{{Whitney} {et~al.}(2003{\natexlab{a}}){Whitney}, {Wood}, {Bjorkman},
  \& {Cohen}}]{2003ApJ...598.1079W}
{Whitney}, B.~A., {Wood}, K., {Bjorkman}, J.~E., \& {Cohen}, M.
  2003{\natexlab{a}}, \apj, 598, 1079

\bibitem[{{Whitney} {et~al.}(2003{\natexlab{b}}){Whitney}, {Wood}, {Bjorkman},
  \& {Wolff}}]{2003ApJ...591.1049W}
{Whitney}, B.~A., {Wood}, K., {Bjorkman}, J.~E., \& {Wolff}, M.~J.
  2003{\natexlab{b}}, \apj, 591, 1049

\bibitem[{{Wolf}(1920)}]{1920AN....211..119W}
{Wolf}, M. 1920, Astronomische Nachrichten, 211, 119

\bibitem[{{Woods}(1921)}]{1921BHarO.753R...2W}
{Woods}, I. 1921, Harvard College Observatory Bulletin, 753, 2

\bibitem[{{Woods}(1928)}]{1928BHarO.855...22W}
{Woods}, I.~E. 1928, Harvard College Observatory Bulletin, 855, 22

\bibitem[{{Wright} {et~al.}(2010){Wright}, {Eisenhardt}, {Mainzer}, {Ressler},
  {Cutri}, {Jarrett}, {Kirkpatrick}, {Padgett}, {McMillan}, {Skrutskie},
  {Stanford}, {Cohen}, {Walker}, {Mather}, {Leisawitz}, {Gautier}, {McLean},
  {Benford}, {Lonsdale}, {Blain}, {Mendez}, {Irace}, {Duval}, {Liu}, {Royer},
  {Heinrichsen}, {Howard}, {Shannon}, {Kendall}, {Walsh}, {Larsen}, {Cardon},
  {Schick}, {Schwalm}, {Abid}, {Fabinsky}, {Naes}, \&
  {Tsai}}]{2010AJ....140.1868W}
{Wright}, E.~L., {et~al.} 2010, \aj, 140, 1868

\bibitem[{{Yamamura} {et~al.}(2009){Yamamura}, {Makiuti}, {Ikeda}, {Fukuda},
  {Yamauchi}, {Hasegawa}, {Nakagawa}, {Narumi}, {Baba}, {Takagi}, {Jeong},
  {Oh}, {Lee}, {Savage}, {Rahman}, {Thomson}, {Oliver}, {Figueredo},
  {Serjeant}, {White}, {Pearson}, {Wang}, {Rowan-Robinson}, {Kester}, {van der
  Wolk}, {Barthel}, {Salama}, {Alfageme}, {Garc{\'{\i}}a-Lario}, {Stephenson},
  {Cohen}, \& {Mueller}}]{2009ASPC..418....3Y}
{Yamamura}, I., {et~al.} 2009, in Astronomical Society of the Pacific
  Conference Series, Vol. 418, AKARI, a Light to Illuminate the Misty Universe,
  ed. T.~{Onaka}, G.~J. {White}, T.~{Nakagawa}, \& I.~{Yamamura}, 3

\bibitem[{{Young} {et~al.}(1993{\natexlab{a}}){Young}, {Phillips}, \&
  {Knapp}}]{1993ApJ...409..725Y}
{Young}, K., {Phillips}, T.~G., \& {Knapp}, G.~R. 1993{\natexlab{a}}, \apj,
  409, 725

\bibitem[{{Young} {et~al.}(1993{\natexlab{b}}){Young}, {Phillips}, \&
  {Knapp}}]{1993ApJS...86..517Y}
---. 1993{\natexlab{b}}, \apjs, 86, 517

\end{thebibliography}

\clearpage

\begin{deluxetable}{lcccccccc}
\tablecaption{Stellar Properties and New Observations of Sample RCB \& HdC Stars}
\tablenum{1}
\scriptsize
\tablehead{\colhead{Name}&\colhead{Right Ascension}&\colhead{Declination}&\colhead{[m$_V$]$_{\rm max}$}&\colhead{D$_{\rm Modeled}$}&\colhead{D$_{Gaia}$}&\colhead{L$_{\rm Modeled}$}&\colhead{T$_{eff}$}&\colhead{Observations\tablenotemark{1}} \\ \colhead{} & \colhead{(J2000)} & \colhead{(J2000)} & \colhead{} & \colhead{(kpc)} & \colhead{(kpc)} & \colhead{(L$_\odot$)} & \colhead{ (K)} & \colhead{}}
\startdata
%\begin{tabular}{  c | c | c | c | c | c | c | c }
%\hline
%Star        &     l            &       b         & [m$_V$]$_{\rm max}$ &  D             &   L                 & T$_{eff}$ & Observations  \\
 %              & ($^\circ$)  &  ($^\circ$) &                                     &    (kpc)     &  (L$_\odot$)  &        (K)    &                   \\ \hline \hline
MV~Sgr  &  18:44:31.97     &    $-$20:57:12.77   &                12.0              &    11.5      & $9.14\substack{+275.0 \\ -4.50}$ &      5200         &   16000   &   M      \\
%\hline
R~CrB    & 15:48:34.41 & $+$28:09:24.26      &                 5.8              &    1.40      &  $1.31\substack{+0.24 \\ -0.18}$     & 9150        &   6750     &   P        \\
%\hline
RY~Sgr  &   19:16:32.76 & $-$33:31:20.43    &                 6.5             &     1.50      &   $1.97\substack{+0.54 \\ -0.35}$    & 8900        &   7250 & M,P,S    \\
%\hline
SU~Tau &   05:49:03.73 & $+$19:04:22.00     &                 9.5              &     3.30      &  $1.57\substack{+0.74 \\ -0.38}$  & 10450         &   6500 & M,P,S   \\
%\hline
UW~Cen &  12:43:17.18 & $-$54:31:40.72       &                 9.6             &    3.50      &  $7.28\substack{+25.8 \\ -3.19}$  &  7320          &   7500 & M,P   \\
%\hline
V854~Cen  & 14:34:49.41 & $-$39:33:19.18      &                 7.0             &    2.28      &  ---\tablenotemark{2}   & 11760           &   6750 & M   \\
%\hline
V~CrA     &   18:47:32.30 & $-$38:09:32.32   &               9.4               &    5.50      &  ---\tablenotemark{2}    & 6550           &   6250 & M,P,S   \\
HD~173409   &   18:46:26.63 & $-$31:20:32.07   &               9.5               &    ---\tablenotemark{3}      & $2.00\substack{+0.55 \\ -0.35}$ &      ---\tablenotemark{3}           &   7000 & P,S   \\
%\hline
%\hline
\enddata
\tablenotetext{1}{M: MIPS; P: PACS; S: SPIRE}
\tablenotetext{2}{No parallax in the Gaia DR2.}
\tablenotetext{3}{The HD~173409 SED was not modeled in this work (see Section 5.8.3).}
%\end{tabular}
%\smallskip
%\end{center} 
\end{deluxetable}

\clearpage

\begin{deluxetable}{lcc}%[h]
\tablecaption{MV~Sgr Photometry}
\tablenum{2}
\scriptsize
\tablehead{\colhead{Band}&\colhead{Flux}&\colhead{$\sigma$} \\
                  \colhead{}        &\colhead{(Jy)}&\colhead{(Jy)}}
\startdata                
$U$ (0.365)       &         0.013     &       1.20e-04    \\
$B$ (0.433)       &         0.015     &       1.40e-04    \\
$V$ (0.550)       &        0.018     &        1.60e-04      \\
$R_{\rm C}$ (0.640) &   0.015   &   1.4e-04  \\
$I_{\rm C}$ (0.790)   &   0.025    &  2.3e-04  \\
2MASS$/J$ (1.235)  &    0.060   &   0.001   \\
$J$  (1.25)       &       0.075    &      0.007   \\  
$H$ (1.60)       &        0.130    &     0.012   \\
2MASS$/H$ (1.66)   &  0.117  &     0.003  \\
2MASS$/K_{\rm S}$ (2.16) & 0.188   &    0.004 \\
$K$ (2.20)       &       0.213     &     0.020   \\
$L$ (3.40)       &        0.344    &      0.032  \\ % Kilkenny & Whittet
WISE$/$3.4 &     0.180       &     0.004  \\
WISE$/$4.6 &    0.199       &    0.004  \\
$M$ (4.80)      &       0.874    &      0.402  \\  % Kilkenny & Whittet
$M$ (4.80)      &       0.551     &       3$\sigma$ upper limit \\ % Goldsmith et al. 
AKARI$/$9        &      0.330   &           0.019  \\
$N$ (10.2)     &        0.814    &      0.150  \\    % Kilkenny & Whittet
$N$ (10.2)    &         0.742    &      0.068  \\ % Goldsmith et al. 
IRAS$/$12     &        0.597    &         0.130  \\
WISE$/$12    &  0.409     &      0.006 \\
AKARI$/$18    & 1.00      &        0.008  \\
WISE$/$22       &  1.11        &      0.017 \\
MIPS$/$24       &  1.00     &   0.004   \\
IRAS$/$25        &   1.57    &         0.140  \\
IRAS$/$60        &    0.777     &      0.078  \\
AKARI$/$65      &   0.257     &           3$\sigma$ upper Limit   \\
MIPS$/$70        &    0.286   &     0.009  \\
AKARI$/$90       &   0.496   &       0.103  \\
IRAS$/$100       &    3.47       &        3$\sigma$ upper Limit  \\
\enddata
\end{deluxetable}

\clearpage

\begin{deluxetable}{lcc}%[h]
\tablecaption{R~CrB Photometry}
\tablenum{3}
\scriptsize
\tablehead{\colhead{Band}&\colhead{Flux}&\colhead{$\sigma$} \\
                  \colhead{}        &\colhead{(Jy)}&\colhead{(Jy)}}
\startdata                
$U$ (0.365)       &         4.53     &        0.039        \\
$B$ (0.433)       &        11.60    &         0.107	       \\
$V$ (0.550)       &       17.90     &        0.166            \\
$R_{\rm C}$ (0.640)       &        20.50    &   0.190  \\
$I_{\rm C}$ (0.790)       &        20.70    &     0.191  \\
$J$  (1.25)       &        17.70    &         0.165  \\
$H$ (1.60)       &        14.30    &         0.100  \\
$K$ (2.20)       &        14.30    &         0.133  \\
$L$ (3.40)       &        25.60    &         0.236  \\
AKARI$/$9        &        53.00    &         2.440  \\ 
IRAS$/$12        &        38.90    &         1.550  \\
AKARI$/$18        &        21.50    &         0.029  \\
IRAS$/$25        &        17.10    &         0.684  \\
IRAS$/$60        &         3.94     &        0.315  \\
MIPS$/$70        &         2.03   &         0.034  \\
PACS$/$70            &       2.13     &       0.003  \\
AKARI$/$90          &        1.49     &         0.114  \\
IRAS$/$100            &       2.00       &       0.160  \\
PACS$/$100           &       1.04     &        0.0023  \\
MIPS$/$160            &       0.297     &        0.00936  \\
PACS$/$160            &       0.335     &        0.00211	\\
SPIRE$/$250           &        0.0781   &         0.01170 \\
SPIRE$/$350           &        0.0340   &         0.00510 \\
SPIRE$/$500          &        0.0125   &         0.00434 \\
\enddata
\end{deluxetable}

\clearpage

\begin{deluxetable}{lcc}%[h]
\tablecaption{RY~Sgr Photometry}
\tablenum{4}
\scriptsize
\tablehead{\colhead{Band}&\colhead{Flux}&\colhead{$\sigma$} \\
                  \colhead{}        &\colhead{(Jy)}&\colhead{(Jy)}}
\startdata                
$U$ (0.365)       &         4.31      &         0.040       \\
$B$ (0.433)       &       9.77       &          0.090       \\
$V$ (0.550)       &       13.3     &           0.122          \\
$R_{\rm C}$ (0.640)  &      14.30  &    0.132 \\
$I_{\rm C}$ (0.790)    &   14.10     &    0.130  \\
$J$  (1.25)       &       13.5       &         0.124 \\
$H$ (1.60)       &        15.2      &          0.140  \\
$K$ (2.20)       &       23.8       &         0.219  \\
$L$ (3.40)       &       54.0       &         0.497  \\
WISE$/$3.4 &   18.9       &     2.72  \\
WISE$/$4.6 &    46.2      &      8.20  \\
AKARI$/$9   &    48.0   &        3.66  \\ 
IRAS$/$12    &    77.2    &       5.40  \\
WISE$/$12     &   36.2     &       0.966 \\
AKARI$/$18    &   20.2     &      1.02  \\
WISE$/$22    &    14.0    &        0.232 \\
IRAS$/$25     &    26.2    &        1.048  \\
IRAS$/$60     &     5.43     &        0.489  \\
AKARI$/$65  &    3.50     &         0.104 \\
MIPS$/$70    &    2.92     &        0.021  \\
PACS$/$70    &   4.39    &        0.008  \\
AKARI$/$90    &   2.61   &      0.139  \\
IRAS$/$100     &  4.60       &       0.414  \\
PACS$/$100    &  3.31    &     0.007  \\
AKARI$/$140   & 2.08   &  3$\sigma$ upper Limit \\
MIPS$/$160     & 1.34     &        0.030 \\
AKARI$/$160   &  2.60     &  3$\sigma$ upper Limit \\
PACS$/$160     &  1.79    &    0.005   \\
SPIRE$/$250    &  0.766   &    0.010 \\
SPIRE$/$350    &  0.324    &   0.007\\
SPIRE$/$500    &  0.126   &     0.006 \\
\enddata
\end{deluxetable}

\clearpage

\begin{deluxetable}{lcc}%[h]
\tablecaption{SU~Tau Photometry}
\tablenum{5}
\scriptsize
\tablehead{\colhead{Band}&\colhead{Flux}&\colhead{$\sigma$} \\
                  \colhead{}        &\colhead{(Jy)}&\colhead{(Jy)}}
\startdata     
$B$ (0.433)       &        0.241    &     0.004     \\
$V$ (0.550)       &       0.560     &      0.010      \\
$R_{\rm C}$ (0.640)  &        0.827    &   0.015  \\
$I_{\rm C}$ (0.790)  &        0.977    &     0.018  \\
$J$  (1.25)       &        1.60    &         0.044  \\
$H$ (1.60)       &        1.70    &         0.047  \\
$K$ (2.20)       &        1.94    &         0.054  \\
$L$ (3.40)       &        3.57     &         0.099  \\
WISE$/$3.4  &      3.62    &   0.260  \\
WISE$/$4.6  &     11.1   &    0.675  \\
AKARI$/$9    &      14.7    &         0.050  \\ 
IRAS$/$12    &        9.48    &         0.759  \\
WISE$/$12   &       7.77     &    0.079  \\
AKARI$/$18  &        6.16    &         0.046  \\
WISE$/$22   &        3.38    &     0.037 \\
MIPS$/$24   &        3.07    &      0.037 \\
IRAS$/$25    &        4.12    &         0.288  \\
IRAS$/$60    &         1.54     &        0.139  \\
AKARI$/$65   &          0.351  &        3$\sigma$ upper Limit \\
MIPS$/$70     &         0.322   &         0.003  \\
PACS$/$70    &       0.523     &       0.002  \\
AKARI$/$90   &         1.18     &         0.080  \\
IRAS$/$100   &       2.87       &        0.315  \\
PACS$/$100   &       0.318     &        0.002  \\
MIPS$/$160   &       0.142     &        0.007  \\
PACS$/$160   &       0.133     &        0.001  \\
SPIRE$/$250 &        0.117   &          3$\sigma$ upper Limit \\
SPIRE$/$350   &        0.064   &         3$\sigma$ upper Limit  \\
SPIRE$/$500  &        0.028   &          3$\sigma$ upper Limit \\
\enddata
\end{deluxetable}

\clearpage

\begin{deluxetable}{lcc}%[h]
\tablecaption{UW~Cen Photometry}
\tablenum{6}
\scriptsize
\tablehead{\colhead{Band}&\colhead{Flux}&\colhead{$\sigma$} \\
                  \colhead{}        &\colhead{(Jy)}&\colhead{(Jy)}}
\startdata     
$U$ (0.365)       &         0.234     &        0.002        \\
$B$ (0.433)       &        0.594    &         0.005	       \\
$V$ (0.550)       &       1.020     &        0.009            \\
$R_{\rm C}$ (0.640)       &        1.170    &   0.011  \\
$I_{\rm C}$ (0.790)       &        1.340    &     0.012  \\
$J$  (1.25)       &        1.350    &         0.025  \\
$H$ (1.60)       &        1.090    &         0.020  \\
$K$ (2.20)       &        0.879    &         0.016  \\
$L$ (3.40)       &        1.070    &         0.049  \\
WISE$/$3.4 &  2.150        &      0.105 \\
IRAC$/$3.6 &   5.260      &  0.026   \\
IRAC$/$4.5 &   6.690       &     0.034  \\
WISE$/$4.6   & 8.660      &     0.263 \\
IRAC$/$5.8    &         7.690    &        0.054 \\
IRAC$/$8.0    &          9.050    &           0.060 \\
AKARI$/$9        &        9.760     &         0.070  \\ 
IRAS$/$12        &        7.850    &         0.471  \\
WISE$/$12      &           6.820  &         0.044  \\
AKARI$/$18        &        5.700    &       0.047  \\
WISE$/$ 22      &         4.570      &     0.046  \\
MIPS$/$24      &         4.350     &       0.011 \\
IRAS$/$25        &        5.750    &         0.345  \\
IRAS$/$60        &         9.220     &        0.737  \\
AKARI$/$65      &       6.650    &      0.413 \\
MIPS$/$70        &        5.570     &     0.007 \\
PACS$/$70            &       2.130     &       0.003  \\
AKARI$/$90          &       7.300     &     0.332   \\
IRAS$/$100            &       5.940       &       0.594  \\
PACS$/$100           &      4.950   &       0.006  \\
AKARI$/$140      &      4.180       &      0.349   \\
MIPS$/$160            &       2.530     &       0.055 \\
AKARI$/$160      &      2.810           &   1.020    \\
PACS$/$160            &     2.470     &     0.004	\\
\enddata
\end{deluxetable}

\clearpage

\begin{deluxetable}{lcc}%[h]
\tablecaption{V854~Cen Photometry}
\tablenum{7}
\scriptsize
\tablehead{\colhead{Band}&\colhead{Flux}&\colhead{$\sigma$} \\
                  \colhead{}        &\colhead{(Jy)}&\colhead{(Jy)}}
\startdata     
$U$ (0.365)       &         1.450     &        0.036        \\
$B$ (0.433)       &        3.820    &         0.095	       \\
$V$ (0.550)       &       5.500     &        0.137            \\
$R_{\rm C}$ (0.640)       &        6.067    &   0.152  \\
$I_{\rm C}$ (0.790)       &        6.512    &     0.163  \\
2MASS$/J$ (1.24) & 5.756     &     0.095 \\
$J$  (1.25)       &        6.998    &      0.193  \\
$H$ (1.60)       &        8.268    &      0.229  \\
2MASS$/H$ (1.66) & 5.399   &       0.085 \\
2MASS$/K_{\rm S}$ (2.16) & 7.480    &      0.124 \\
$K$ (2.20)       &        12.50  &        0.345  \\
$L$ (3.40)       &        27.50   &       0.761  \\
AKARI$/$9        &        23.00   &          1.170  \\ 
IRAS$/$12        &        23.00    &         1.150  \\
AKARI$/$18        &        7.364   &          0.033  \\
MIPS$/$24      &    4.944     &     0.001  \\
IRAS$/$25        &        7.820    &         0.469  \\
IRAS$/$60        &         1.510     &        0.136  \\
AKARI$/$65    &        0.940      &        3$\sigma$ upper Limit   \\
MIPS$/$70        &        0.641    &      0.001  \\
PACS$/$70            &       2.132     &       0.003  \\
AKARI$/$90          &         0.705     &        0.036 \\
IRAS$/$100            &       1.030       &       3$\sigma$ upper Limit  \\
AKARI$/$140         &       0.185       &      3$\sigma$ upper Limit \\
MIPS$/$160            &      0.068    &     0.001  \\
AKARI$/$160      &         1.040        &    3$\sigma$ upper Limit \\
\enddata
\end{deluxetable}

\clearpage

\begin{deluxetable}{lcc}%[h]
\tablecaption{V~CrA Photometry}
\tablenum{8}
\scriptsize
\tablehead{\colhead{Band}&\colhead{Flux}&\colhead{$\sigma$} \\
                  \colhead{}        &\colhead{(Jy)}&\colhead{(Jy)}}
\startdata     
$U$ (0.365)       &         0.138     &        0.003        \\
$B$ (0.433)       &        0.364      &         0.007	\\
$V$ (0.550)       &       0.444       &        0.008         \\
$J$  (1.25)       &        0.681	  &    0.019  \\
$H$ (1.60)       &        0.804     &          0.022  \\
$K$ (2.20)       &        1.180    &            0.033  \\
$L$ (3.40)       &       3.080      &          0.085  \\
WISE$/$3.4 &     1.400       &     0.139  \\
WISE$/$4.6 &       3.490     &       0.286  \\
AKARI$/$9        &        3.610    &         0.210  \\ 
IRAS$/$12        &        5.660    &         0.226  \\
WISE$/$12       &        3.820     &       0.053  \\
AKARI$/$18        &        2.170    &         0.013  \\
WISE$/$22        &         1.960    &        0.025 \\
MIPS$/$24        &         1.520    &    0.003 \\
IRAS$/$25        &        2.460    &         0.172  \\
IRAS$/$60        &         0.405     &        0.036  \\
MIPS$/$70        &        0.272   &         0.004  \\
PACS$/$70            &       0.263     &       0.003  \\
AKARI$/$90          &         1.490     &         0.114  \\
IRAS$/$100            &       1.320       &       3$\sigma$ upper Limit  \\
PACS$/$100           &       0.135     &        0.003  \\
MIPS$/$160            &       0.117     &        3$\sigma$ upper Limit  \\
PACS$/$160            &       0.051     &        0.005	\\
SPIRE$/$250           &        0.050   &         3$\sigma$ upper Limit \\
SPIRE$/$350           &        0.013   &         3$\sigma$ upper Limit \\
SPIRE$/$500          &        0.006   &         3$\sigma$ upper Limit \\
\enddata
\end{deluxetable}

\clearpage

\begin{deluxetable}{lcc}%[h]
\tablecaption{V605~Aql Photometry}
\tablenum{9}
\scriptsize
\tablehead{\colhead{Band}&\colhead{Flux}&\colhead{$\sigma$} \\
                  \colhead{}        &\colhead{(Jy)}&\colhead{(Jy)}}
\startdata     
$J$  (1.25)       &        3.21e-05    &         1.50e-05  \\
$H$ (1.60)       &        2.27e-04    &         3.00e-05  \\
$K$ (2.20)       &        8.56e-04    &         8.00e-05  \\
WISE$/$3.4 &	   1.44e-02     &         3.05e-04 \\	
WISE$/$4.6 &	    1.13e-01    &         2.07e-03 \\	
AKARI$/$9  &        2.85e+00    &      1.68e-02  \\ 
IRAS$/$12   &      4.99e+00    &       2.00e-01   \\
WISE$/$12   &     8.56e+00   &        5.52e-02  \\
AKARI$/$18  &     1.53e+01    &        1.20e-01  \\
WISE$/$22   &      2.22e+01   &         1.02e-01 \\
MIPS$/$24    &     1.57e+01   &         8.01e-02 \\
IRAS$/$25     &    2.95e+01    &        1.18e+00   \\
IRAS$/$60     &    4.07e+01   &          4.10e+00   \\
AKARI$/$65  &    2.67e+01   &         2.50e+00   \\
MIPS$/$70     &   1.78e+01     &       1.85e-01 \\
AKARI$/$90    &  2.08e+01     &        9.87e-01  \\
IRAS$/$100    &   1.83e+01    &       2.00e+00  \\
MIPS$/$160    &    2.82e+00    &      1.12e-01  \\
\enddata
\end{deluxetable}

\clearpage

\begin{deluxetable}{lcc}%[h]
\tablecaption{HD~173409 Photometry}
\tablenum{10}
\scriptsize
\tablehead{\colhead{Band}&\colhead{Flux}&\colhead{$\sigma$} \\
                  \colhead{}        &\colhead{(Jy)}&\colhead{(Jy)}}
\startdata     
$U$ (0.365)       &         0.095     &        0.002        \\
$B$ (0.433)       &        0.324    &         0.006	       \\
$V$ (0.550)       &       0.626     &        0.012         \\
$R_{\rm C}$ (0.640)       &        0.733    &   0.014 \\
$I_{\rm C}$ (0.790)       &        0.770    &     0.014  \\
2MASS$/J$ (1.24) & 0.698   &  0.017 \\
$J$  (1.25)       &        0.739    &         0.020  \\
$H$ (1.60)       &        0.531    &         0.015  \\
2MASS$/H$ (1.66)   &   0.475  &   0.020 \\
2MASS$/K_{\rm S}$ (2.16) &    0.354  &     0.012 \\
$K$ (2.20)       &        0.356    &         0.010  \\
$L$ (3.40)       &        0.190    &         0.009  \\
WISE$/$3.4 &      0.185    &          0.004   \\
WISE$/$4.6 &      0.102   &           0.002    \\
WISE$/$12 &      0.020    &       0.0004   \\
WISE$/$25 &    0.004      &       0.001    \\
PACS$/$70   &       7.06e-05     &      3$\sigma$ upper Limit  \\
PACS$/$100  &       1.47e-04      &      3$\sigma$ upper Limit  \\
PACS$/$160  &       1.48e-04     &       3$\sigma$ upper Limit  \\
SPIRE$/$250  &        0.113   &       3$\sigma$ upper Limit  \\
SPIRE$/$350  &        0.001  &      3$\sigma$ upper Limit   \\
SPIRE$/$500 &        0.002   &      3$\sigma$ upper Limit    \\
\enddata
\end{deluxetable}

\clearpage

\begin{table}[h]
\tablenum{11}
\begin{center}
\caption{Derived MOCASSIN Properties and Measured Outer Radii}
\begin{tabular}{  c | c | c | c | c  }
\hline
Star & R$_{\rm in}$ & R$_{\rm out}$ &  M$_{\rm Dust}$ & R$_{\rm out-Measured}$ \\
       &    (cm)         &        (cm)            &    (M$_\odot$)       &    (cm) \\ \hline \hline
MV~Sgr     & \begin{tabular}{c} $3.25 \times 10^{14}$ \\ $3.25 \times 10^{16}$\end{tabular} & 
                    \begin{tabular}{c} $9.45 \times 10^{15}$ \\ $9.45 \times 10^{17}$\end{tabular} &   
                    \begin{tabular}{c} $7.59 \times 10^{-8}$ \\  $3.27 \times 10^{-4}$\end{tabular} &    $1.03 \times 10^{18}$ \\
\hline
R~CrB       & \begin{tabular}{c} $1.00 \times 10^{15}$ \\ $3.40 \times 10^{17}$\end{tabular} & 
                    \begin{tabular}{c} $3.00 \times 10^{16}$ \\ $1.00 \times 10^{19}$\end{tabular} &  
                    \begin{tabular}{c} $9.09 \times 10^{-7}$ \\  $2.42 \times 10^{-4}$\end{tabular} &    $1.23 \times 10^{19}$  \\
\hline
RY~Sgr     & \begin{tabular}{c} $8.62 \times 10^{14}$ \\ $5.15 \times 10^{17}$\end{tabular} & 
                    \begin{tabular}{c} $2.50 \times 10^{16}$ \\ $4.50 \times 10^{18}$\end{tabular} & 
                    \begin{tabular}{c} $8.90 \times 10^{-7}$ \\  $7.25 \times 10^{-4}$\end{tabular} &    $1.36 \times 10^{18}$ \\
\hline
SU~Tau     & \begin{tabular}{c} $2.10 \times 10^{15}$ \\ $1.00 \times 10^{18}$\end{tabular} & 
                    \begin{tabular}{c} $4.25 \times 10^{16}$ \\ $9.00 \times 10^{18}$\end{tabular} &  
                    \begin{tabular}{c} $2.27 \times 10^{-6}$ \\  $6.80 \times 10^{-4}$\end{tabular} &   $6.91 \times 10^{18}$ \\
\hline
UW~Cen   & \begin{tabular}{c} $1.55 \times 10^{15}$ \\ $7.00 \times 10^{17}$\end{tabular} & 
                    \begin{tabular}{c} $4.50 \times 10^{16}$ \\ $2.50 \times 10^{18}$\end{tabular} & 
                    \begin{tabular}{c} $2.40 \times 10^{-6}$ \\  $5.14 \times 10^{-3}$\end{tabular} &    
                    \begin{tabular}{c} $4.01 \times 10^{17}$ \\ $2.62 \times 10^{18}$ \end{tabular} \\
\hline
V854~Cen  & \begin{tabular}{c} $4.90 \times 10^{14}$ \\ $3.45 \times 10^{16}$\end{tabular} & 
                    \begin{tabular}{c} $1.00 \times 10^{16}$ \\ $1.00 \times 10^{18}$\end{tabular} &   
                    \begin{tabular}{c} $3.08 \times 10^{-7}$ \\  $2.60 \times 10^{-5}$\end{tabular} &    $1.48 \times 10^{18}$ \\
\hline
V~CrA        & \begin{tabular}{c} $1.70 \times 10^{15}$ \\ $1.00 \times 10^{17}$\end{tabular} & 
                    \begin{tabular}{c} $4.90 \times 10^{16}$ \\ $1.00 \times 10^{18}$\end{tabular} &  
                    \begin{tabular}{c} $4.00 \times 10^{-6}$ \\  $5.90 \times 10^{-5}$\end{tabular} &     $1.39 \times 10^{18}$  \\
\hline
\hline
\end{tabular}
%\smallskip
\end{center} 
\end{table}

\clearpage

\begin{figure}%[h!]
\figurenum{1}
\begin{center}
\includegraphics[scale=0.7]{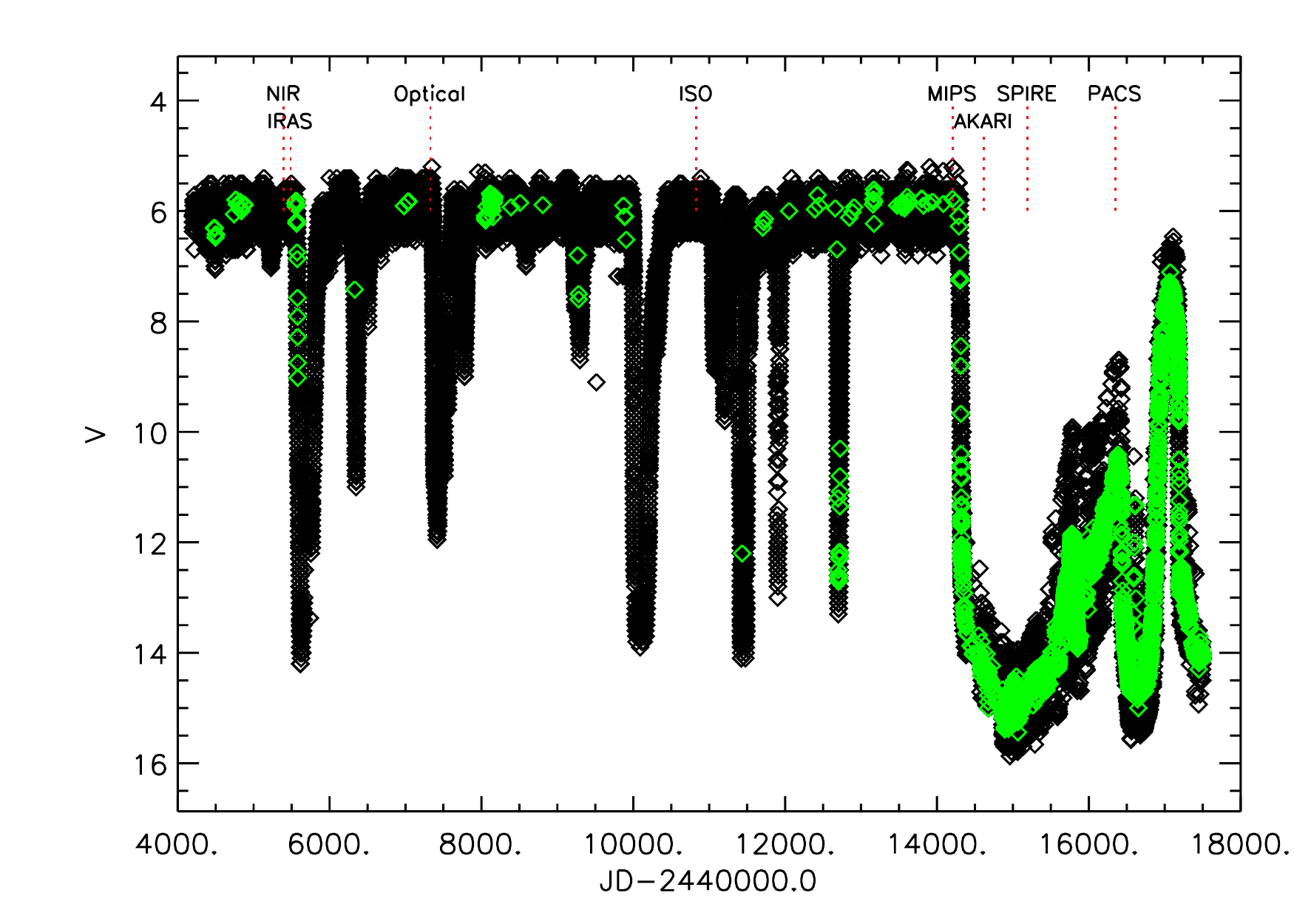}
\includegraphics[scale=0.7]{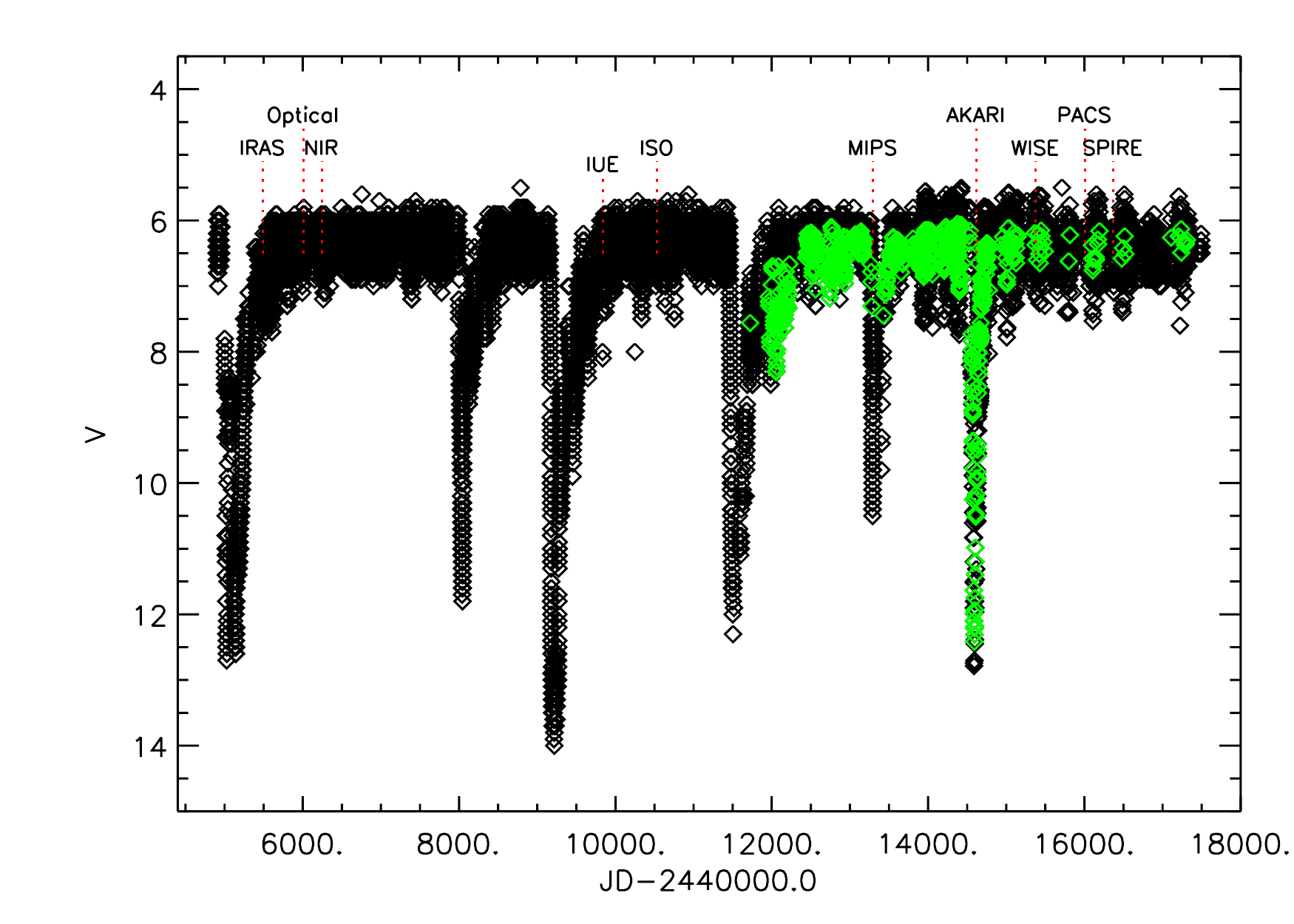}
\end{center}
\caption{AAVSO observations of R~CrB (top) and RY~Sgr (bottom) since November 1979 and October 1981, respectively. Black diamonds are visual observations and green diamonds are Johnson $V$ observations. The dates of the observations that went into my SED analysis are marked by the red,dashed vertical lines.}
\end{figure}

\begin{figure}%[h!]
\figurenum{2}
\begin{center}
\includegraphics[scale=0.7]{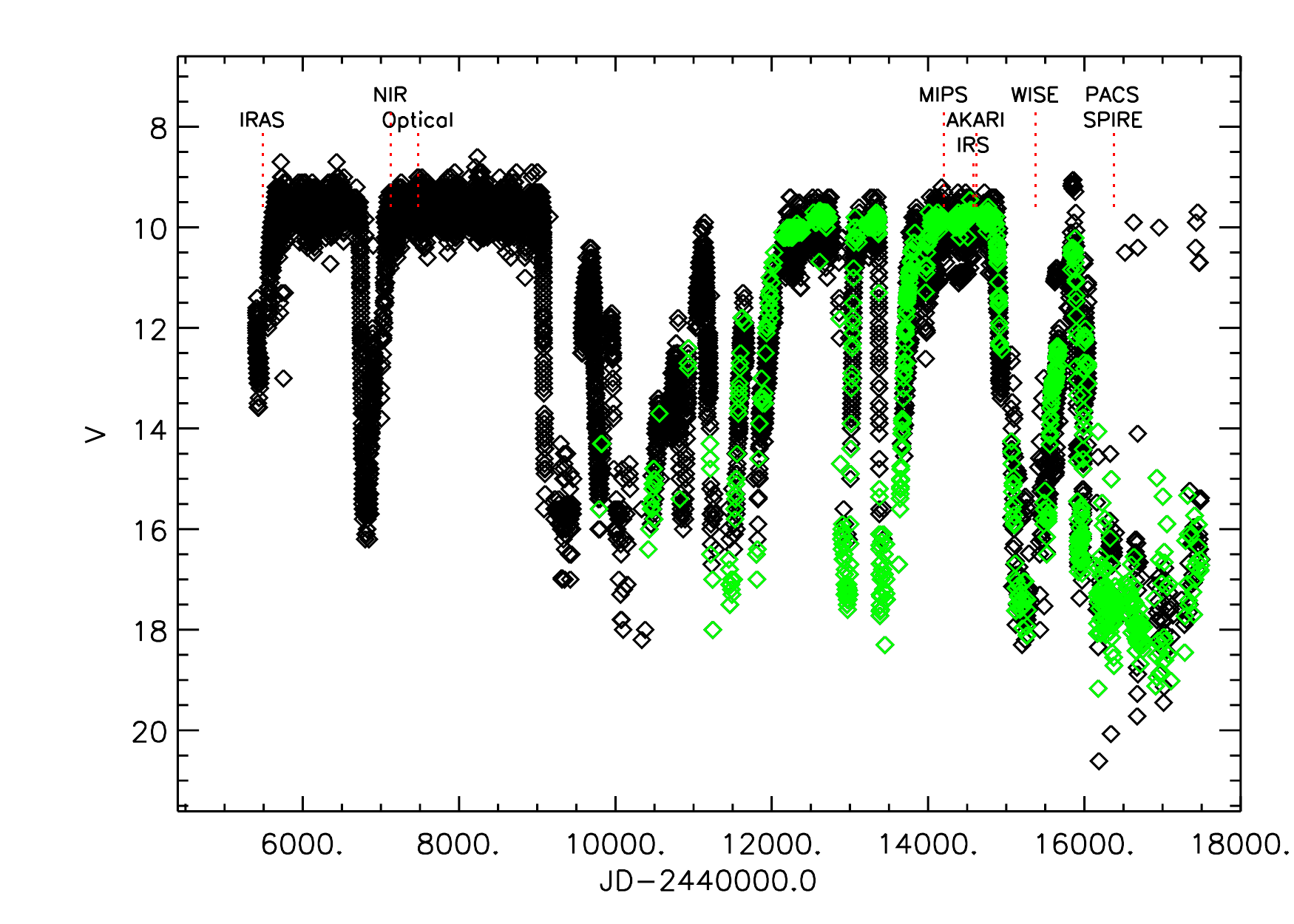}
\includegraphics[scale=0.7]{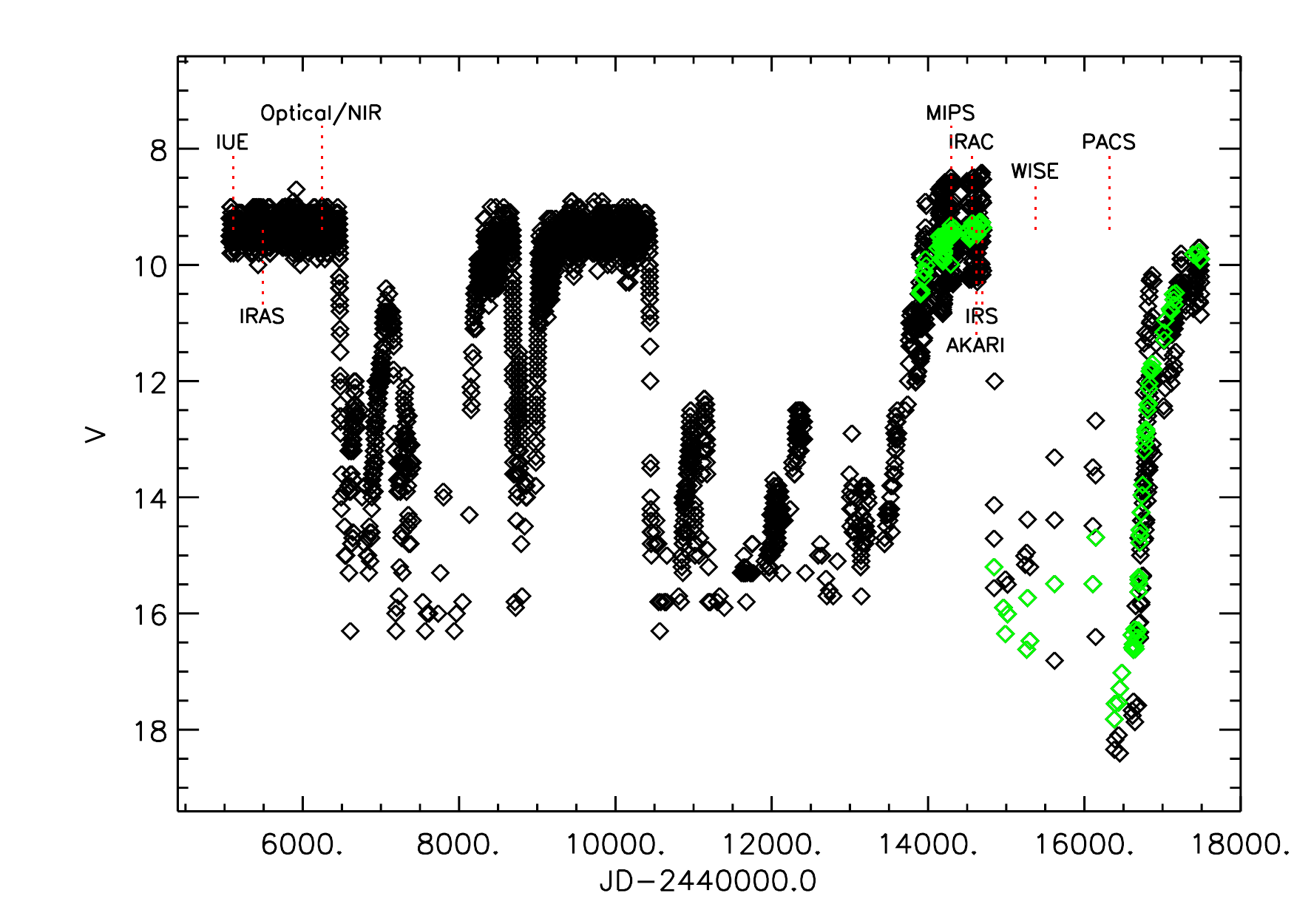}
\end{center}
\caption{AAVSO observations of SU~Tau (top) and UW~Cen (bottom) since March 1983 and April 1982, respectively. Black diamonds are visual observations and green diamonds are Johnson $V$ observations. The dates of the observations that went into my SED analysis are marked by the red,dashed vertical lines.}
\end{figure}

\begin{figure}%[h!]
\figurenum{3}
\begin{center}
\includegraphics[scale=0.7]{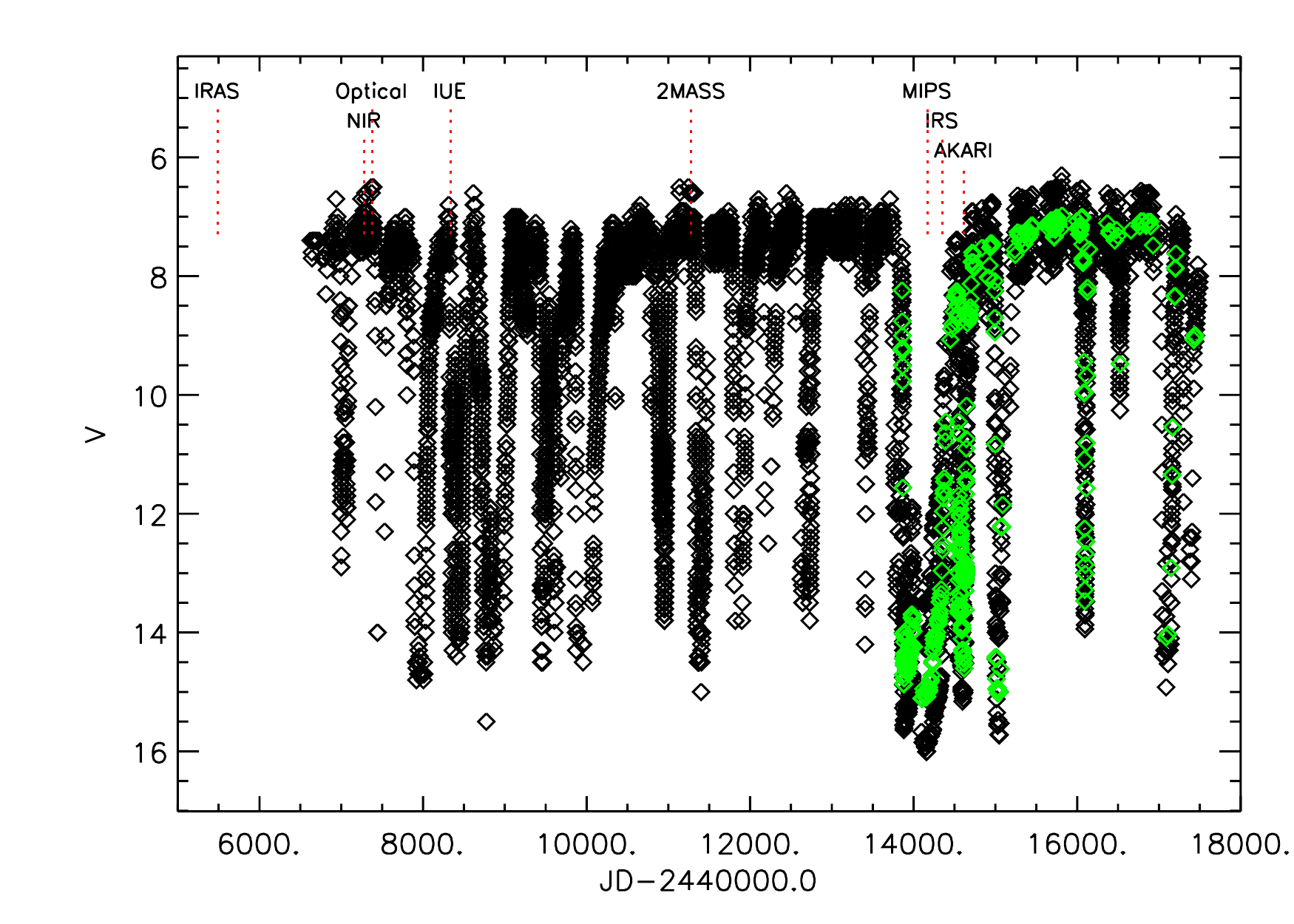}
\includegraphics[scale=0.7]{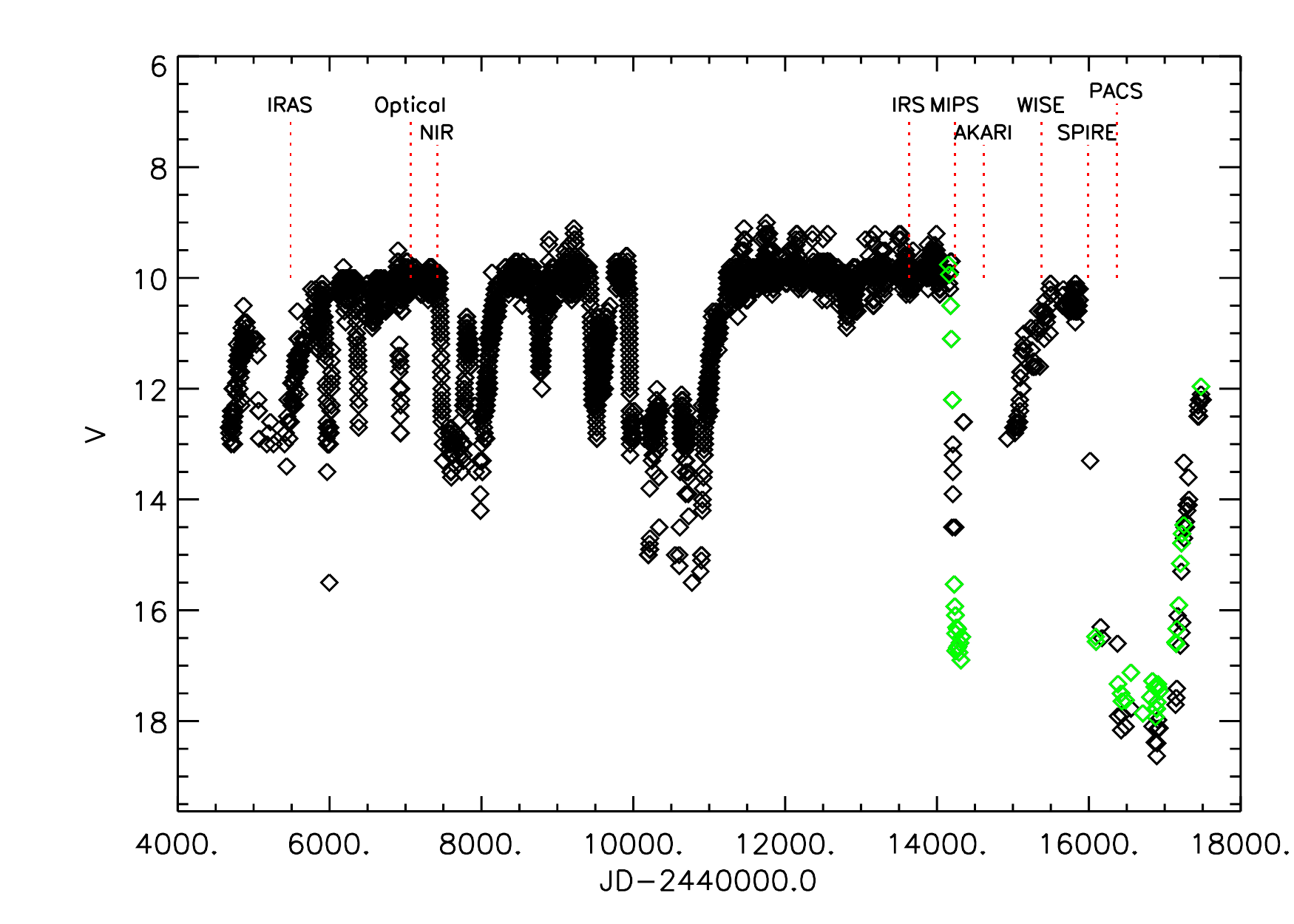}
\end{center}
\caption{AAVSO observations of V854~Cen (top) and V~CrA (bottom) since July 1986 and November 1979, respectively. Black diamonds are visual observations and green diamonds are Johnson $V$ observations. The dates of the observations that went into my SED analysis are marked by the red,dashed vertical lines.}
\end{figure}

\begin{figure}%[h!]
\figurenum{4}
\begin{center}
\includegraphics[scale=0.6]{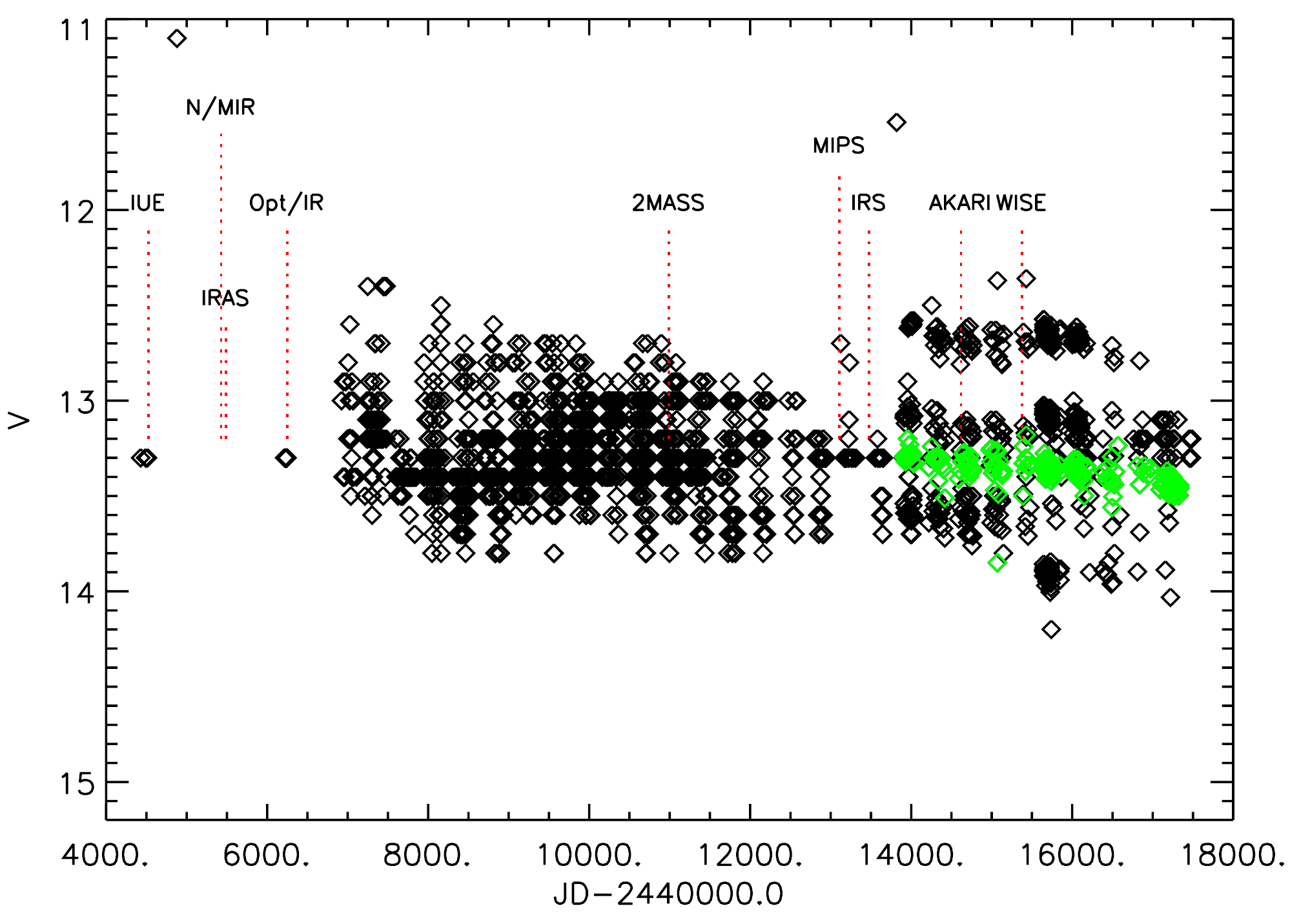}
\end{center}
\caption{AAVSO observations of MV~Sgr since October 1980. Black diamonds are visual observations and green diamonds are Johnson $V$ observations. The time that the observations that went into my SED analysis are marked by the red,dashed vertical lines.}
\end{figure}

\clearpage

\begin{figure}%[h!]
\figurenum{5}
\begin{center}
\includegraphics[scale=0.3]{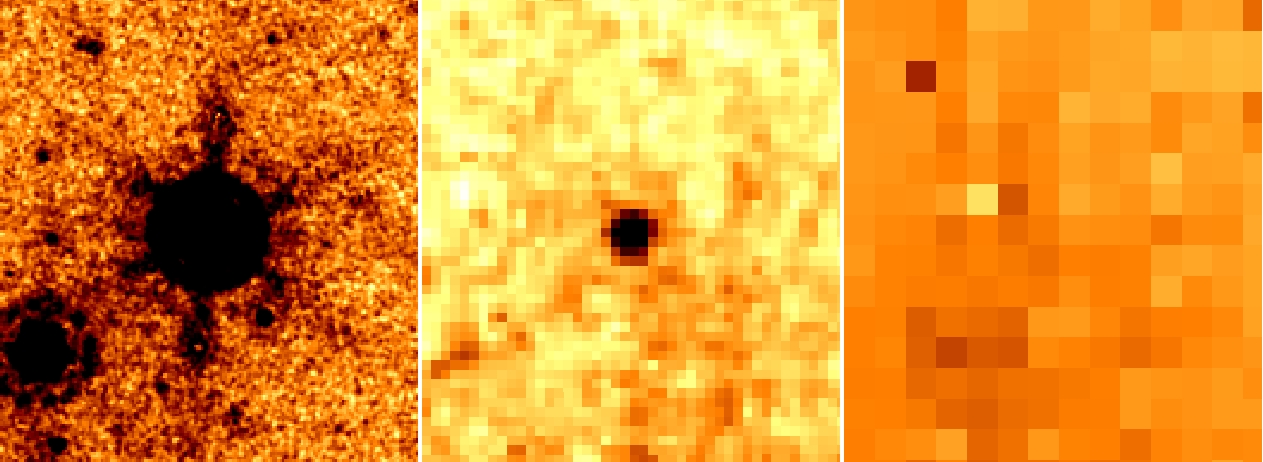}
\end{center}
\caption{The {\it Spitzer}$/$MIPS view of MV~Sgr. The panels are (left to right) 24, 70, and 160~$\mu$m, respectively, and the field--of--view is $3.6' \times 4.0'$. North is up and East is left.}
\end{figure}

\begin{figure}%[h!]
\figurenum{6}
\begin{center}
\includegraphics[scale=0.7]{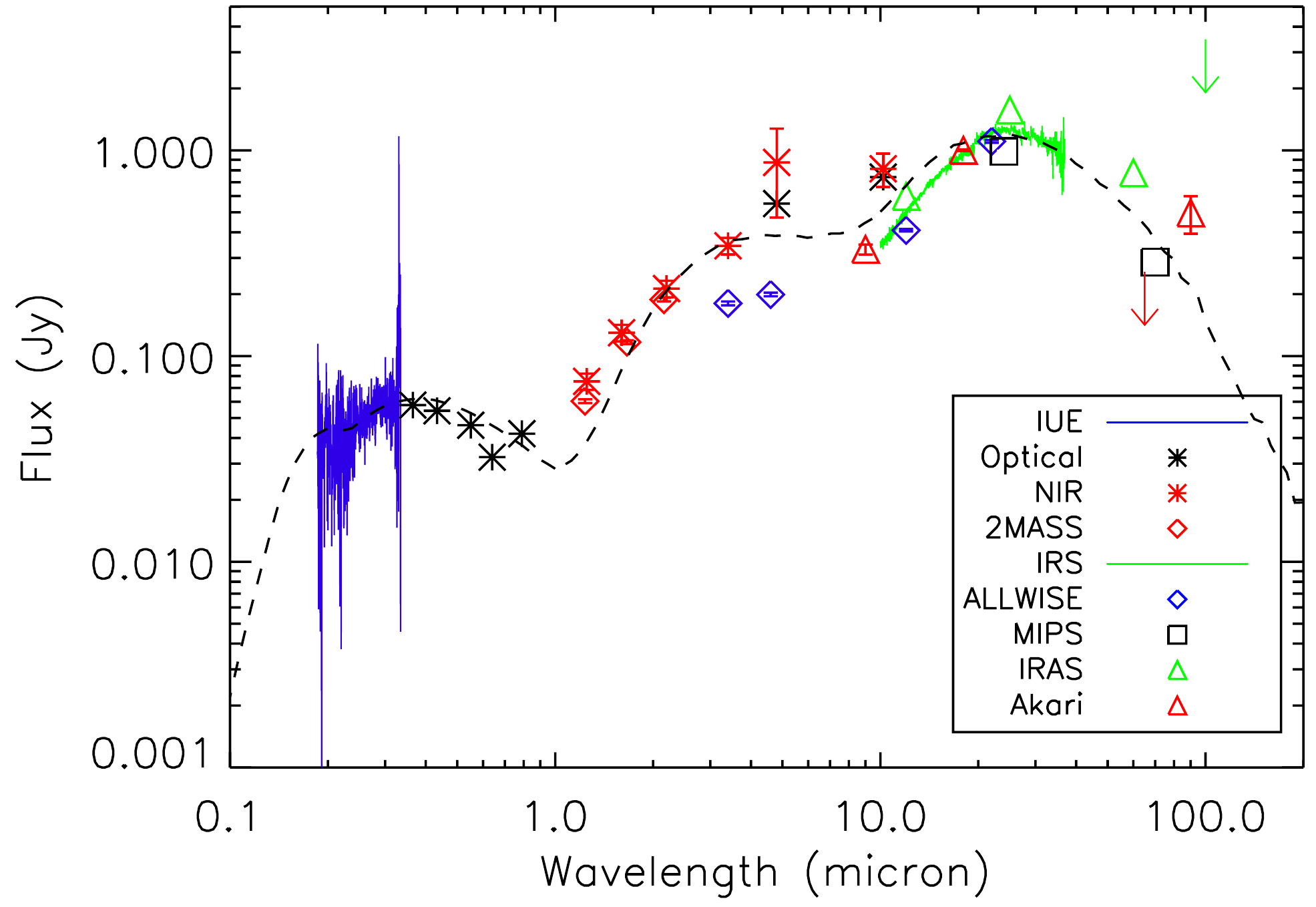}
\end{center}
%\vspace{-30mm}
\caption{The maximum--light SED of MV~Sgr. Blue line: IUE spectrum; black asterisks: $UBVR_{\rm C}I_{\rm C}MN$; red asterisks: $JHKLMN$; open red diamonds: 2MASS $JHK_{\rm S}$; open blue diamonds: WISE (3.4, 4.6, 12.0, 22.0~$\mu$m); green line: {\it Spitzer}$/$IRS spectrum; open black squares: {\it Spitzer}$/$MIPS (24 and 70~$\mu$m); open green triangles and arrow (3$\sigma$): IRAS (12, 25, 60, 100~$\mu$m);  open red triangle and arrow (3$\sigma$): AKARI (60 and 100~$\mu$m). The sum of the best--fit MOCASSIN models for the central source, warm, and cold dust shells is represented by the dashed black line.}
\end{figure}

\clearpage

\begin{figure}%[h]
\figurenum{7}
\begin{center}
\includegraphics[scale=0.4,angle=90,origin=c]{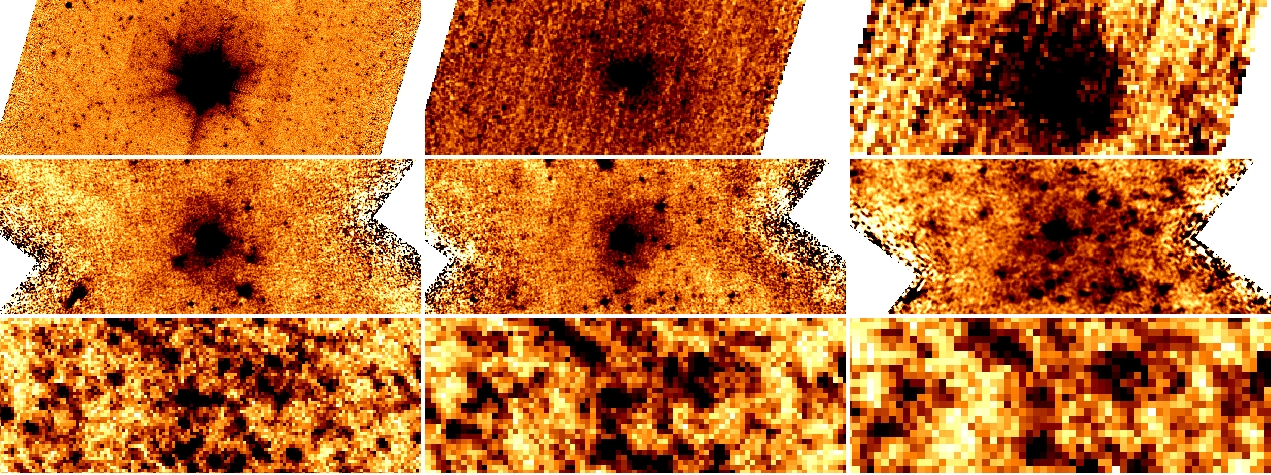}
\end{center}
\caption{First row (beginning lower left corner):{\it Spitzer}$/$MIPS observations of R~CrB 24, 70, 160~$\mu$m, respectively. The field--of--view shown for all three bands is $25' \times 10'$. Second row: {\it Herschel}$/$PACS observations of R~CrB at 70, 100, and 160~$\mu$m, respectively. The field--of--view shown for all three bands is $12' \times 5'$. Third row: {\it Herschel}$/$SPIRE observations of R~CrB at 250, 350, and 500~$\mu$m, respectively. The field--of--view shown for all three bands is $13.5' \times 5.0'$. North is left and East is down.}
\end{figure}

\begin{figure}%[h!]
\figurenum{8}
\begin{center}
\includegraphics[scale=0.7]{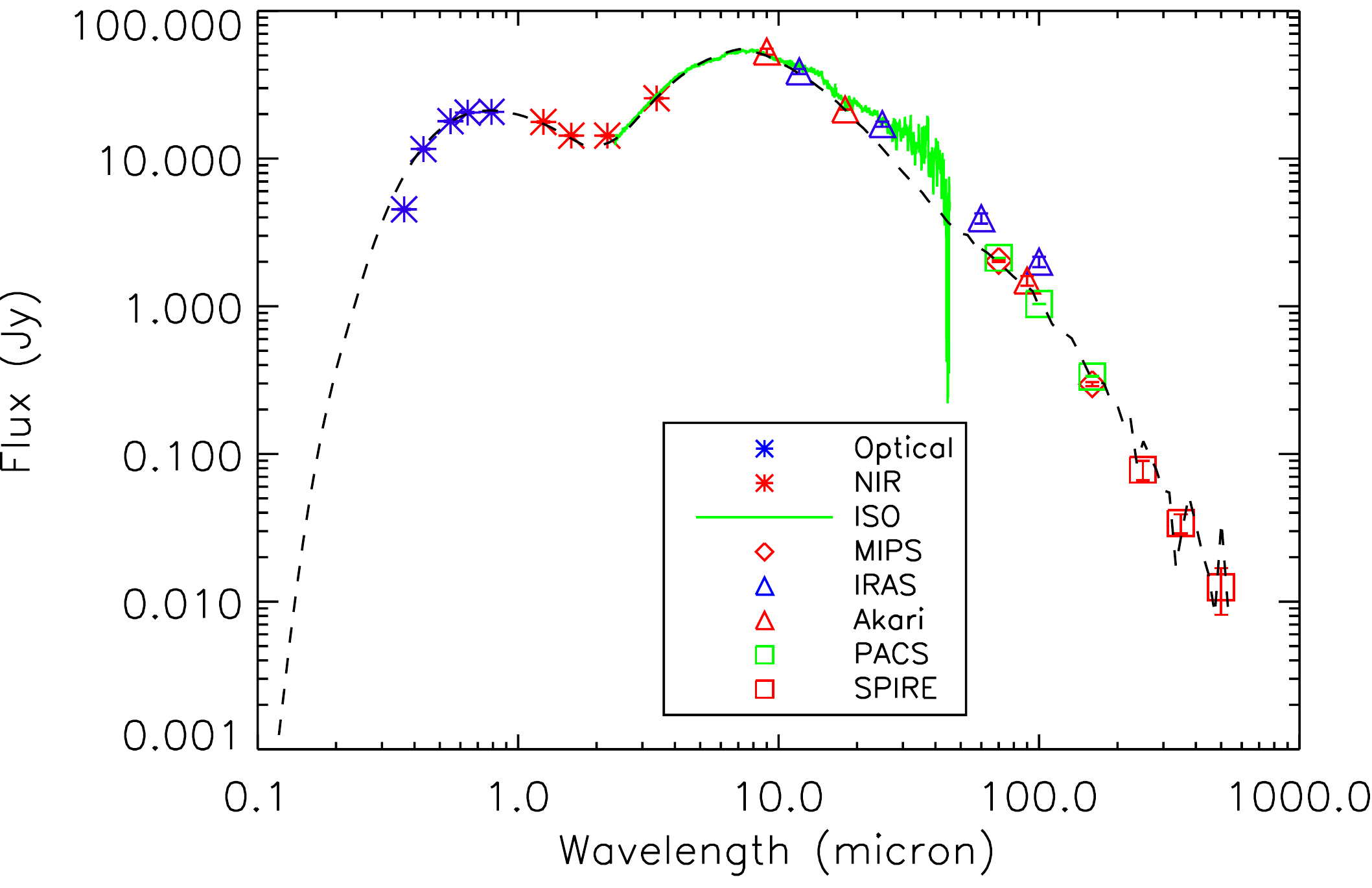}
\end{center}
%\vspace{-30mm}
\caption{The maximum--light SED of R~CrB. Blue asterisks: $UBVR_{\rm C}I_{\rm C}$; red asterisks: $JHKL$; green line: ISO spectrum; open red triangles: AKARI (9, 18, 90~$\mu$m); open red diamonds: {\it Spitzer}$/$MIPS (24 and 70~$\mu$m); open blue triangles: IRAS (12, 25, 60, 100~$\mu$m);  open green squares: {\it Herschel}$/$PACS (70, 100, 160~$\mu$m); open red squares: {\it Herschel}$/$SPIRE (250, 350, 500~$\mu$m). The sum of the best--fit MOCASSIN models for the central source, warm, and cold dust shells is represented by the dashed black line.}
\end{figure}

\clearpage

\begin{figure}%[h!]
\figurenum{9}
\begin{center}
\includegraphics[scale=0.4,angle=90,origin=c]{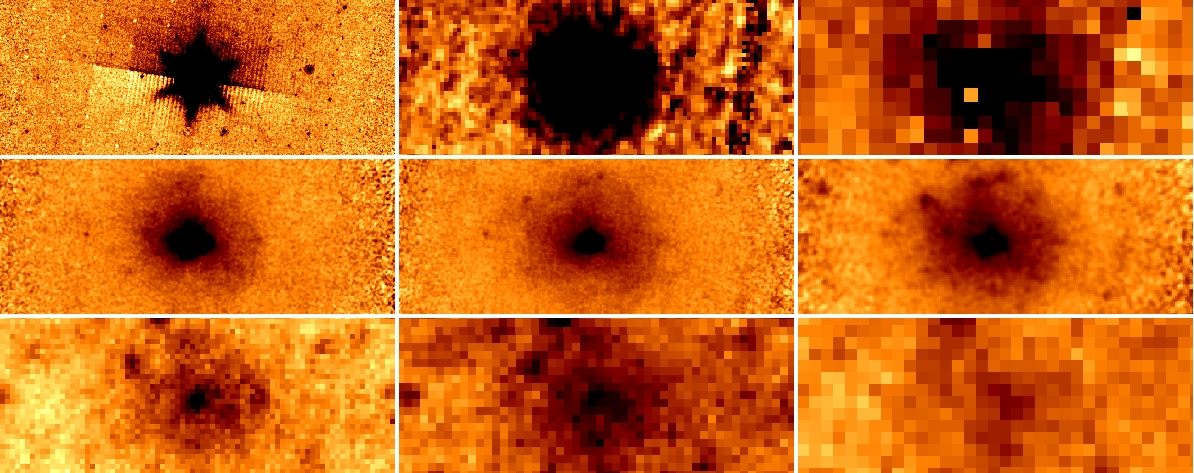}
\end{center}
\caption{First row (beginning lower left corner): {\it Spitzer}$/$MIPS observations of RY~Sgr 24, 70, 160~$\mu$m, respectively. Second row: {\it Herschel}$/$PACS observations of RY~Sgr at 70, 100, and 160~$\mu$m, respectively. Third row: {\it Herschel}$/$SPIRE observations of RY~Sgr at 250, 350, and 500~$\mu$m, respectively. The field--of--view in the {\it Spitzer}$/$MIPS 24~$\mu$m is $15.5' \times 6.1'$, while the remaining panels are all $7.6' \times 3.2'$. North is left and East is down.}
\end{figure}

\begin{figure}%[h!]
\figurenum{10}
\begin{center}
\includegraphics[scale=0.7]{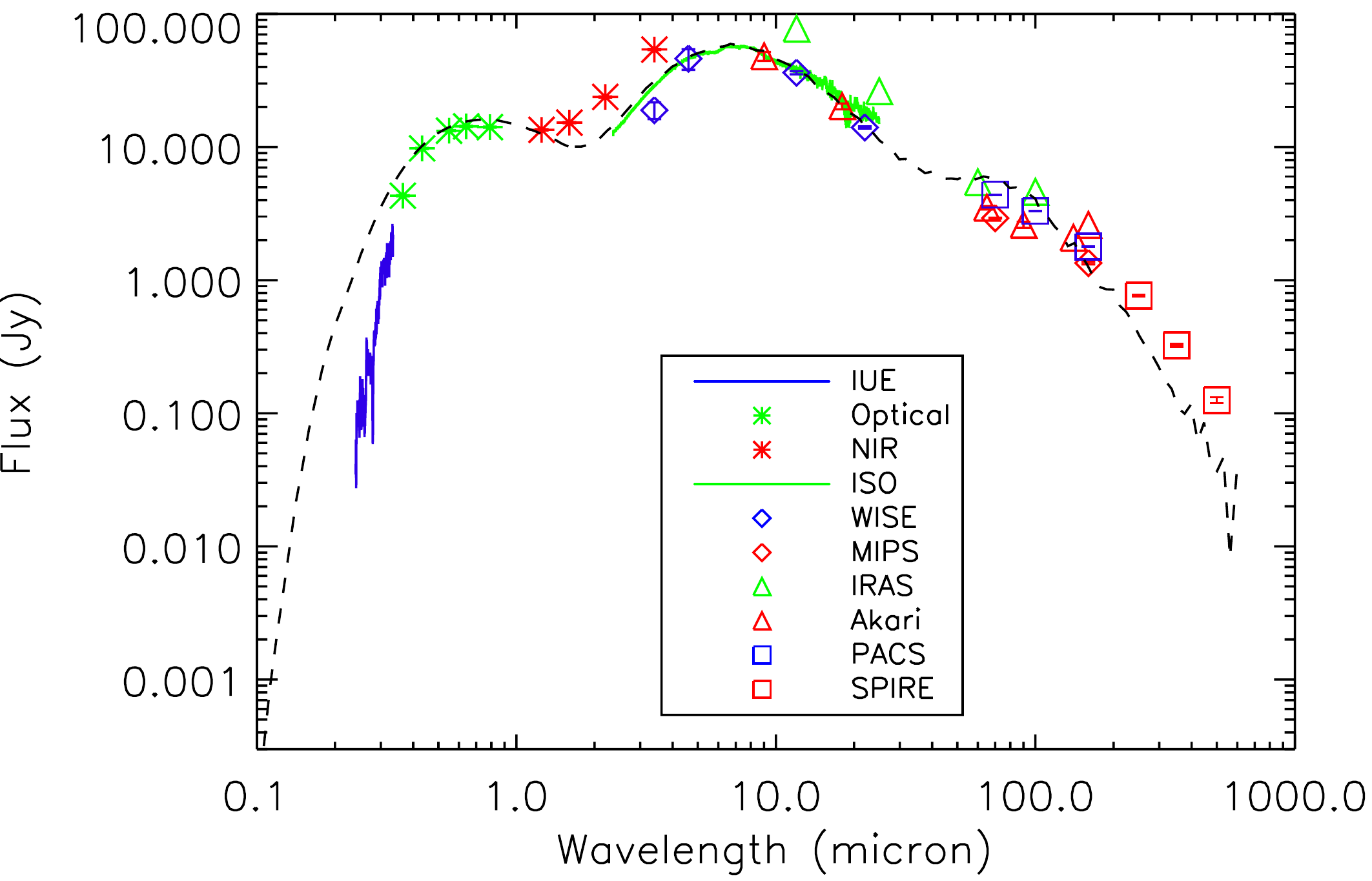}
\end{center}
%\vspace{-30mm}
\caption{The maximum--light SED of RY~Sgr. Blue line: IUE spectrum; green asterisks: $UBVR_{\rm C}I_{\rm C}$; red asterisks: $JHKL$; green line: ISO spectrum; open blue diamonds: WISE (3.4, 4.6, 12.0, 22.0~$\mu$m ; open red triangles: AKARI (9, 18, 90, 140, 160~$\mu$m); open red diamonds: {\it Spitzer}$/$MIPS (70 and 160~$\mu$m); open green triangles: IRAS (12, 25, 60, 100~$\mu$m);  open blue squares: {\it Herschel}$/$PACS (70, 100, 160~$\mu$m); open red squares: {\it Herschel}$/$SPIRE (250, 350, 500~$\mu$m). The sum of the best--fit MOCASSIN models for the central source, warm, and cold dust shells is represented by the dashed black line.}
\end{figure}

\clearpage

\begin{figure}%[h!]
\figurenum{11}
\begin{center}
\includegraphics[scale=0.4,angle=90,origin=c]{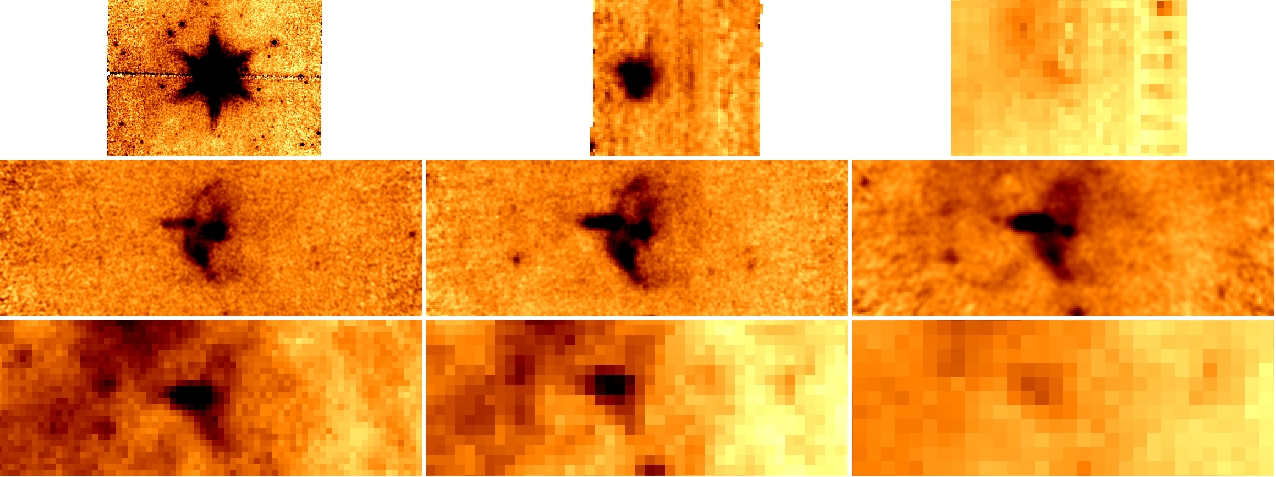}
\end{center}
\caption{First row (starting lower left corner):{\it Spitzer}$/$MIPS observations of SU~Tau 24, 70, 160~$\mu$m, respectively. The field--of--view of the observations (not including white space) are: $8.5' \times 5.75'$, $5.45' \times 5.0'$, and $8.4' \times 5.25'$, respectively. Second row: {\it Herschel}$/$PACS observations of SU~Tau at 70, 100, and 160~$\mu$m, respectively. Third row: {\it Herschel}$/$SPIRE observations of SU~Tau at 250, 350, and 500~$\mu$m, respectively. The field--of--view of the {\it Herschel} observations are all $7' \times 2.5'$. North is left and East is down.}
\end{figure}

\begin{figure}%[h!]
\figurenum{12}
\begin{center}
\includegraphics[scale=0.35]{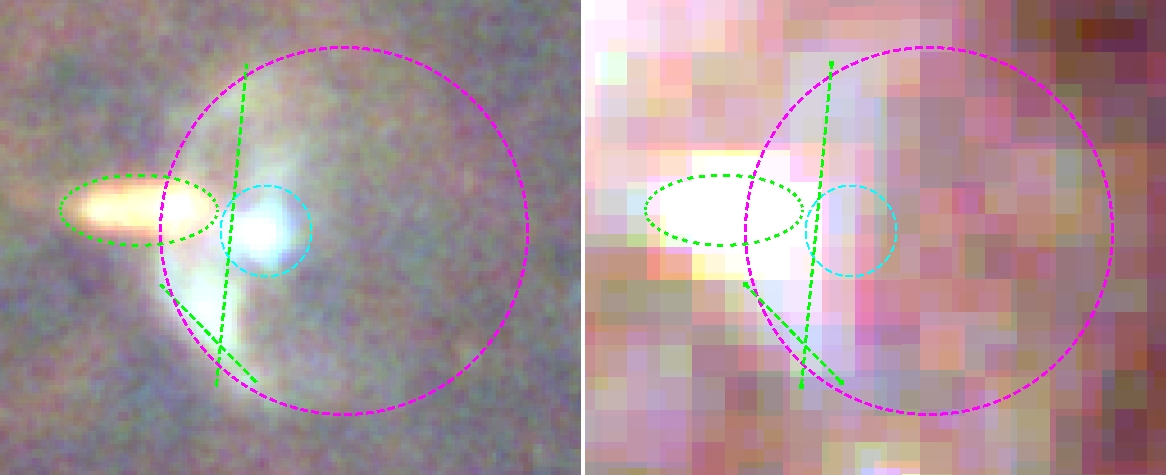}
\end{center}
\caption{Left: Three color {\it Herschel}$/$PACS image with the 70, 100, and 160~$\mu$m observations being represented by blue, green, and red, respectively. Right: Three color {\it Herschel}$/$SPIRE image with 250, 350, and 500~$\mu$m observations being represented by blue, green, and red, respectively. The field--of--view in both frames is $3' \times 2.4'$. The cyan circle (radius $= 14"$) is centered on the position of SU~Tau, the green ellipse shows the background galaxy separate from SU~Tau, and the green lines (1.66' \& 0.7' in length) are there to guide the eye to the bow shock feature being associated with SU~Tau. The magenta circle (radius $= 0.95'$ is centered to have a section of its arc pass through the bow shock and to highlight the diffuse emission west of SU~Tau. These observations are now oriented in the traditional astronomical sense: North is up and East is left.}
\end{figure}

\begin{figure}%[h!]
\figurenum{13}
\begin{center}
\includegraphics[scale=0.35]{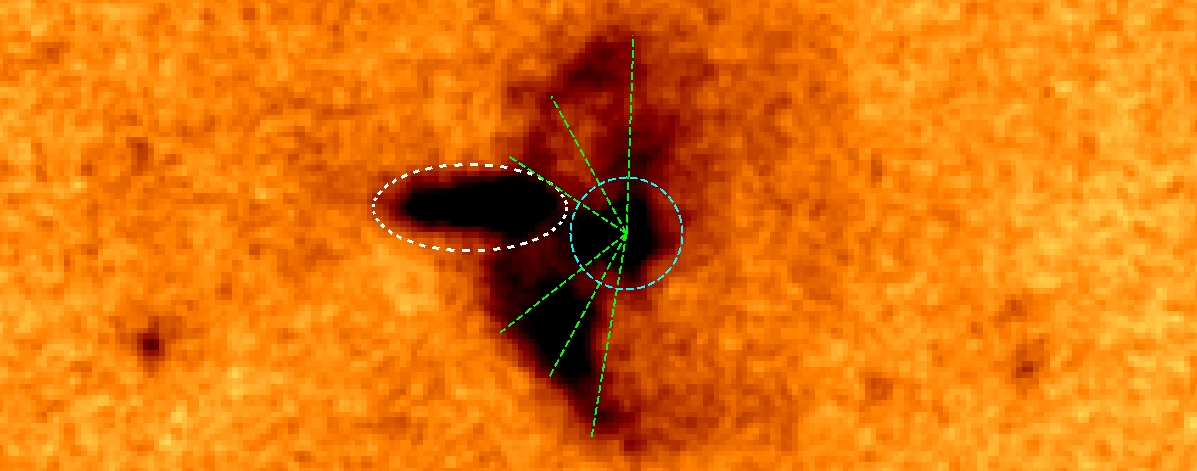}
\end{center}
\caption{A zoom--in ($5.0' \times 2.0'$) of the {\it Herschel}$/$PACS 100~$\mu$m observation of SU~Tau. The white ellipse marks the background galaxy, 2MFGC 4715, while the cyan circle (radius $= 14''$) is centered on SU~Tau. The six rays beginning from the central coordinates of SU~Tau mark extend out to the bow shock are about 30$''$ to 50$''$ in length. North is up and East is left.}
\end{figure}

\begin{figure}%[h!]
\figurenum{14}
\begin{center}
\includegraphics[scale=0.35]{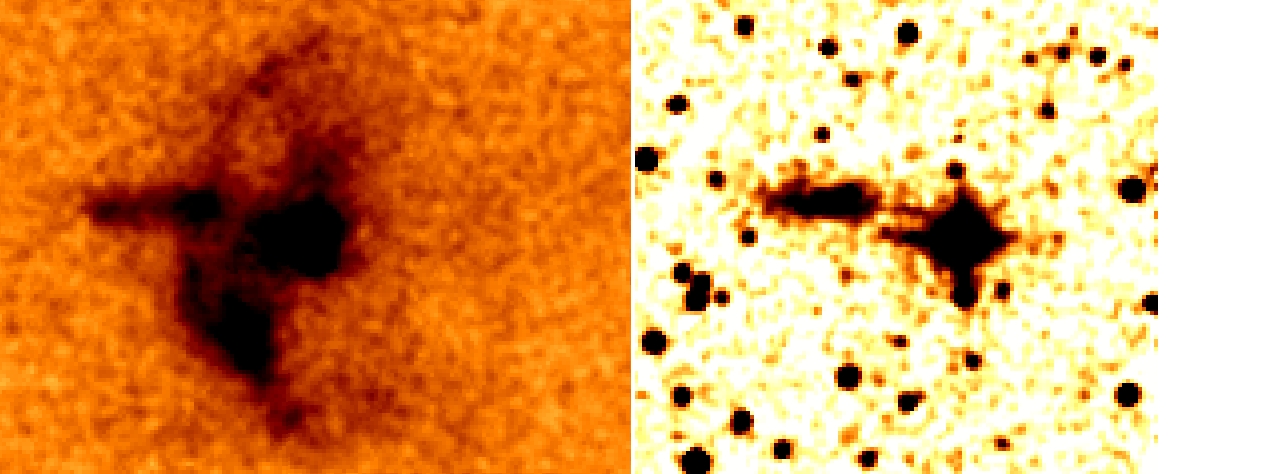}
\end{center}
\caption{Left: A zoom--in ($2.6' \times 2.0'$) of the {\it Herschel}$/$PACS 70~$\mu$m observation of SU~Tau. Right: A $2.0' \times 2.0'$ field of the 2MASS $J$--band tile containing SU~Tau. The bow shock that is prominently seen on the left can be partly seen as a 0.5$'$ vertical line to the East of SU~Tau in the 2MASS observation. North is up and East is left.}
\end{figure}

\begin{figure}%[h!]
\figurenum{15}
\begin{center}
\includegraphics[scale=0.7]{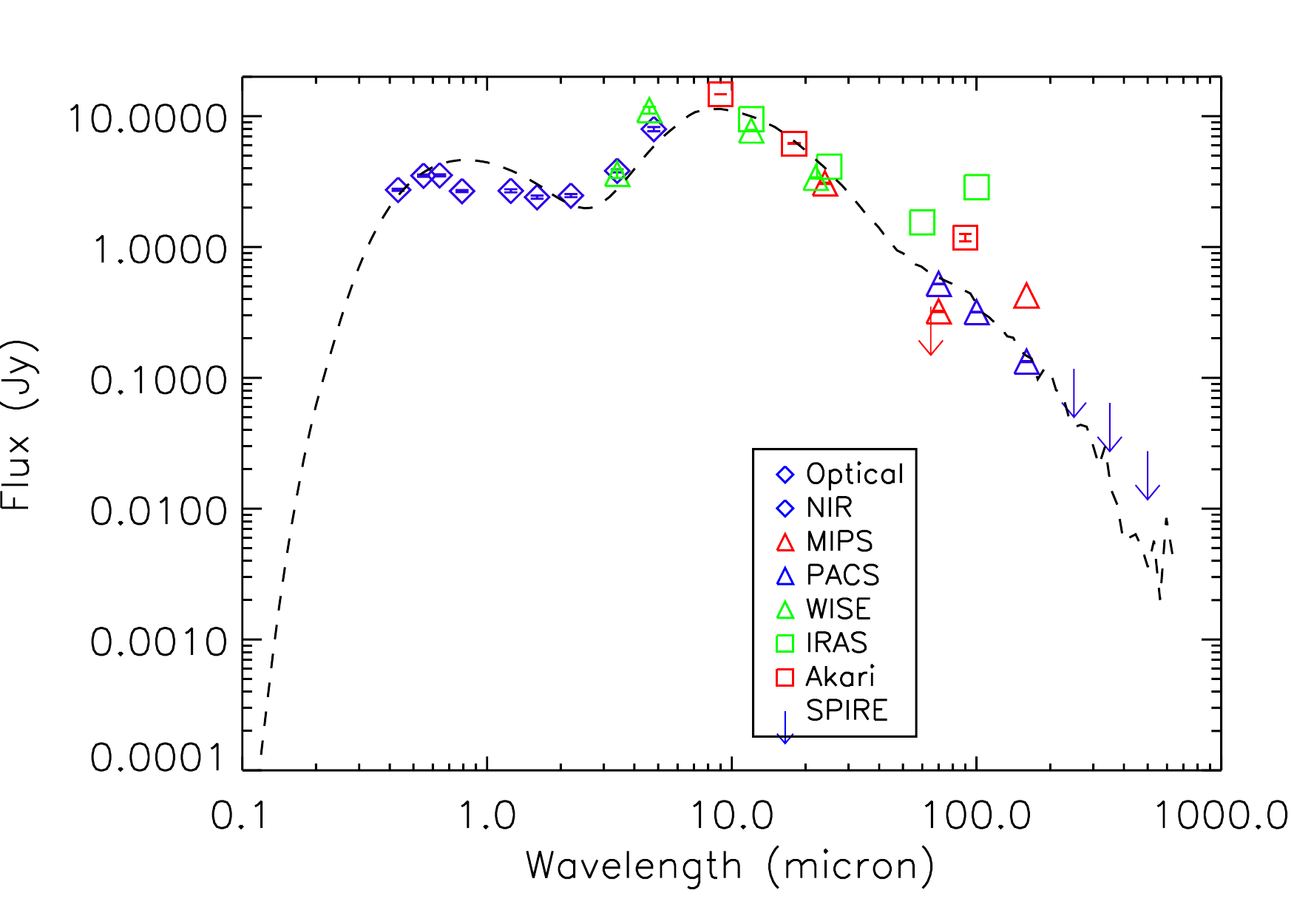}
\end{center}
%\vspace{-30mm}
\caption{The maximum--light SED of SU~Tau. Open blue diamonds: $BVR_{\rm C}I_{\rm C}JHKL$; open green triangles: WISE (3.4, 4.6, 12.0, 22.0~$\mu$m ; open red squares: AKARI (9, 18, 90~$\mu$m), a red arrow represents the AKARI 65~$\mu$m upper limit; open red triangles: {\it Spitzer}$/$MIPS (24, 70, 160~$\mu$m); open green squares: IRAS (12, 25, 60, 100~$\mu$m);  open blue triangles: {\it Herschel}$/$PACS (70, 100, 160~$\mu$m); blue arrows (3$\sigma$): {\it Herschel}$/$SPIRE (250, 350, 500~$\mu$m). The sum of the best--fit MOCASSIN models for the central source, warm, and cold dust shells is represented by the dashed black line.}
\end{figure}

\clearpage

\begin{figure}%[h!]
\figurenum{16}
\begin{center}
\includegraphics[scale=0.4]{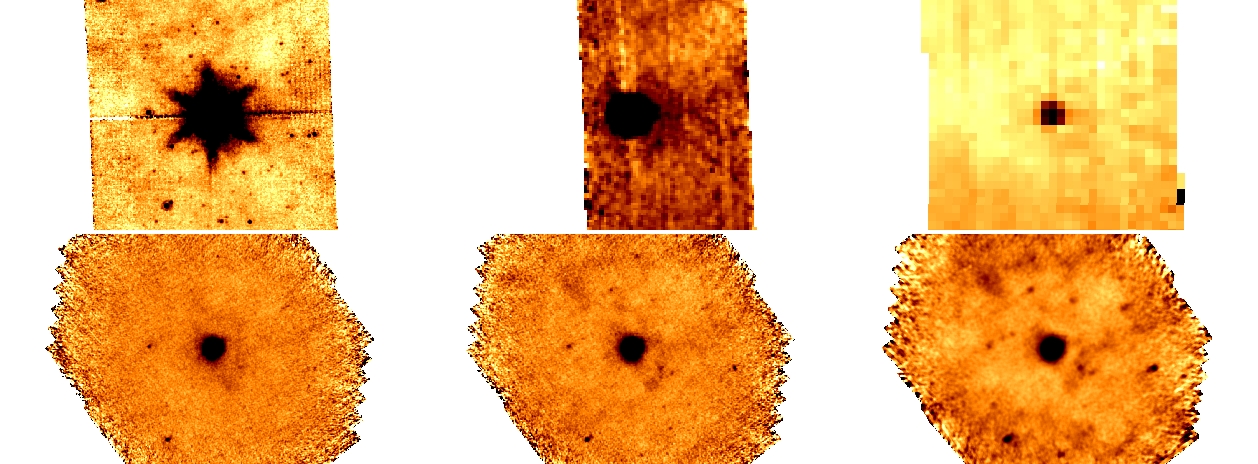}
\end{center}
\caption{Top row:{\it Spitzer}$/$MIPS observations of UW~Cen 24, 70, 160~$\mu$m, respectively. Bottom row: {\it Herschel}$/$PACS observations of UW~Cen at 70, 100, and 160~$\mu$m, respectively. The field--of--view in the {\it Spitzer} images are all $8.0' \times 7.5'$ and in the {\it Herschel} images are all $10.6' \times 7.5'$. North is up and East is left.}
\end{figure}

\begin{figure}%[h!]
\figurenum{17}
\begin{center}
\includegraphics[scale=0.7]{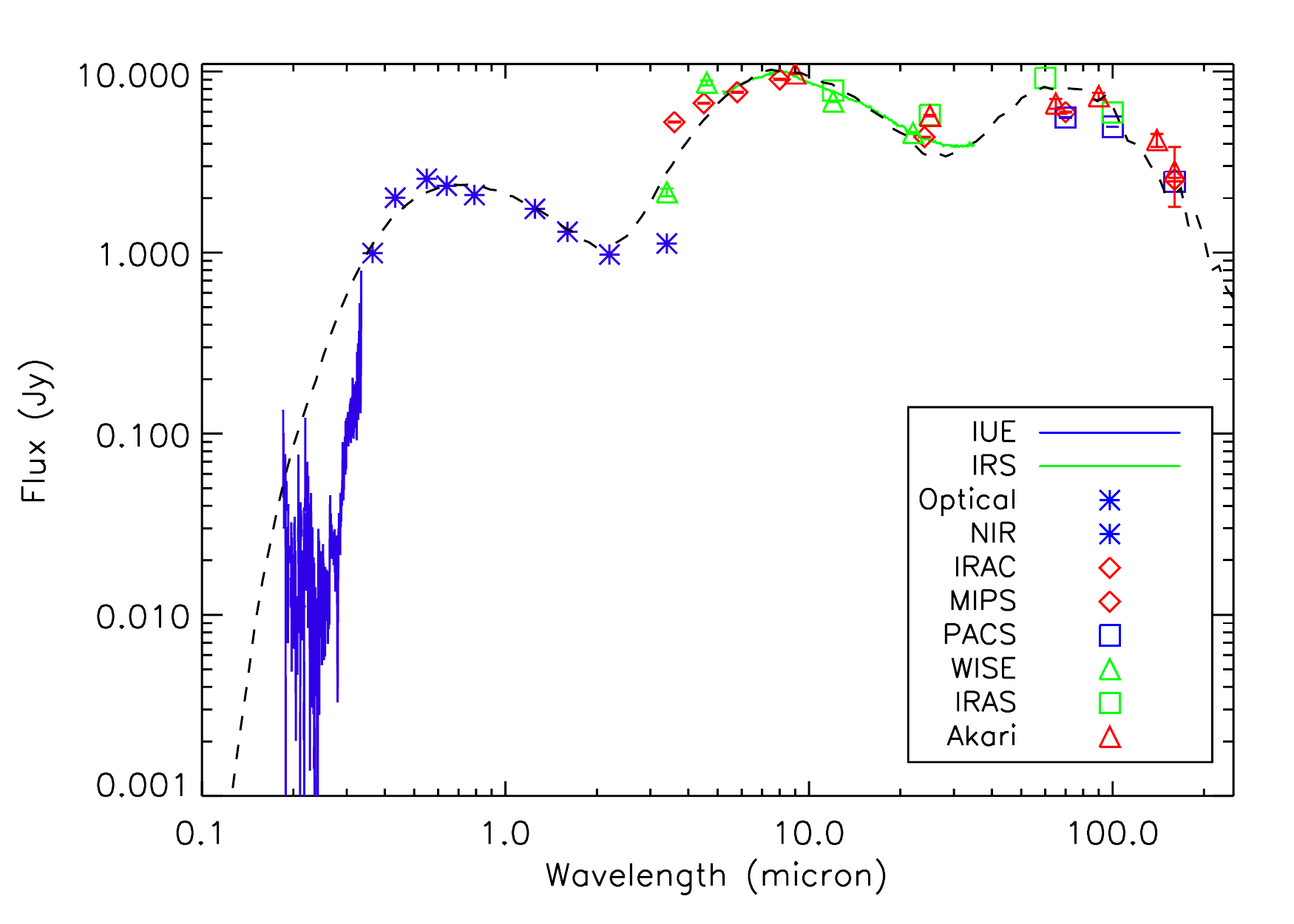}
\end{center}
%\vspace{-30mm}
\caption{The maximum--light SED of UW~Cen. Blue line: IUE spectrum; blue asterisks: $UBVR_{\rm C}I_{\rm C}JHKL$; open red diamonds: {\it Spitzer}$/$IRAC (3.6, 4.5, 5.8, 8.0~$\mu$m) and {\it Spitzer}$/$MIPS (24, 70, 160~$\mu$m); open green triangles: WISE (3.4, 4.6, 12.0, 22.0~$\mu$m); green line: {\it Spitzer}$/$IRS spectrum; open green squares: IRAS (12, 25, 60, 100~$\mu$m); open red triangles: AKARI (9, 25, 65, 90, 140 and 160~$\mu$m); open blue squares: {\it Herschel}$/$PACS (70, 100, 160~$\mu$m). The sum of the best--fit MOCASSIN models for the central source, warm, and cold dust shells is represented by the dashed black line.}
\end{figure}

\clearpage

\begin{figure}%[h!]
\figurenum{18}
\begin{center}
\includegraphics[scale=0.35]{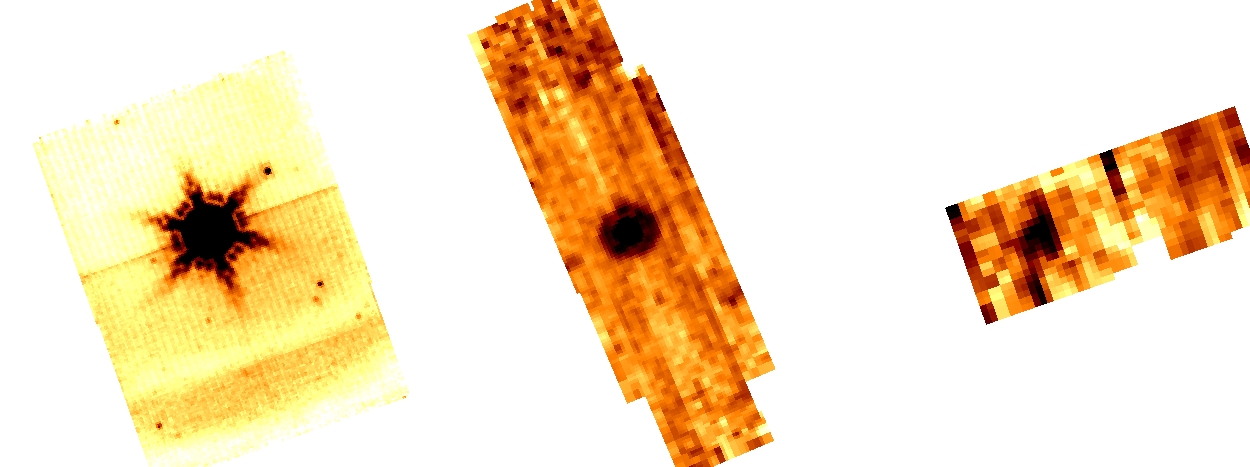}
\end{center}
\caption{The {\it Spitzer}$/$MIPS observations of V854~Cen at 24, 70, 160~$\mu$m, respectively. The field--of--views shown (not accounting for white space) are $6.0' \times 7.8'$, $3.4' \times 7.0'$, and $5.4' \times 2.7'$, respectively. North is up and East is left.}
\end{figure}

\begin{figure}%[h!]
\figurenum{19}
\begin{center}
\includegraphics[scale=0.7]{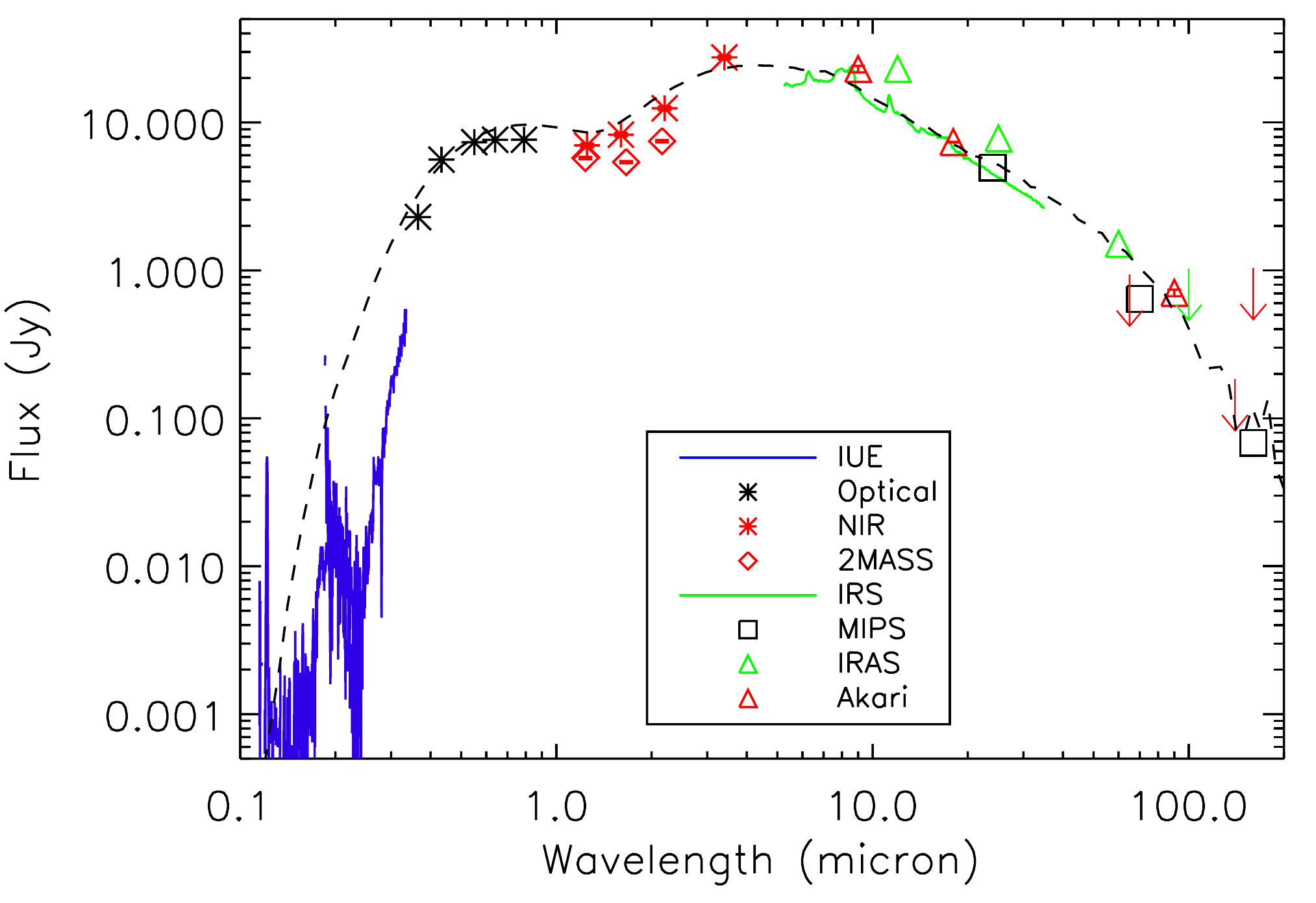}
\end{center}
%\vspace{-30mm}
\caption{The maximum--light SED of V854~Cen. Blue line: IUE spectrum; black asterisks: $UBVR_{\rm C}I_{\rm C}$; red asterisks: $JHKL$; open red diamonds: 2MASS $JHK_{\rm S}$; green line: {\it Spitzer}$/$IRS spectrum; open black squares: {\it Spitzer}$/$MIPS (24, 70, 160~$\mu$m); open green triangles and arrow (3$\sigma$): IRAS (12, 25, 60, 100~$\mu$m);  open red triangles and arrow (3$\sigma$): AKARI (9, 18, 65, 90, 140, 160~$\mu$m). The sum of the best--fit MOCASSIN models for the central source, warm, and cold dust shells is represented by the dashed black line.}
\end{figure}

\clearpage

\begin{figure}%[h!]
\figurenum{20}
\begin{center}
\includegraphics[scale=0.4,angle=90,origin=c]{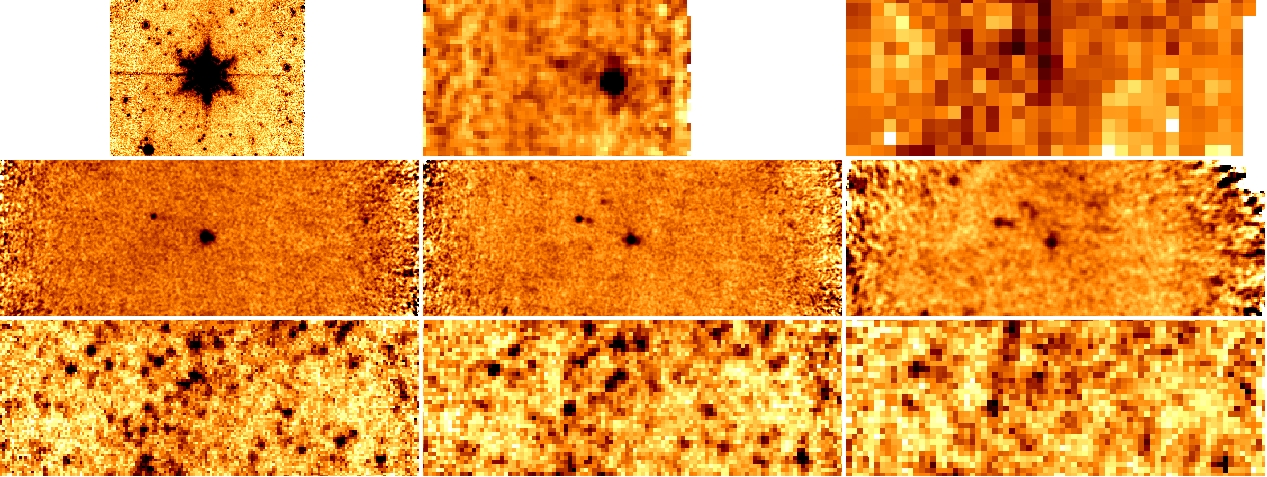}
\end{center}
\caption{First row (starting lower left corner):{\it Spitzer}$/$MIPS observations of V~CrA 24, 70, 160~$\mu$m, respectively. The field--of--view, ignoring white spaces, displayed at 24~$\mu$m is $8.5' \times 6.25'$, at 70~$\mu$m $5.5' \times 3.2'$, and at 160~$\mu$m $8.0' \times 3.1'$. for all three bands is $25' \times 10'$. Middle row: {\it Herschel}$/$PACS observations of V~CrA at 70, 100, and 160~$\mu$m, respectively. The field--of--view in all three 3 columns is $8.5' \times 3.2'$. Bottom row: {\it Herschel}$/$SPIRE observations of V~CrA at 250, 350, and 500~$\mu$m, respectively. The field--of--view for all three bands is $17.1' \times 6.3'$. North is left and East is down.}
\end{figure}

\begin{figure}%[h!]
\figurenum{21}
\begin{center}
\includegraphics[scale=0.7]{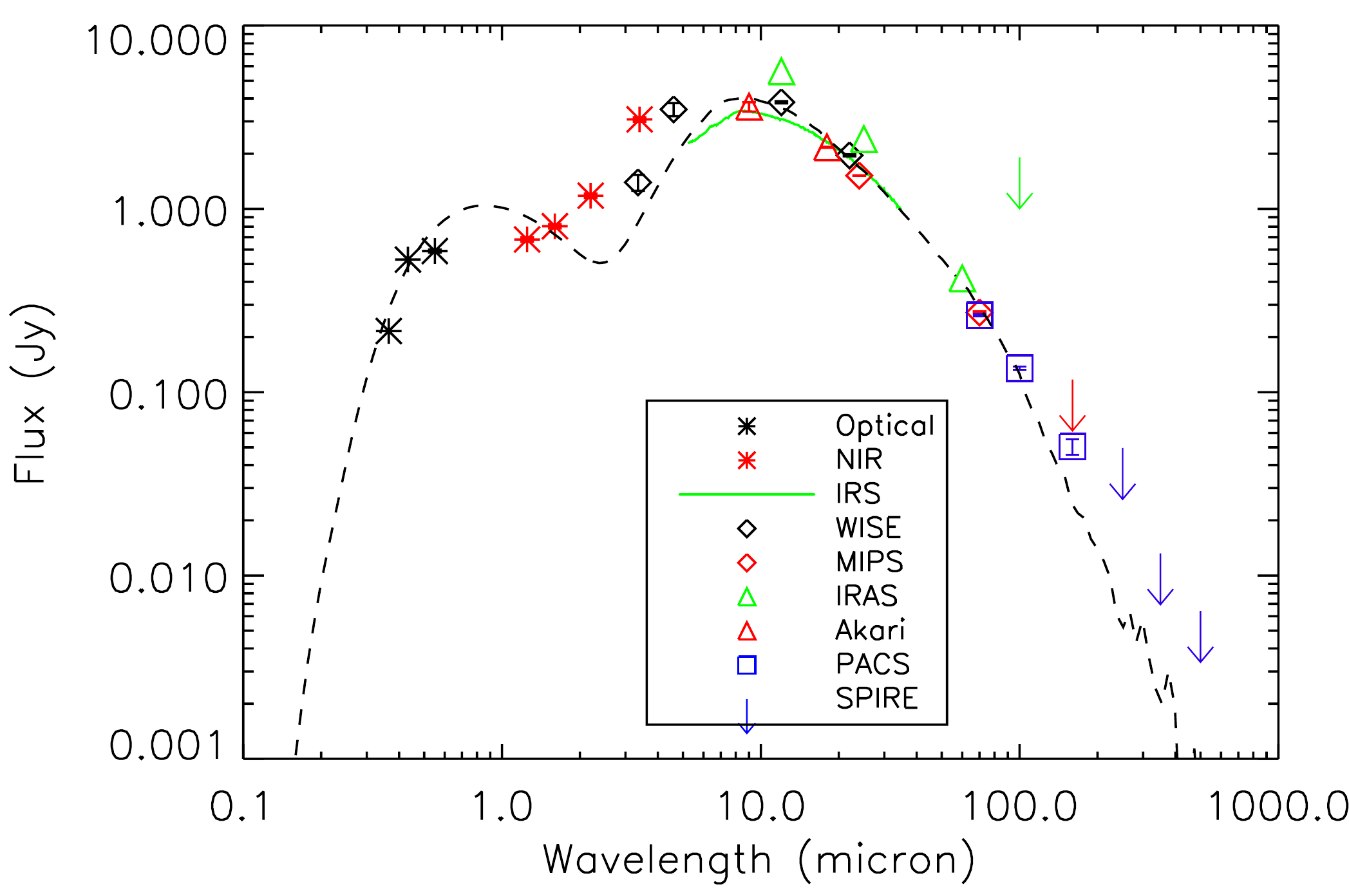}
\end{center}
%\vspace{-30mm}
\caption{The maximum--light SED of V~CrA. Black asterisks: $UBV$; red asterisks: $JHK$; open black diamonds: WISE (3.4, 4.6, 12.0, 22.0~$\mu$m); green line: {\it Spitzer}$/$IRS spectrum; open red diamonds: {\it Spitzer}$/$MIPS (24 and 70~$\mu$m); open green triangles and arrow (3$\sigma$): IRAS (12, 25, 60, 100~$\mu$m);  open red triangle and arrow (3$\sigma$): AKARI (9, 18, 65, 90~$\mu$m); open blue squares: {\it Herschel}$/$PACS (70, 100, 160~$\mu$m); blue arrows (3$\sigma$): {\it Herschel}$/$SPIRE (250, 350, 500~$\mu$m). The sum of the best--fit MOCASSIN models for the central source, warm, and cold dust shells is represented by the dashed black line.}
\end{figure}

\clearpage

\begin{figure}
\figurenum{22}
\begin{center}
\includegraphics{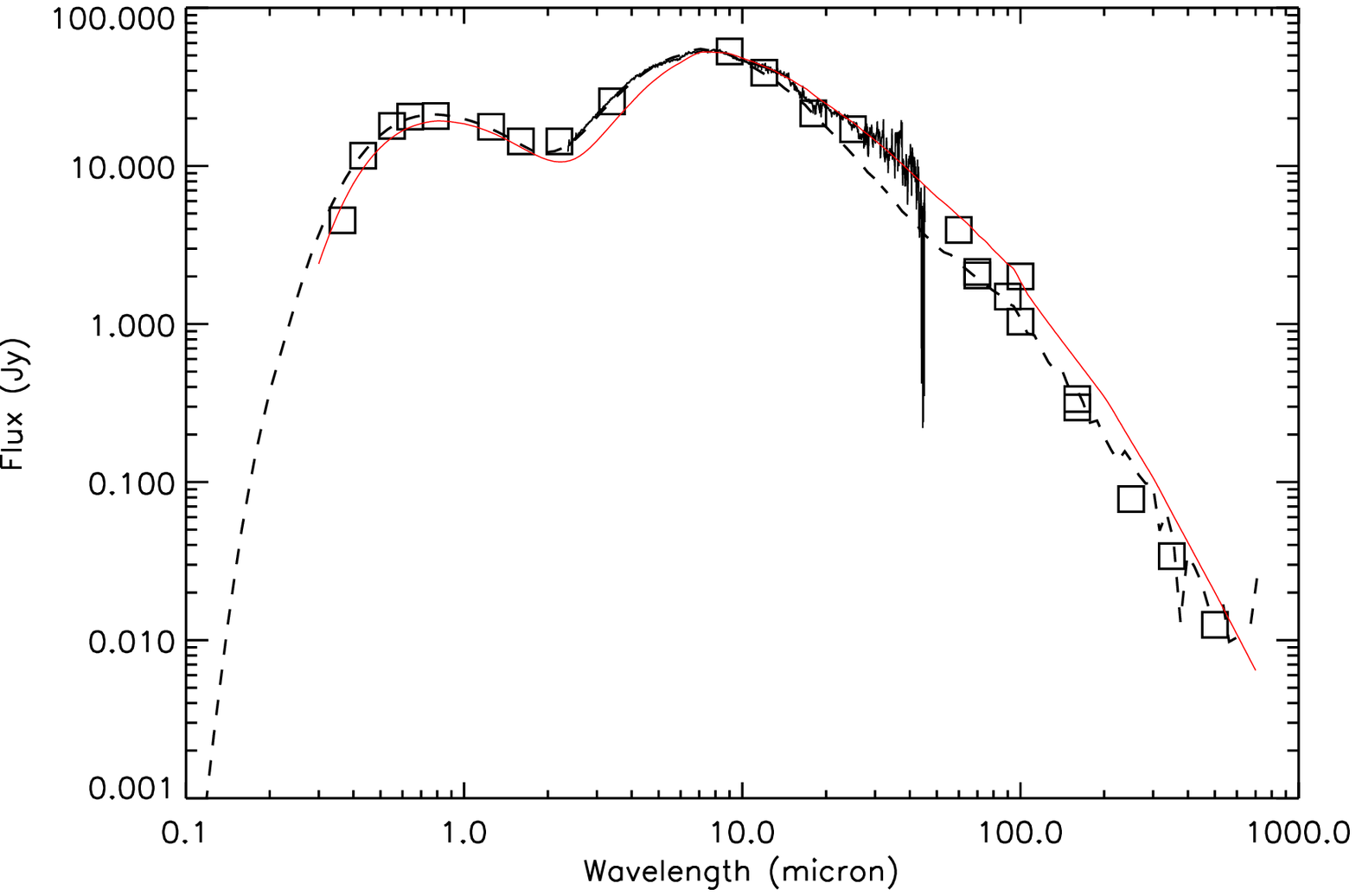}
\end{center}
\caption{The maximum--light SED of R~CrB with the same configuration as Figure 8. The best fit MOCASSIN model is represented by the dashed black line and the corresponding single shell model from QuickSAND is represented by the solid red line. The QuickSAND model does not account for the warmest material and overestimates the contribution of the coldest material.}
\end{figure}

\clearpage

\begin{figure}
\figurenum{23}
\begin{center}
\includegraphics{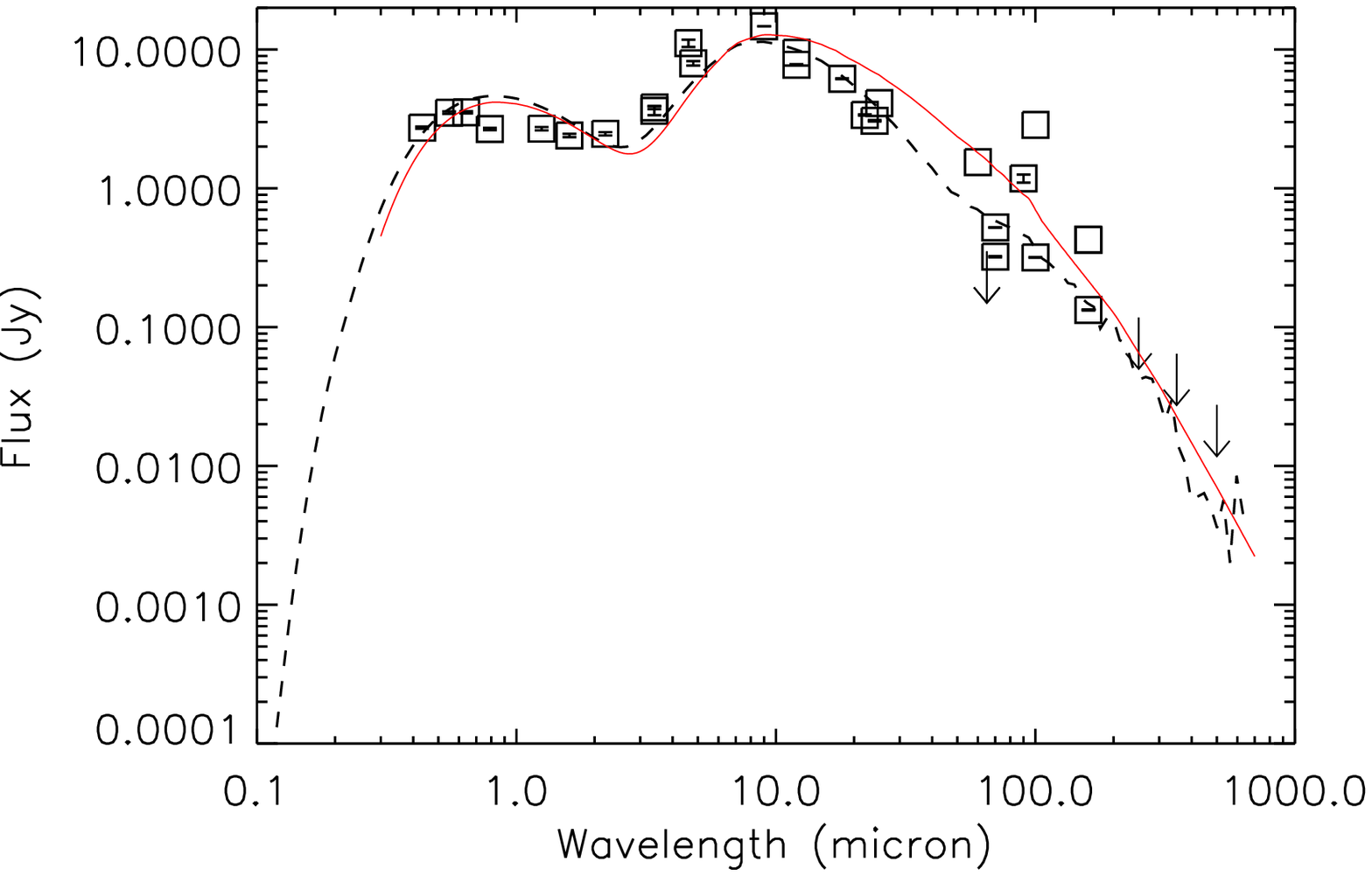}
\end{center}
\caption{The maximum--light SED of SU~Tau with the same configuration as Figure 15. The best fit MOCASSIN model is represented by the dashed black line and the corresponding single shell model from QuickSAND is represented by the solid red line. The QuickSAND model does not account for the warmest material and overestimates the contribution of the coldest material.}
\end{figure}

\clearpage

\begin{figure}
\figurenum{24}
\begin{center}
\includegraphics{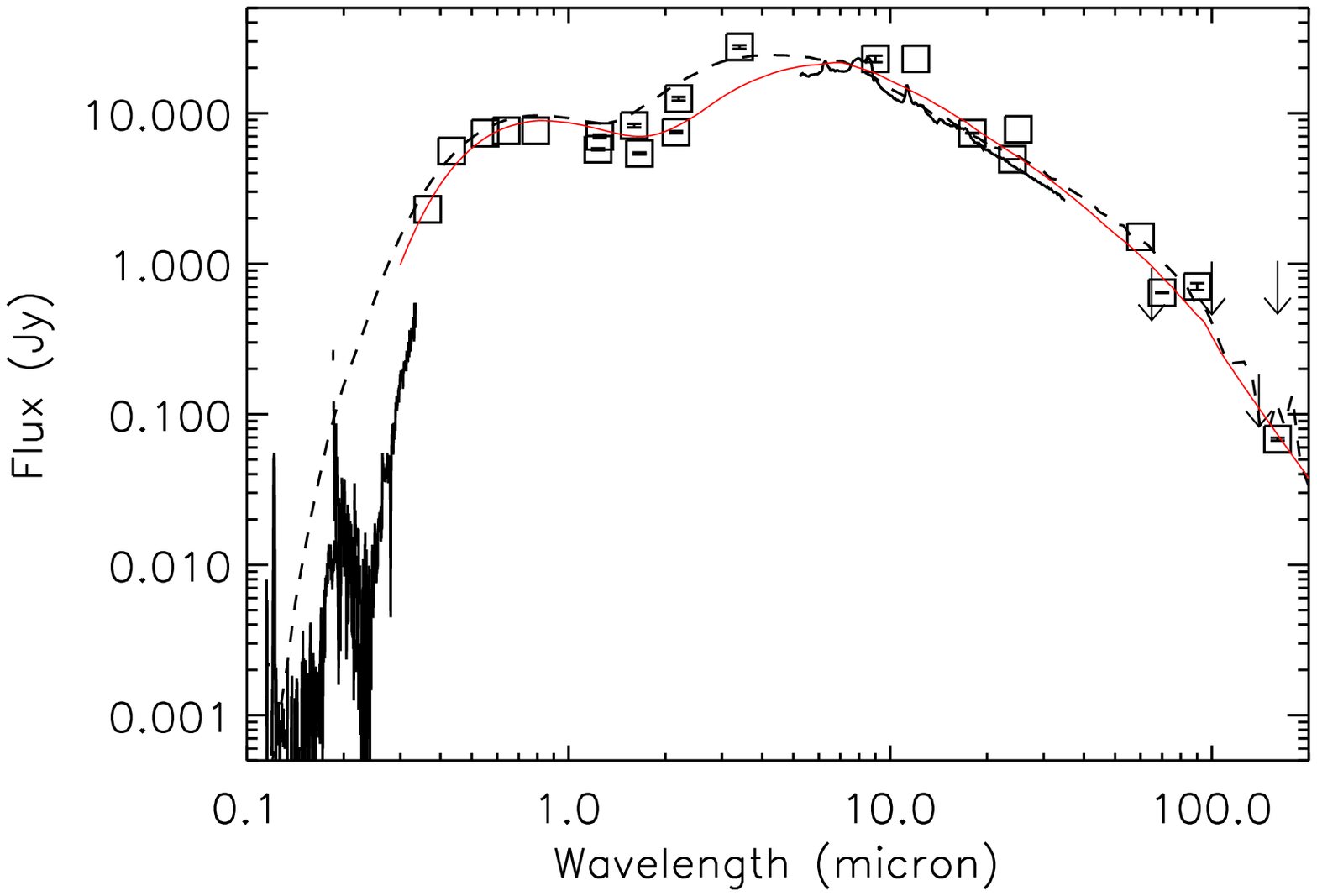}
\end{center}
\caption{The maximum--light SED of V854~Cen with the same configuration as Figure 19. The best fit MOCASSIN model is represented by the dashed black line and the corresponding single shell model from QuickSAND is represented by the solid red line. The QuickSAND model does not account for the warmest material and overestimates the contribution of the coldest material.}
\end{figure}

\clearpage

\begin{figure}
\figurenum{25}
\begin{center}
\includegraphics{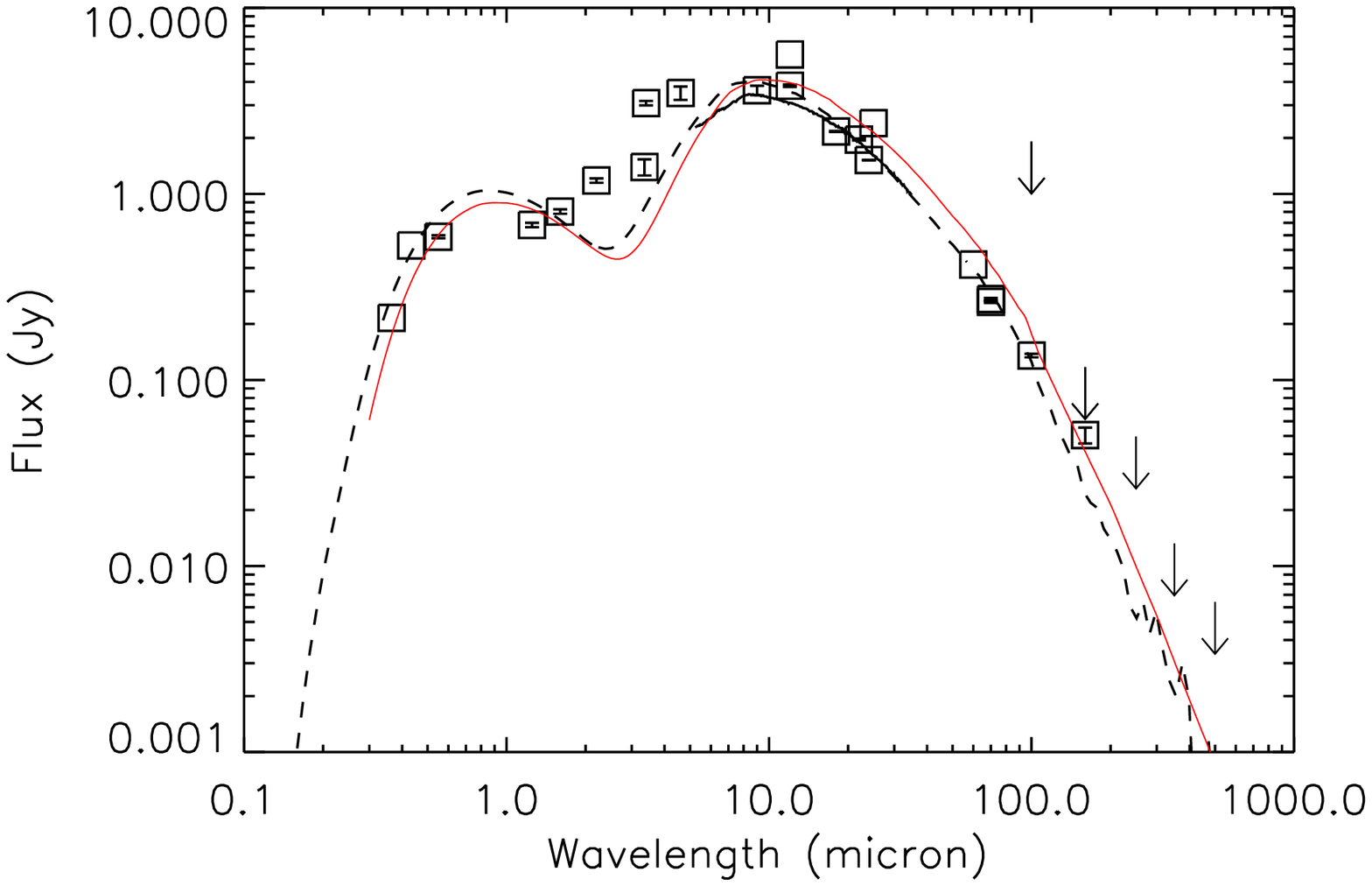}
\end{center}
\caption{The maximum--light SED of V~CrA with the same configuration as Figure 21. The best fit MOCASSIN model is represented by the dashed black line and the corresponding single shell model from QuickSAND is represented by the solid red line. The QuickSAND model does not account for the warmest material and overestimates the contribution of the coldest material.}
\end{figure}

\clearpage

\begin{figure}%[h!]
\figurenum{26}
\begin{center}
\includegraphics[scale=0.8]{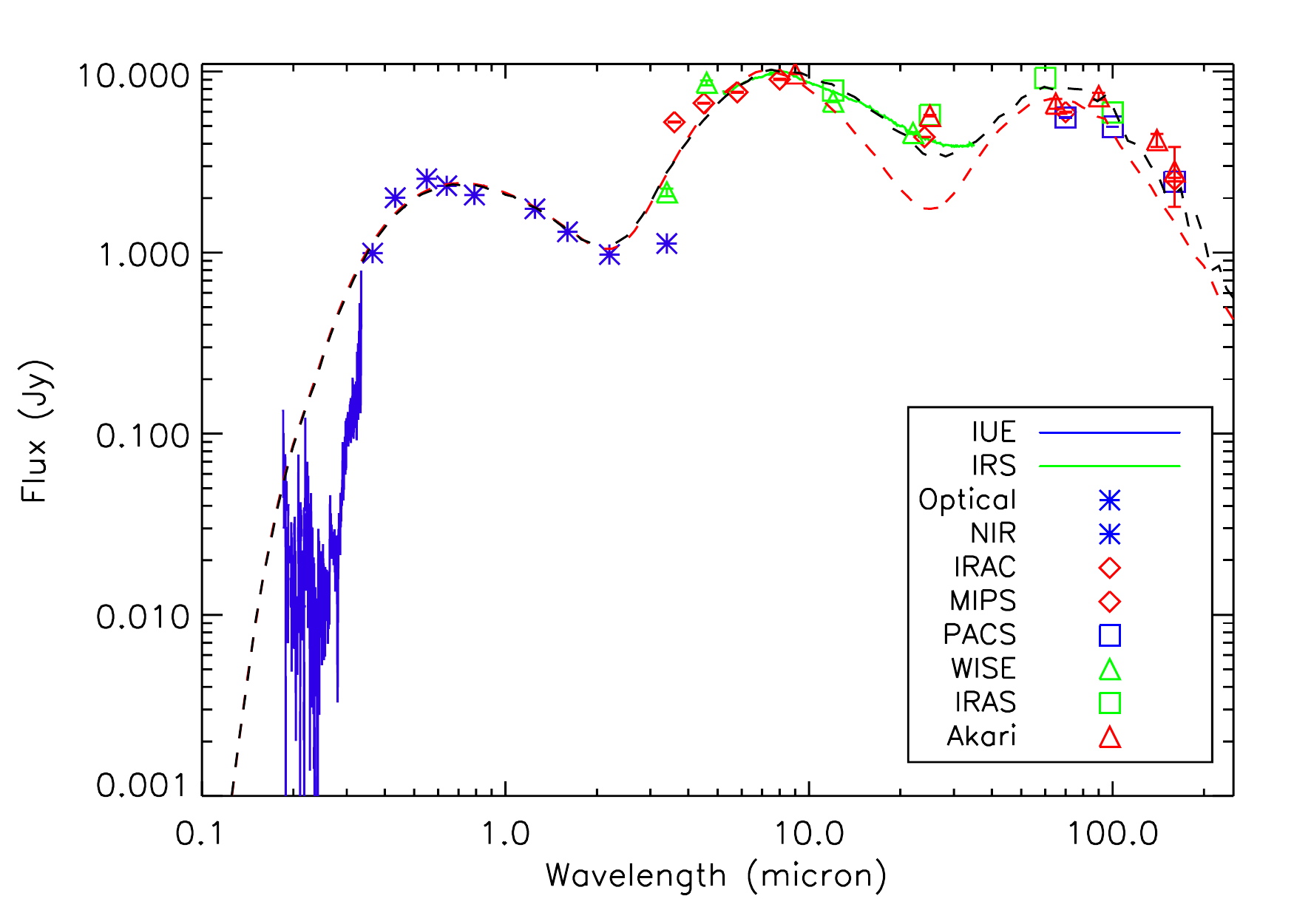}
\end{center}
%\vspace{-30mm}
\caption{The maximum--light SED of UW~Cen with the same configuration as Figure 17. The sum of the best--fit MOCASSIN models for the central source, warm, and cold dust shells is still represented by the dashed black line. The MOCASSIN fit for thin shells (R$_{\rm out} = 2$R$_{\rm in}$ is represented by the red dashed line. The need for thicker shells is apparent in that the SED beyond 10~$\mu$m is a poorer match to the thin shell model.}% larger reservoirs of ``cold'' dust in both the inner and outer envelopes is apparent with The need for larger reservoirs of ``cold'' dust in both the inner and outer envelopes is apparent with the fits to the right of the local maxima being well below the observations.
\end{figure} 

\clearpage

\begin{figure}%[h!]
\figurenum{27}
\begin{center}
\includegraphics[scale=0.7]{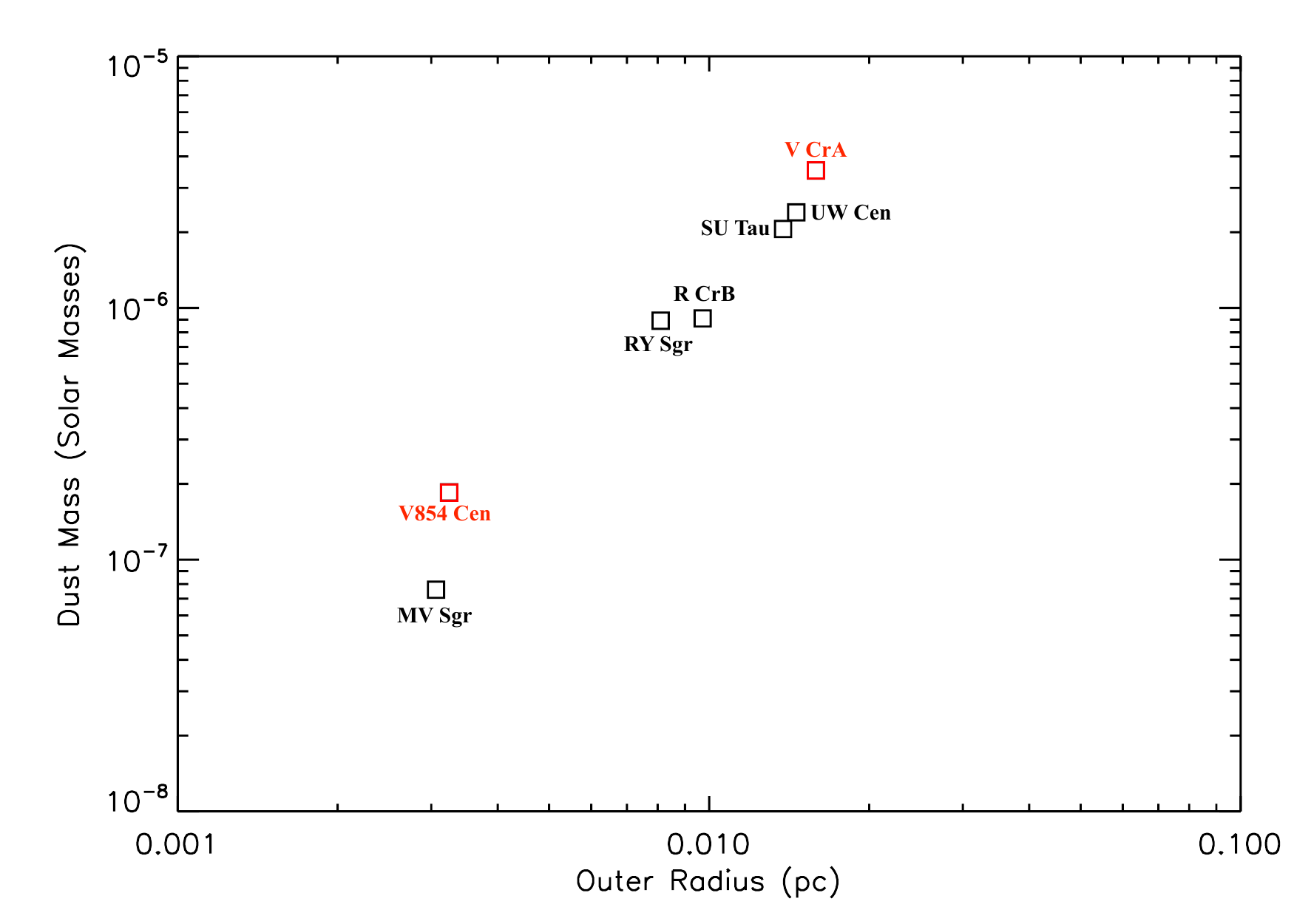}
\includegraphics[scale=0.7]{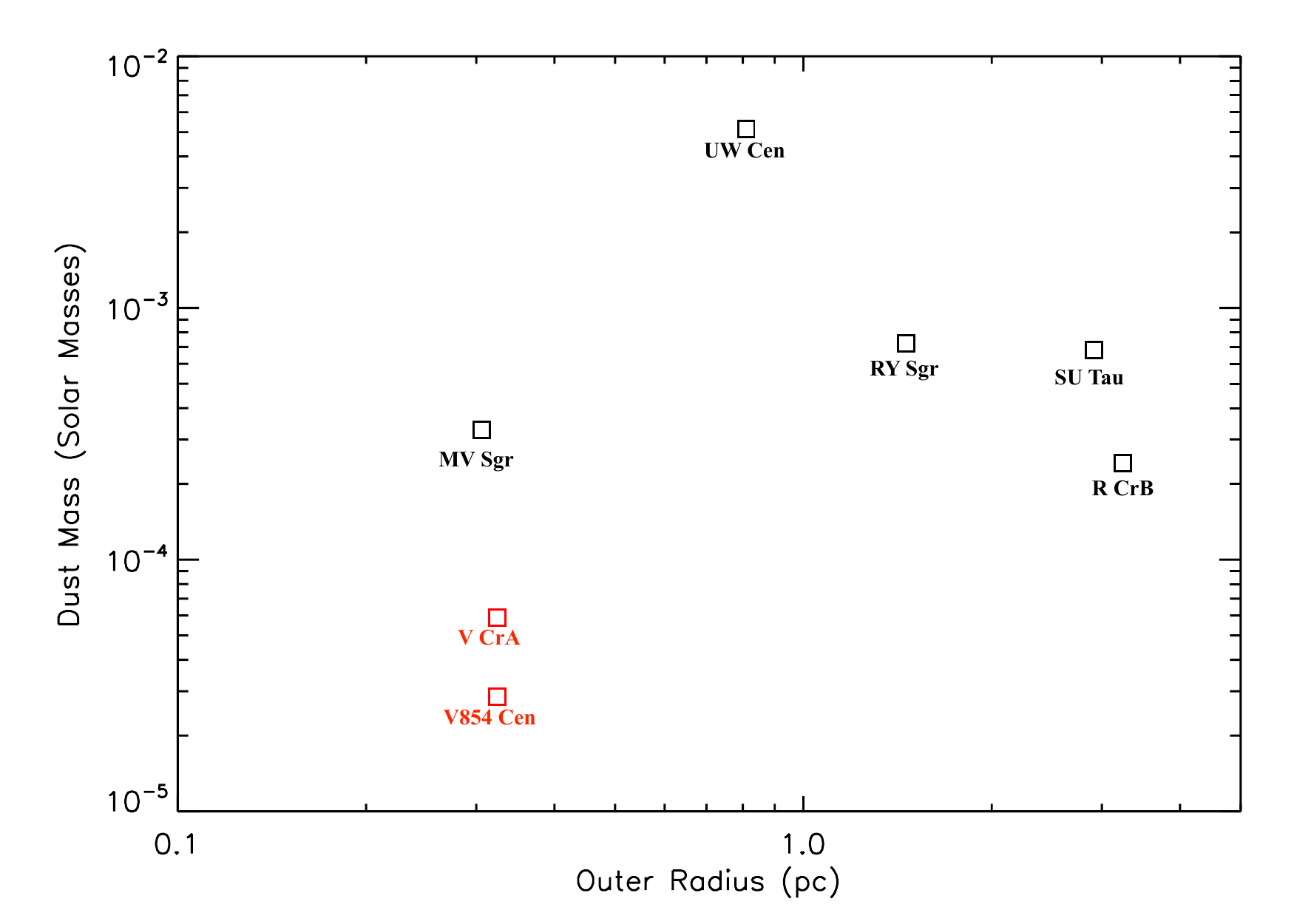}
\end{center}
%\vspace{-30mm}
\caption{Top: Plot comparing the derived dust masses to the outer radii of the modeled warm dust envelope for the sample of RCB stars. Bottom: The same as the first plot but for the cold dust envelopes. Majority RCB stars are represented by black squares while minority RCB stars by red squares. The apparent linear trend for the inner envelopes is tied more to the outer radii than a true relationship between dust mass and shell size. No obvious trend arises for the outer envelopes. The minority RCB stars were best fit by smaller, less massive shells than the majority RCB stars for cold envelopes.} %comparing the derived dust masses to the outer radii of the modeled cold dust envelope for the sample of RCB stars. 
\end{figure} 

\clearpage

\begin{figure}%[h!]
\figurenum{28}
\begin{center}
\includegraphics[scale=0.7]{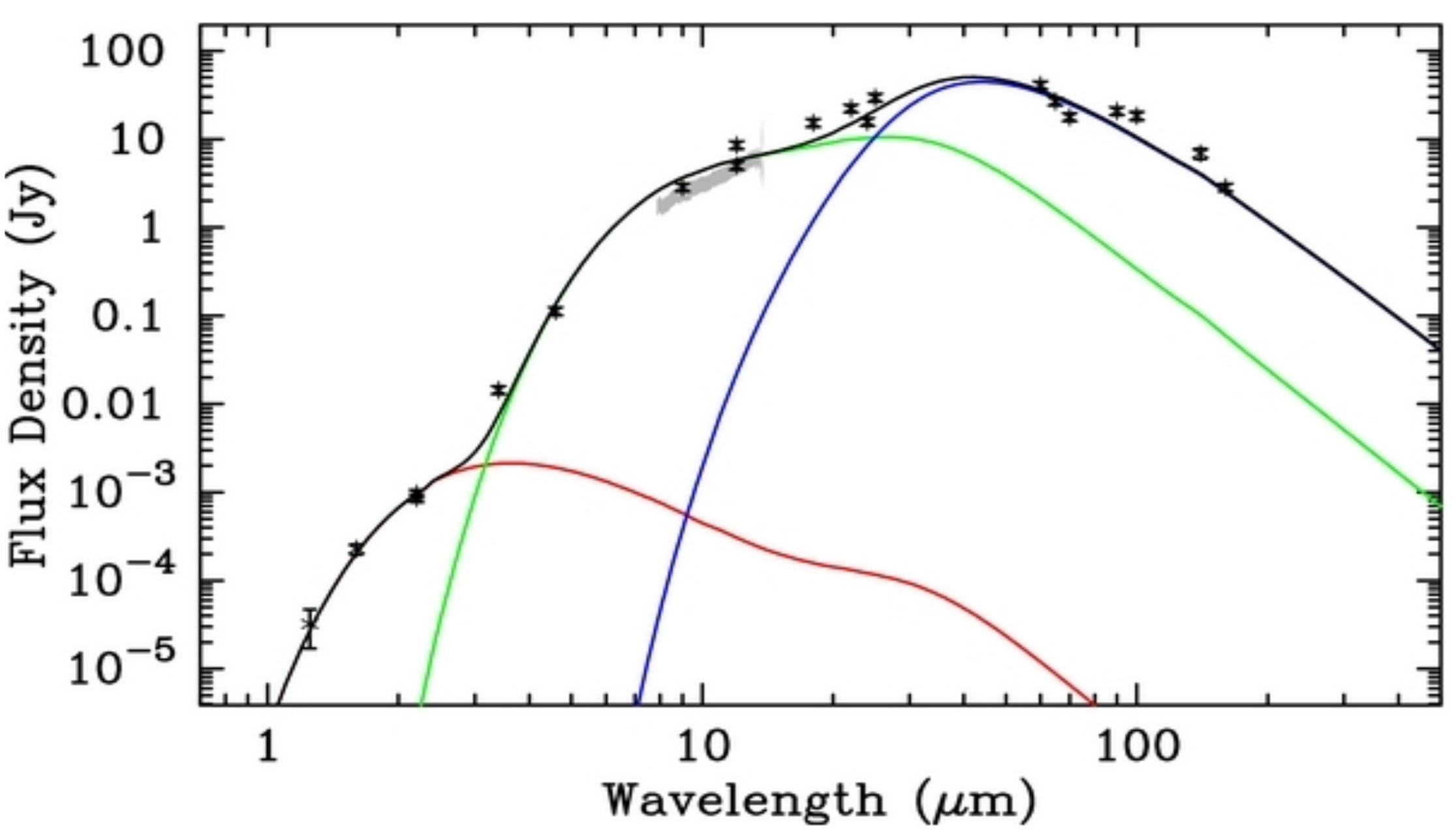}
\end{center}
%\vspace{-30mm}
\caption{The V605~Aql SED presented by \citet{2013ApJ...771..130C}. The black symbols represent photometry from ground--based $JHK$ \citep{2008A&A...479..817H}, WISE, AKARI, IRAS, and {\it Spitzer}$/$MIPS. The gray line is a spectrum observed with the MID--infrared Interferometric instrument (MIDI) at the Very Large Telescope Interferometer (VLTI). The photometry presented by Clayton et al. are provided Table 4.9. The red, green, and blue lines are individual fits using the emission curves of amorphous carbon dust with temperatures of 810 K, 235 K, and 75 K, respectively. The black line is the sum of the three fits.}
\end{figure}

\clearpage

\begin{figure}%[h!]
\figurenum{29}
\begin{center}
\includegraphics[scale=0.4]{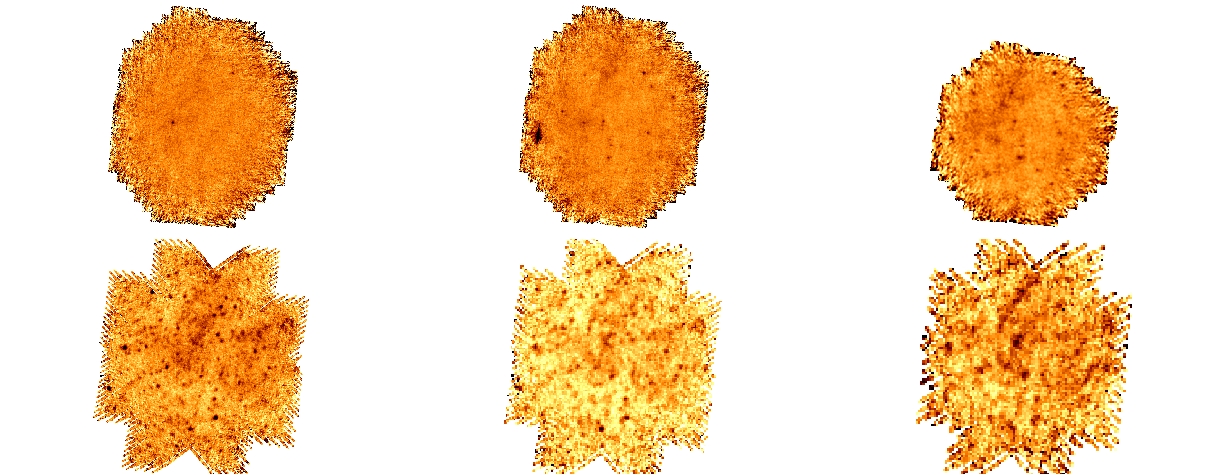}
\end{center}
\caption{Top row: {\it Herschel}$/$PACS observations of HD~173409 at 70, 100, and 160~$\mu$m, respectively. The 70 and 100~$\mu$m fields are both $9' \times 11'$, while the 160~$\mu$m field is $9' \times 9'$. Bottom row: {\it Herschel}$/$SPIRE observations of HD~173409 at 250, 350, and 500~$\mu$m, respectively. The fields are all $20' \times 20'$. The lack of a point source or any nebulosity centered on the position of HD~173409 is consistent with HdC stars having no IR excess.}
\end{figure}

\begin{figure}%[h!]
\figurenum{30}
\begin{center}
\includegraphics[scale=0.7]{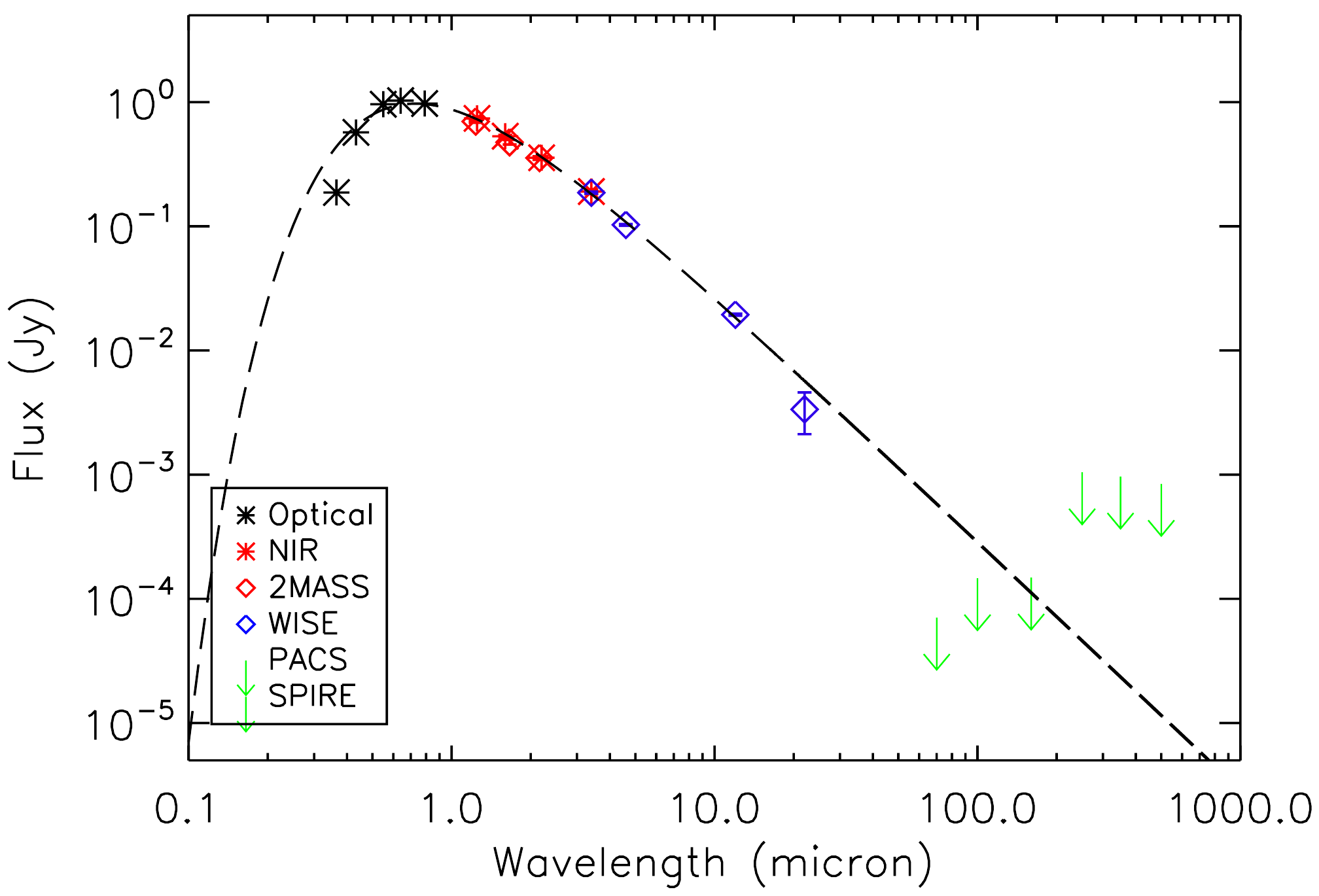}
\end{center}
\caption{The HD~173409 SED. Black asterisks: $UBVR_{\rm C}I_{\rm C}$; red asterisks: $JHKL$; open blue diamonds: WISE (3.4, 4.6, 12.0, 22.0~$\mu$m); green arrows: {\it Herschel} PACS and SPIRE 3$\sigma$ upper limits.}
\end{figure}

\end{document}